%% file: thesis.tex
\documentclass[a4paper,12pt,numbered,onesided,print,index,fourier,custombib]{PhDThesisPSnPDF}

\input{Preamble/preamble}

\input{thesis-info}

\ifdefineAbstract
\fi

\begin{document}

\frontmatter

\maketitle

\include{Dedication/dedication}


\include{Acknowledgement/acknowledgement}

\include{Statement/statement}

\include{Abstract/abstract}


\tableofcontents

\listoffigures



\printnomenclature

\mainmatter

\part{Background material}
\include{Chapter1/chapter1}
\include{Chapter2/chapter2}

\part{Research work}
\include{Chapter3/chapter3}
\include{Chapter4/chapter4}
\include{Chapter5/chapter5}
\include{Chapter6/chapter6}


\begin{spacing}{0.9}


\bibliographystyle{habbrv}
\cleardoublepage
\bibliography{References/references} 



\end{spacing}

\part{Extra material}

\begin{appendices} 

\include{Appendix1/appendix1}
\include{Appendix3/appendix3}

\end{appendices}

\printthesisindex 

\end{document}

%% file: thesis-info.tex
\title{Primordial black hole formation processes with full numerical relativity}

\author{Eloy de Jong}

\dept{Department of Physics}

\university{King's College London}
\crest{\includegraphics[width=0.45\textwidth]{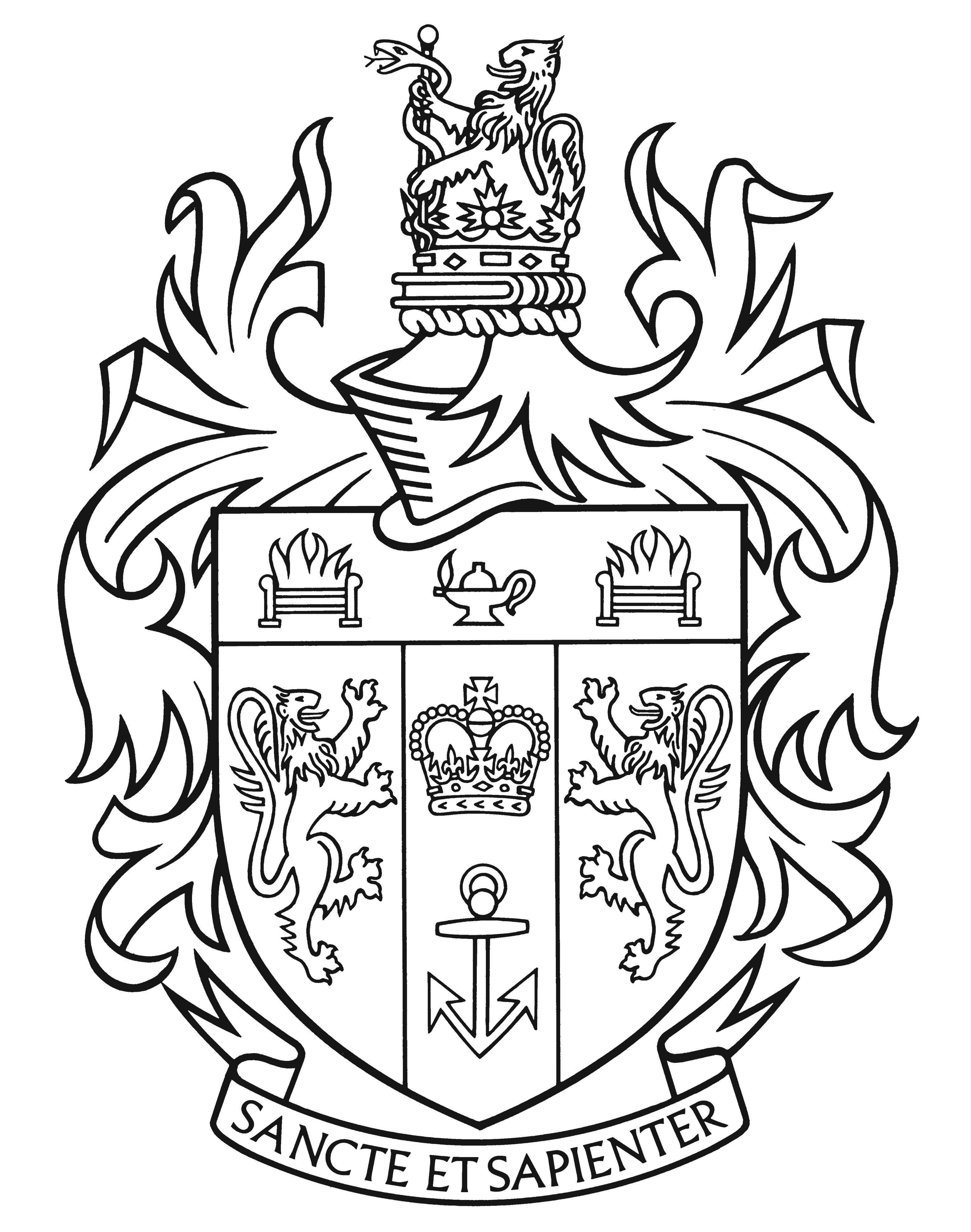}}

\supervisor{Prof. Eugene A. Lim}

\degreetitle{Doctor of Philosophy}

\degreedate{8 February 2024} 

\subject{LaTeX} \keywords{{LaTeX} {PhD Thesis} {Physics} {King's College London}}

%% file: Dedication/dedication.tex

\begin{dedication} 

\vspace*{\stretch{0.5}}
\itshape             
\raggedleft          

The more I learn, the less I know.

\vspace{\stretch{3}}

\end{dedication}

%% file: Acknowledgement/acknowledgement.tex

\begin{acknowledgements}
    \hspace{15pt} My sincerest gratitude goes out to Eugene Lim, my supervisor, for believing in me and for mentoring and guiding me throughout my PhD. 
    I admire you as a researcher and as a person, and 
    I am so grateful for your support. Thank you. \newline

    I cannot thank my friend and collaborator Josu Aurrekoetxea enough for taking me under his wings, for his endless patience and meticulousness in answering my continuous barrage of questions and for helping me become the researcher I am today. \newline

    Thank you to the past and present members of TPPC, Alex, Angelo, Ankit, Ansh, Bo-Xuan, Charlie, Claire, Damon, Drew, Giuseppe, James, John, Josu, Katarina, Leonardo, Liina, Louis, Matt, Nikos, Nicole, Panos, Shiqian, Silvia, Wenyuan and other members of faculty, postdocs and PhDs, for making King's both an intellectually stimulating and enjoyable place to do research.\newline

    I am indebted to the entire GRChombo collaboration, whose members have been such a pleasure to interact with and have collectively shaped my understanding of numerical relativity. I cannot imagine a more supportive and kinder group of researchers. \newline

    I am grateful for the close friends I have made during my time at Gymnasium Beekvliet, Amsterdam University College, the University of Cambridge and King's College London. They have kept me happy and sane throughout my entire academic journey. 
    
    Thank you to Bas, Ben, David, Eva, Flip, Frankie, Guus, Harry, Jari, Jef, Jelmer, Jelmer, Kasper, Kees, Lara, Laurens, Lia, Louise, Mark, Niels, Renske, Victor, Willem, Yoli and others.\newline

    Finally, thank you to my parents Kees en Annemarie and my sisters Nuria and In\'es, for their never-ending support and encouragement.
\end{acknowledgements}

%% file: Statement/statement.tex

\begin{statement}
    \hspace{15pt} I declare that the thesis has been composed by myself and that the work has not be submitted for any other degree or professional qualification. I confirm that the work submitted is my own, except where work that has formed part of jointly-authored publications has been included. My contribution and those of the other authors to this work have been explicitly indicated below. I confirm that appropriate credit has been given within this thesis where reference has been made to the work of others.

    The work presented in Chapter \ref{Chapter3} was previously published in
    \textit{Journal of Cosmology and Astroparticle Physics (JCAP)} as \textit{Primordial black hole formation with full numerical relativity} by Eloy de Jong, Josu C. Aurrekoetxea and Eugene A. Lim. I wrote this work myself. 

    The work presented in Chapter \ref{Chapter4} was previously published in
    \textit{Journal of Cosmology and Astroparticle Physics (JCAP)} as \textit{Spinning primordial black holes formed during a matter-dominated era} by Eloy de Jong, Josu C. Aurrekoetxea, Eugene A. Lim. and Tiago Fran\c{c}a. I wrote this work myself, except for the discussion of the apparent horizon (AH) in section \ref{Ssect::diagnostics} and appendix \ref{Sapp::AH}, which was written by Tiago Fran\c{c}a.
\end{statement}

%% file: Abstract/abstract.tex
\begin{abstract}
\addcontentsline{toc}{chapter}{Abstract}

Primordial black holes (PBHs) can form in the early universe, and there are several mass windows in which their abundance today may be large enough to comprise a significant part of the dark matter density. Additionally, numerical relativity (NR) allows one to investigate the formation processes of PBHs in the fully nonlinear strong-gravity regime. In this thesis, we will describe the use of NR methods to study PBH formation, motivated in particular by open questions about the nonspherical effects PBH formation in a matter-dominated early universe. 

We demonstrate that superhorizon non-linear perturbations can collapse and form PBHs via the direct collapse or the accretion collapse mechanisms in a matter-dominated universe. The heaviest perturbations collapse via the direct collapse mechanism, while lighter perturbations trigger an accretion process that causes a rapid collapse of the ambient DM. From the hoop conjecture we propose an analytic criterion to determine whether a given perturbation will collapse via the direct or accretion mechanism and we compute the timescale of collapse. Independent of the formation mechanism, the PBH forms within an efold after collapse is initiated and with a small initial mass compared to the Hubble horizon, $\mbh H_0\sim 10^{-2}\mpl^2$. Finally, we find that PBH formation is followed by extremely rapid growth $M_\textrm{BH}\propto H^{-\beta}$ with $\beta\gg 1$, during which the PBH acquires most of its mass.

Furthermore, we study the formation of spinning primordial black holes during an early matter-dominated era. Using non-linear 3+1D general relativistic simulations, we compute the efficiency of mass and angular momentum transfer in the process -- which we find to be $\mathcal{O}(10\%)$. We show that subsequent evolution is important due to the seed PBH accreting non-rotating matter from the background, which decreases the dimensionless spin.  Unless the matter era is short, we argue that the final dimensionless spins will be negligible. 

Finally, we discuss the high computational cost of NR simulations and how these can be remedied in specific scenarios using dimensional reduction. We extend the modified cartoon method for the BSSN formalism by adding matter fields, specifically a real scalar field, and give explicit cartoon expressions for the evolution equations. Additionally, we give cartoon expressions for dimensional reduction of the CCTK method for finding NR initial conditions. We discuss the true vacuum bubble collision in the context of first-order phase transitions as a specific application of this work and show that our method provides continued stable numerical evolution of this process. 

\vspace{-0.75cm}
\end{abstract}

%% file: Chapter1/chapter1.tex

\chapter{Introduction}\label{Chapter1}  

\graphicspath{{Chapter1/Figs/}}

There are three main parts to this thesis: background material, research work and extra material. The background material consists of chapters \ref{Chapter1} and \ref{Chapter2}, in which we present key background information. In chapter \ref{Chapter1}, we review general relativity (GR) \cite{Einstein:1916} and how it is used to describe black holes (BHs) and cosmological spacetimes. We also introduce primordial black holes (PBHs). In chapter \ref{Chapter2}, we discuss the key technical details of numerical relativity (NR), the main method used for the research presented in this work. Our research work is discussed in chapters \ref{Chapter3}, \ref{Chapter4} and \ref{Chapter5} and these chapters make up the core of this thesis. In chapter \ref{Chapter3}, we study the collapse of spherically symmetric sub- and superhorizon overdensities in a matter-dominated early universe and the subsequent collapse into PBHs. In chapter \ref{Chapter4}, we study the collapse of overdensities with angular momentum in a matter-dominated early universe. In chapter \ref{Chapter5}, we discuss dimensionally reduced NR simulations with matter fields. Finally, the extra material contains supplementary material that has been omitted from the main text.

\section{Primordial black holes with NR}

The demand for NR simulations of PBH formation mechanisms is motivated by various scientific questions. Firstly, NR is the most accurate method of determining the PBH formation threshold, discussed in section \ref{section::PBH_threshold}, regardless of the exact equation of state of the universe at PBH formation. Conventionally, simulations investigating the threshold focus on spherically symmetric scenarios in a radiation-dominated universe, since the high overdensities needed to form PBHs in this setting tend to be spherically symmetric, as mentioned in section \ref{sect::PBHs}. 

Another mechanism studied using NR is PBH critical collapse, referenced in section \ref{sect::PBHs}, which concerns the study of the collapse of overdensities that just satisfy the collapse threshold. Close to the threshold, there is a scaling law such that $M_\textrm{BH} \propto (\delta - \delta_c)^\gamma$, where $\gamma \approx 0.36$. This means e.g. that it becomes possible to form PBHs much smaller than the Hubble horizon at formation, even in a radiation-dominated universe, where that is usually forbidden because of the large Jeans length. 

PBH critical collapse, like collapse threshold determination, is usually simulated under the assumption of spherical symmetry, using dimensionally reduced codes. However, in both cases it is interesting to consider how the picture changes when deviations away from spherical symmetry are taken into account. For instance, the effect of ellipticity on PBH formation in a radiation-dominated universe is expected to be small \cite{Yoo:2020lmg}, but when the equation of state becomes sufficiently soft so that pressure effects become negligible, nonspherical effects start to dominate over the Jeans criterion. In this case, a small deviation from spherical symmetry can be amplified during collapse \cite{Lin:1965,Saenz:1978}, causing collapse into a ``pancake'' or ``line'' rather than a point. Spherically symmetric simulations are not suitable to model this type of collapse and it is unclear how to apply the hoop conjecture, see \eqn{eqn::hoop_conjecture}, in such a scenario, so that full 3+1D NR simulations are required to shed light on this topic. Full 3+1D NR is also needed to investigate the effects of non-sphericity on critical collapse, see e.g. \cite{Baumgarte:2016xjw} and references therein. 

Furthermore, as alluded to in section \ref{sect::PBHs}, it is important to predict the spin distribution of PBHs to connect current GW observations and PBH formation scenarios. Spins of PBHs formed during a radiation-dominated era are expected to be small, but the softer the equation of state, the more likely it becomes that nonspherical effects lead to nonzero PBH spins at formation. The transfer of angular momentum from an overdensity to the PBH can only be investigated in full generality in full 3+1D NR. 

The research in chapters \ref{Chapter3} and \ref{Chapter4} is mostly motivated by this last point. In chapter \ref{Chapter3}, we use full 3+1D NR to study the collapse of a spherically symmetric overdensity in a matter-dominated universe, studying in detail the mechanics of collapse and the properties of the resulting PBHs. Using full 3+1D NR here facilitates straightforward extensions to nonspherical initial conditions and formation processes, of which we give an example in chapter \ref{Chapter4}, where we study the collapse of overdensities with angular momentum in a matter-dominated universe, and study the efficiency of mass and angular momentum transfer from the overdensity to the resulting PBH.

In doing so, we show that studying nonspherical effects on PBH formation using NR is feasible and because we run our simulations using the code $\grchombo$ \cite{Clough:2015sqa,Andrade2021}, which is open-source and publicly available, it is possible for any member of the PBH community to use our method to study PBH spin in more detail, or apply the method to any other questions regarding nonspherical effects on PBH formation, such as those mentioned above. 

Another way to generalise NR simulations of PBH collapse is to assume axisymmetry in lieu of full spherical symmetry, by means of which it becomes possible to study PBH formation via e.g. vacuum bubble collisions in an early-universe first-order phase transition. This question is exactly what motivates the work in chapter \ref{Chapter5}, in which we develop a method to run dimensionally reduced axisymmetric NR simulations with matter. For the case where the matter fields are real scalar fields, we show explicity how one obtains expressions for the constraint and evolution equations and we show good numerical convergence of our code. We apply the method by preliminarily investigating the collision of vacuum bubbles and we note that this code is also based on $\grchombo$, making the extension of our work less complicated. For instance, it might be possible to study the formation of spinning PBHs using the method presented, since a Kerr black hole spacetime possesses an axisymmetry, although one would have to relax the twist-free assumption that we employ in chapter \ref{Chapter5}. 

In the rest of this chapter, we give a brief review of GR in section \ref{section::GR1}, in which we cover the most important concepts such as metric and connection and we mention BHs and gravitational waves (GWs). We discuss the Friedmann-Lemaître-Robertson-Walker expanding universe and dark matter in section \ref{section::expanding_universe}, after which we introduce PBHs in section \ref{sect::PBHs}. In this last section, we specifically discuss the PBH formation threshold and we highlight some of the experimental constraints on today's PBH abundance. 

\section{General Relativity}\label{section::GR1}

Before GR, Albert Einstein published the theory of special relativity \cite{Einstein:1905}, which is groundbreaking in itself and introduced the concepts of time dilation, length contraction and a finite speed of light. In this theory already, Einstein did away with the concept of global time, meaning clocks of two different observers could run at entirely different speeds and time measurements became entirely dependent on the observer's reference frame. The acceptance of this theory asked for a reconsideration of the main theory of gravity at the time, Newtonian gravity. Newtonian gravity is an accurate theory for relatively light bodies that are moving relatively slowly, but the fact that it does not incorporate any of special relativity's abovementioned characteristics means it can at most be a low energy limit of a more general theory. 

Indeed, Einstein published his new gravity formalism some 11 years later, presenting the theory of \textit{der Allgemeinen Relativit\"atstheorie}, or General Relativity \cite{Einstein:1916}. Building heavily on concepts from differential geometry, it rephrases the laws of gravity in terms of manifolds and tensors. There is an intricate interplay between the curvature of space and the distribution of matter, governed by a complicated set of partial differential equations (PDEs), the Einstein field equations (EFE), and it is incredibly succesful at describing a wide range of phenomena, such as the prediction of the precession of Mercury's orbit and more recently, the first detection of GWs \cite{PhysRevLett.116.061102} and the image of (light bent around) the BH at the centre of the M87 galaxy \cite{EventHorizonTelescope:2019dse,EventHorizonTelescope:2019uob,
EventHorizonTelescope:2019jan,EventHorizonTelescope:2019ths,
EventHorizonTelescope:2019pgp,EventHorizonTelescope:2019ggy}. A more elaborate review can be found in e.g. \cite{Will:2005va}.

\vspace{2mm}\textit{The influence that GR has had on our understanding of the universe cannot in any way be understated.} \vspace{2mm}

GR is built on two main principles: the principle of general covariance and the principle of equivalence. The principle of general covariance dictates that the laws of physics must be the same for all observers. This motivates the use of tensors to describe gravity, since tensors are objects that are invariant under coordinate transformations (even though e.g. a vector's components will generally change when one changes coordinates, the vector itself still points in the same direction). 

The equivalence principle states that inertial masses (as in Newton's second law $F = ma$) and gravitational masses (as in $F = GMm/r^2$) are the same. This means that all objects fall the with the same acceleration in a gravitational field, regardless of their mass. The strongest form of this statement is the Einstein Equivalence Principle (EEP), which states that all physical laws reduce to those of special relativity in any inertial (freely falling) frame, which comes with the caveat that the region of space in which these laws are considered should be small compared to the length scale on which the gravitational field varies. 

We will now discuss the main ingredients of GR. In GR, we consider spacetime as a four-dimensional differentiable manifold $\mathcal{M}$, which can be mapped one-to-one to $\mathbb{R}^4$, although one may need more than one coordinate chart to cover the entire manifold, like in the two-dimensional case of the sphere $S^2$. 

\subsection{Metric}

To find the distance from one point to another, we may define a notion of distance on the manifold $M$ by introducing the metric tensor field, i.e. the metric is a map $g:T_p(M) \times T_p(M) \xrightarrow{} \mathbb{R}$, linear in each argument, where $T_p(M)$ is the vector space of vectors tangent to $M$ at point $p \in M$ and $\mathbb{R}$ is the set of real numbers. Furthermore, we require that the metric tensor is:
\begin{enumerate}
    \item symmetric: $g(X,Y) = g(Y,X)$ for all $X,Y \in T_p(M)$,
    \item non-degenerate: $g(X,Y) = 0$ for all $Y \in T_p(M)\leftrightarrow{} X = 0$. 
\end{enumerate}
By choosing a suitable coordinate basis, one can always (locally) diagonalize the metric. In any such orthormal basis, the number of of positive and negative elements on the metric's diagonal is the same and is referred to as the metric's signature. GR deals with Lorentzian metrics specifically, meaning the sign of one dimension (the temporal one in our conventions) is negative, whilst all other (the spatial) dimensions have a positive sign \footnote{This is the convention for the Minkowski metric that will be used throughout this thesis.}. A common way to specify a spacetime's metric is to write
\be \label{eqn::line_element}
    ds^{2} = = \sum_\mu \sum_\nu g_{\mu\nu}(x^\mu)dx^\mu dx^\nu \equiv g_{\mu\nu}(x^\mu)dx^\mu dx^\nu,
\ee
where $ds^2$ is the squared distance between two points infinitesimally close to one another and $g_{\mu\nu}(x^\mu)$ are the metric components, dependent on spacetime coordinates. In \eqn{eqn::line_element}, the indices $\mu, \nu$ sum over all spacetime indices and  we employ the Einstein summation convention, which implies repeated indices are summed over, and we will omit explicit summation symbols in the remainder of this thesis. The range of these indices is $\mu = 0, \ldots, D - 1$, where $D$ is the dimensionality of the spacetime. In special relativity in four spacetime dimensions, this metric is given by the diagonal $g_{\mu\nu} = (-1,1,1,1)$, the Minkowski metric, so the line element becomes $ds^2 = -dt^2 + dx^2 + dy^2 + dz^2$, which is independent of the coordinates. In GR, the metric can be much more complicated, in the sense that it generally includes off-diagonal terms and has non-trivial dependence on the various coordinates. However, the EEP requires that coordinates exist in which the metric reduces to the Minkowski metric when expanded around any given point on $\mathcal{M}$. 

Using the metric, we can define the square of the length of a vector $X^{\mu}$ as:
\be 
    |X^{\mu}|^{2} = g_{\mu\nu}X^{\mu}X^{\nu}.
\ee
Depending on the sign of the length, vectors are either spacelike, null or timelike:
\begin{subequations}
\begin{align}
    |X^{\mu}|^{2} = 0 &\to \q \text{$X^{a}$ is null},\\
    |X^{\mu}|^{2} < 0 &\to \q \text{$X^{a}$ is timelike},\\
    |X^{\mu}|^{2} > 0 &\to \q \text{$X^{a}$ is spacelike}.
\end{align}
\end{subequations}

\subsection{Covariant derivative}

Apart from the notion of distance introduced by the metric, one needs a consistent notion of a tensor's derivative on $\mathcal{M}$, since physical laws will generally involve such derivatives. One such derivative is the covariant derivative, and we note that this object is a priori completely independent of the metric. In fact, one does not need a metric to define a covariant derivative, and a manifold with just a covariant derivative and nothing else is a perfectly sound mathematical object (althought the covariant derivative is arguably quite useless if there are no tensors to apply it to). In any case, for a vector field $Y \in T_p(M)$, a covariant derivative $\nabla Y: T_p(M) \to T_p(M)$ is any linear map that satisfies the following properties:
\begin{subequations}
\begin{align}
    &\nabla_{fX + gY}Z = f\nabla_{X}Z + g\nabla_{Y}Z\\
    &\nabla_{X}(Y + Z) = \nabla_{X}Y + \nabla_{X}Z\\
    &\nabla_{X}(fY) = f\nabla_{X}Y + (\nabla_{X}f)Y.
\end{align}    
\end{subequations}
For functions $f: M \to \mathbb{R}$, the covariant derivative acts simply as a partial derivative, i.e. $\nabla_\mu f = \partial_\mu f$, and the Leibniz rule can be used to define the action of $\nabla$ on tensors of rank higher than one. 

In GR, one uses a particular covariant derivative (or connection), which is directly related to the spacetime metric. On a manifold $M$ with metric $g$, the Levi-Civita connection is a unique connection $\nabla$ such that
\begin{enumerate}
    \item the metric is covariantly constant, i.e. $\nabla$g = 0,
    \item $\nabla$ is torsion-free, i.e. $\nabla_{\mu}\nabla_{\nu}f = \nabla_{\mu}\nabla_{\nu}f$ for any function $f$,
\end{enumerate}
and we will assume that any reference to a connection is one to the Levi-Civita connection from here onwards. It can be shown that the way covariant and partial derivatives act on a vector are related as follows:
\be 
    \nabla_{X}Y^{\mu} = X^{\nu}e_{\nu}Y^{\mu} + \Gamma^{\mu}_{\rho\nu}Y^{\rho},
\ee
where $\{e^{\nu}\}$ forms a coordinate basis and $\Gamma^{\mu}_{\rho\nu}$ are the Christoffel symbols, related to the metric components by:
\be 
    \Gamma^{\mu}_{\rho\nu} = \frac{1}{2}g^{\mu\sigma}\big(g_{\sigma\nu, \rho} + g_{\sigma\rho, \nu} - g_{\nu\rho, \sigma}\big), \label{eqn::chris_symbols}
\ee
where indices after a comma indicate partial differentiation with respect to a coordinate. 

The Levi-Civita connection possesses the same coordinate-independent properties as tensors, unlike regular partial derivatives with respect to the coordinates.

\subsection{Geodesics}

Geodesics are an important concept in GR. They are curves that extremize the proper time between two points of a spacetime. The proper time between points $p$ and $q$ can be defined as follows: for any timelike curve $\lambda$ between $p$ and $q$, parametrized by $s$, the proper time between these points along this curve is 
\begin{equation}
    \tau[\lambda] = \int ~ds \sqrt{-g_{\mu\nu}(x(s))\dot{x}^\mu\dot{x}^\nu},
\end{equation}
where the dot signifies a derivative with respect to $s$. It can be shown that the proper time is minimized by the path $x^\mu(\tau)$, where $\tau$ is now the proper time along the curve, that satisfies the geodesic equation
\begin{equation}\label{eqn::geodesic_1}
    \frac{d^2 x^\mu}{d\tau^2} + \Gamma^\mu_{\nu\rho}\frac{dx^\nu}{d\tau}\frac{dx^\rho}{d\tau} = 0,
\end{equation} 
where $\Gamma^\mu_{\nu\rho}$ are given by \eqn{eqn::chris_symbols}. Since these symbols also define the action of the covariant derivative on tensors, it is possible to write the geodesic equation as 
\begin{equation}\label{eqn::geodesic_2}
    \nabla_X X = 0,
\end{equation}
where the vector $X^\mu$ is the tangent vector of the curve.

\subsection{Curvature}

A Minkowski spacetime is flat, just like one-, two- or three-dimensional Euclidean space (e.g. a line, flat tabletop or three-dimensional volume) is flat. This is not generally the case, e.g. consider again the case of the two-dimensional sphere $S^2$, whose surface is clearly curved. To quantify the degree to which any spacetime is curved, one uses both the metric and the Levi-Civita connection.

GR governs the interplay between the matter content of a spacetime and its curvature. The relevant notion of curvature is given by the Riemann tensor, which is defined by the equation $R^\mu_{\hspace*{2mm}\nu\rho\sigma}Z^\nu X^\rho Y^\sigma = (R(X,Y)Z)^\mu$, where $X,Y,Z$ are vector fields and $R(X,Y)Z$ is given by the vector field
\begin{equation}\label{eqn::rxyz}
    R(X,Y)Z = \nabla_X \nabla_Y Z - \nabla_Y \nabla_X Z - \nabla_{[X,Y]}Z.
\end{equation}
To understand what the Riemann tensor measures, it is useful to define parallel transport of a tensor along a curve. If a curve $\lambda$ has tangent $X^\mu$, then for a given connection $\nabla$, a tensor field $T$ is parallel transported along $\lambda$ if 
\begin{equation}
    \nabla_X T = 0.
\end{equation}
Parallel transport makes it possible to map the vector space at a point on the manifold to the vector space of any other point on the same manifold, or more generally, it maps the spaces of tensors of any rank at two different points to one another. Given two linearly independent vector fields $X,Y$, whose commutator vanishes everywhere, and a third vector $Z_p \in T_p(M)$, one can obtain a vector $Z_{q1} \in T_q(M)$ by parallel transport for a distance $\delta x$ along $X$ first, then for a distance $\delta y$ along $Y$. One obtains a vector $Z_{q2} \in T_q(M)$ by parallel transport along $Y$ first, then along $X$. The difference $\Delta Z = Z_{q1} -Z_{q2}$ in the limit where $\delta x, \delta y$ go to zero is 
\begin{equation}
    \lim_{\delta x, \delta y \to 0}\frac{\Delta Z}{\delta x \delta y} = R^\mu_{\hspace*{2mm}\nu\rho\sigma}Z_p^\nu X^\rho Y^\sigma.
\end{equation}
Therefore, parallel transport along $X$ and $Y$ only commute when the manifold is flat. 

If the Riemann tensor vanishes everywhere in a spacetime, we call the spacetime flat, as is the case for e.g. Minkowski. If this is not the case, the spacetime is curved. 

In GR, point particles move along geodesics, which are curves that obey the geodesic equation, which is given by \eqn{eqn::geodesic_1}. In a flat spacetime, such as Minkowski, these are just straight lines, and initially parallel geodesics remain parallel but this is not the case in curved spacetimes. This is captured by the geodesic deviation equation 
\begin{equation}
    T^\nu\nabla_\nu(T^\rho\nabla_\rho S^\mu)= R^\mu_{\hspace*{2mm}\nu\rho\sigma}T^\nu T^\rho S^\sigma,
\end{equation}
where the vector $T^\mu$ is tangent to the geodesic and $S^\mu$ is the deviation vector, measuring distance between two geodesics infinitesimally close to one another. If a spacetime is flat, the right hand side vanishes and the above equation implies parallel geodesics remain parallel, unlike in a curved spacetime.

In terms of the components of $\nabla$, the components of the Riemann tensor can be written as 
\begin{equation}\label{eq:riemtens}
    R^\alpha_{\beta\mu\nu}=
    \partial_\mu\Gamma^\alpha_{\beta\nu}-
    \partial_\nu\Gamma^\alpha_{\mu\beta}+
    \Gamma^\alpha_{\lambda\mu}\Gamma^\lambda_{\beta\nu}-
    \Gamma^\alpha_{\lambda\nu}\Gamma^\lambda_{\beta\mu}.
\end{equation}

All this is to explain that the Riemann tensor encodes information about the manifold's curvature and this will be instrumental in describing how the curvature is influenced by the presence of matter and vice versa. To do so, one needs several contractions of the Riemann tensor, namely the Ricci tensor 
\begin{equation}
    R_{\mu\nu} \equiv g^{\rho\sigma}R_{\rho\mu\sigma\nu}
\end{equation}
and the Ricci scalar 
\begin{equation}
    R \equiv g^{\mu\nu}R_{\mu\nu}.
\end{equation}
The Ricci tensor encodes, roughly speaking, how spacetime volume changes as one moves around the manifold. In particular, in a Lorentzian spacetime, it can tell us how spatial volume changes as one moves along the time direction. The Ricci scalar quantifies the difference between the volume of a given spacetime and the corresponding volume in Minkowski space. 

Since the Ricci tensor is a contraction of the Riemann tensor, it has fewer degrees of freedom and can therefore not contain the same amount of information. We can actually consider the Ricci tensor as the trace of the Riemann tensor, as it is contracted with the metric, so the missing degrees of freedom are the traceless part. This traceless part is the Weyl tensor, given by 
\begin{align}
    C_{\rho\sigma\mu\nu} &= R_{\rho\sigma\mu\nu} - \frac{1}{2}(g_{\rho\mu}R_{\nu\sigma} - g_{\rho\nu}R_{\mu\sigma} - g_{\mu\sigma}R_{\nu\rho} + g_{\nu\sigma}R_{\mu\rho}) \\
    &\quad + \frac{1}{6}(g_{\rho\mu}R_{\nu\sigma} - g_{\rho\nu}R_{\mu\sigma})R.
\end{align}
The Weyl tensor stores information on how a spacetime's curvature changes without any volume changes. 

Finally, the matter content of spacetime is described by the stress-energy-momentum tensor (stress-tensor) $T_{\mu\nu}$, which obeys standard energy and momentum conservation laws, captured in the covariant equation 
\begin{equation}
    \nabla_\mu T^{\mu\nu} = 0.
\end{equation}
The interplay between curvature and matter is will be given by relating the Einstein tensor $G_{\mu\nu}$, 
\begin{equation}
    G_{\mu\nu} = R_{\mu\nu} - \frac{1}{2}Rg_{\mu\nu},
\end{equation}
to $T_{\mu\nu}$.

\subsection{GR postulates}\label{section::postulates}

Having discussed the relevant concepts from differential geometry above, we are now in a position to list the four postulates that define GR as a theory. They are:
\begin{enumerate}
    \item Spacetime is a four-dimensional Lorentzian manifold, equipped with the Levi-Civita connection.
    \item Massive/massless free particles follow timelike/null geodesics.
    \item The energy and momentum of matter are described by the stress tensor, $T_{ab}$, which is symmetric and conserved, i.e. $\nabla_{a}T^{ab} = 0$.
    \item Spacetimes curves in response to the presence of matter according to the Einstein field equations (EFE), which are
    \be 
        G_{\mu\nu} + \Lambda g_{\mu\nu} = \frac{8\pi G}{c^4} T_{\mu\nu},
    \ee
\end{enumerate}
where $\Lambda$ is the cosmological constant, $G$ is Newton's gravitational constant and $c$ is the speed of light. The EFE can also be obtained by varying the Einstein-Hilbert action
\begin{equation}
    S_{\textrm{EH}} = \int ~d^dx \sqrt{-g}\left[\frac{c^4}{16\pi G}(R - 2\Lambda) + \mathcal{L}_\textrm{matter}\right],
\end{equation}
where $g$ is the determinant of the metric and $\mathcal{L}_\textrm{matter}$ is the matter Lagrangian, e.g. to couple gravity and electromagnetism one includes $\mathcal{L}_\textrm{matter, EM} = -\frac{1}{4}F^{\mu\nu}F_{\mu\nu}$ in $S_{\textrm{EH}}$, where $F_{\mu\nu}$ is the electromagnetic field strength tensor.

From this point onwards, we will assume that $D = 4$ and $\Lambda = 0$, unless explicitly stated otherwise. 

\vspace{2mm}\textit{The EFE are the main equations in this thesis, which govern all the physics that are discussed.} \vspace{2mm}

It can be shown that for a vacuum spacetime, in which $T_{\mu\nu}$ vanishes, the Ricci tensor and scalar must also vanish. Therefore, vacuum effects like tidal forces and GWs are all described by the Weyl tensor.

\subsection{Black holes}

The existence of BHs is arguably one of the most astonishing predictions of GR and dates back all the way to the first BH solution to the EFE, the Schwarzschild solution \cite{Schwarzschild:1916uq}, which proposed what is now referred to as the Schwarzschild metric, given in Schwarzschild coordinates by 
\begin{equation}\label{eqn::schwarzschild_metric}
    ds^2=-\left(1-\frac{2GM/c^2}{r}\right)c^2 dt^2+\left(1-\frac{2GM/c^2}{r}\right)^{-1}dr^2+r^2\left(d\theta^2+
    \sin^2\theta d\phi^2\right)~,    
\end{equation}
where $M$ is a positive constant that is considered the mass of the BH and $r,\theta,\phi$ are standard spherical coordinates. The metric is time-independent and solves the vacuum EFE equations. In fact, it is the most general spherically symmetric solution to the EFE, and necessarily static and asymptotically flat, which is known as Birkhoff's theoreom \cite{Birkhoff:1923} \footnote{It was pointed out recently that the same theorem had been proven earlier by Jebsen \cite{jebsen1921allgemeinen, VojeJohansen:2005nd}.}. We note that the only way that the Schwarzschild solution can be the most general solution is if it includes Minkowski space, which is indeed obtained by taking $M$ to zero. 

The Schwarzschild metric describes spacetime curvature away from a spherically symmetric central mass distribution and has a physical singularity at the origin, at which the Ricci scalar goes to infinity. We note that the singularity at $r=2M$ in \eqn{eqn::schwarzschild_metric} is due to the choice of coordinates and therefore not physical. The existence of an object like a BH, with its singularity and corresponding event horizon, is certainly a bold proposition, but has become completely accepted and is there is indeed strong experimental evidence that BHs exist, e.g. orbits of stars in the Galactic Center have been used to map the gravitational potential in that region, which was found to correspond to the potential of a supermassive BH \cite{Schodel:2002py,Eisenhauer_2005,Ghez_2003,Ghez:2003qj,Gillessen:2008qv}, whilst the abovementioned GW observations and M87 photo are other strong proofs that BHs exist. 

Kip Thorne has proposed the hoop conjecture \cite{Thorne:1972ji}, which asserts that a self-gravitating configuration of matter will collapse to a BH if and only if \
\begin{equation}\label{eqn::hoop_conjecture}
    M > \frac{c^2}{G} \frac{R}{2},
\end{equation}
where $M$ is the mass of the configuration and one assumes that the region is sufficiently spherical to be described by one radius $R$. 

\subsection{Gravitational waves}

If one assumes that the gravitational field is weak in a given region of spacetime, one can linearize the metric, i.e. decompose it into 
\begin{equation}
    g_{\mu\nu} \approx \eta_{\mu\nu} + h_{\mu\nu},
\end{equation}
where $\eta_{\mu\nu}$ is the Minkowski metric and $h_{\mu\nu}$ a comparatively small metric perturbation. The EFE can then similarly be linearized, in the process of which one obtains a wave equation for $\bar{h}_{\mu\nu} \equiv h_{\mu\nu} - \frac{1}{2}\eta_{\rho\sigma}h^{\rho\sigma}h_{\mu\nu}$ 
\begin{equation}
    \Box \bar{h}_{\mu\nu} = 0.
\end{equation}
Such waves are produced in a variety of physical scenarios but most notably in binary inspirals of BHs and neutron stars. Since the first discovery in 2015 \cite{LIGOScientific:2016aoc}, dozens of such inspirals have been observed \cite{LIGOScientific:2018mvr, LIGOScientific:2020ibl, LIGOScientific:2021djp} and at the time of writing this thesis, LIGO's latest observing run O4 is just underway.

\section{The expanding universe}\label{section::expanding_universe}

In the remainder of this chapter, we will use conventions in which $c = 1$, keeping $G$ explicit.

The EFE are essential to describe strong gravity physics, which implies the presence of high overdensities and therefore a highly inhomogeneous universe. However, the EFE are equally useful to describe an extremely homogeneous universe, i.e. one that is spatially homogeneous and isotropic. We will focus on asymptotically flat universes here, in which case the assumptions of homogeneity and isotropy imply that the spatial line element is simply the flat Euclidean one, $ds_\textrm{spatial}^2 = a(t)^2\left(dx^2 + dy^2 + dz^2\right)$, where $a(t)$ is an arbitrary function of time. The full spacetime line element then becomes 
\begin{equation}\label{eqn:FLRWflat}
    ds^2 = -dt^2 + a(t)^2\sum_i (dx^i)^2,
\end{equation}
where the coordinates $x^i = x, y, z$ are comoving coordinates. This metric is commonly referred to as the Friedmann-Lemaître-Robertson-Walker (FLRW) metric. The physical coordinates $x_\textrm{phys}^i \equiv a(t)x^i$ are the comoving coordinates stretched by the factor $a(t)$, which is commonly referred to as the scale factor and keeps track of how much the universe has expanded at a given point in time. One can obtain the physical velocity of an object by simply taking one time derivative of $x_\textrm{phys}^i$, which yields
\begin{equation}
    v^i_\textrm{phys} = a(t)\frac{dx^i}{dt} + \frac{da}{dt}x^i \equiv v_\textrm{pec}^i + Hx^i_\textrm{phys},
\end{equation}
where one defines the peculiar velocity $v_\textrm{pec}$ and the Hubble parameter 
\begin{equation}
    H \equiv \frac{\dot{a}}{a},
\end{equation}
where the dot denotes a derivative with respect to time. The Hubble parameter thusly gives the velocity of an object that is at rest with respect to the comoving coordinates.

To find the dynamics of such a universe, one subjects it to the EFE under the influence of energy specified by $T_{\mu\nu}$. In fact, one restricts the shape of $T_{\mu\nu}$ to that of a perfect fluid through the assumptions of homogeneity and isotropy, 
\begin{equation}
    T_{\mu\nu} = (\rho + P)U_\mu U_\nu - Pg_{\mu\nu},
\end{equation}
where $\rho$ and $P$ are the energy density and pressure of the fluid respectively, whilst $U^\mu$ is its four-velocity. For a comoving observer, $U^\mu = (1, 0, 0, 0)$, and the $T_{\mu\nu}$ takes the convenient form 
\begin{equation}
    T_{00} = \rho(t), \quad T_{i0} = 0, \quad T_{ij} = -P(t)\delta_{ij}.
\end{equation}
One may use the $\nu = 0$ equation of the conservation law $\nabla_\mu T^\mu_{\hspace*{2mm}\nu} = 0$ to find a relation between $\rho$, $P$ and the Hubble parameter:
\begin{equation}\label{eqn::rhoPrelation}
    \dot{\rho} + 3H(\rho + P) = 0,
\end{equation}
which can be integrated to find a scaling relation between $a$ and $\rho$, provided that $P$ in terms of $\rho$ is known. There are several types of energy one can consider based on this relation, which are characterized by their equation of state $P = w(\rho)$. In the context of cosmology and the expansion of universe, one usually restricts to a linear equation of state, i.e. $P = w\rho$, meaning the value of $w$ specifies the type of matter one is dealing with. Choices include 
\begin{itemize}
    \item \textbf{Matter:} refers to all types of energy for which the pressure is negligible compared to the energy density, i.e. $P \ll \rho$ or for practical purposes, $P = 0$. From \eqn{eqn::rhoPrelation}, this gives 
    \begin{equation}
        \rho_\textrm{matter} \sim a^{-3},
    \end{equation}
    which is a reflection of the fact that as the universe expands, the physical volume grows and the matter mass per volume decreases accordingly. 
    \item \textbf{Radiation:} refers to types of energy for which $P = \rho / 3$, as is the case for a gas of relativistic photons or more generally, relativistic particles whose kinetic energy is much larger than their rest mass. In this case, \eqn{eqn::rhoPrelation} implies 
    \begin{equation}
        \rho_\textrm{radiation} \sim a^{-4},
    \end{equation}
    meaning radiation energy density falls off more quickly than matter energy density as a function of scale factor. This is due to the fact that on top of the volume growing, the growing scale factor stretches the wavelength of the particles, diluting their energy further. 
    \item \textbf{Dark energy:} refers to the relation $P = -\rho$, in which case 
    \begin{equation}
        \rho_\Lambda \sim a^0 \sim \textrm{C},
    \end{equation}
    where we have used the conventional subscript $\Lambda$ to indicate dark energy and the letter C to denote a constant. The energy density is constant even as the scale factor grows. This type of energy explains why the expansion of the universe accelerates at late times. 
\end{itemize}
In summary, 
\begin{equation}\label{eqn::rhoarelation}
    \rho \sim a^{-3(1 + w)}.
\end{equation}

By computing the Einstein tensor and relating it to the stress tensor via the EFE, one obtains the Friedmann equation 
\begin{equation}\label{eqn::friedmann}
    H^2 = \frac{8\pi G}{3}\rho,
\end{equation}
as well as a second equation that couples the matter content and the scale factor 
\begin{equation}
    \frac{\ddot{a}}{a}= -\frac{4\pi G}{3}(\rho + 3P).
\end{equation}
One may rewrite \eqn{eqn::rhoarelation} as $\rho = \rho_0 (a_0/a)^{3(1+w)}$ and use this to integrate \eqn{eqn::friedmann} to obtain 
\begin{equation}
    a(t) = \left(\frac{3(1+w)}2{}\sqrt{\frac{8\pi G}{3}\rho_0}t\right)^{\frac{2}{3(1+w)}},
\end{equation}
giving a direct relation between the energy density and the scale factor. 

One of the consequences is that if the early universe is filled with a mix of different types of energy, e.g. matter and radiation, radiation energy density falls off more quickly than matter energy density. One usually assumes that radiation is the dominant energy component in the early universe, when temperatures are high. Inevitably though, after enough time passes, the matter energy density will become the dominant matter component, since $\rho_\textrm{radiation}/\rho_\textrm{matter} \sim a^{-1}$ from \eqn{eqn::rhoarelation}. This is why the standard history of the universe features an early radiation-dominated epoch, followed by a late matter-dominated one. 

It is worth mentioning that several early universe effects may cause the early radiation-dominated epoch to be interrupted by a matter-dominated one. These effects include inflaton preheating \cite{Khlopov:1985,Carr:2018} and a first-order QCD phase transition \cite{Khlopov:1980mg,Polnarev:1982a,Sobrinho:2016fay}. In this case, the universe briefly becomes matter-dominated before pressure reappears and radiation-domination returns. We note that whilst the early universe's equation of state is poorly constrained \cite{Carroll:2001bv,Hooper:2023brf}, it is observationally required to be radiation-like at the time of BBN to reproduce the correct light-element abundances at that time (see \cite{Wagoner:1966pv} and e.g. \cite{Cyburt:2015mya} for a review).

\subsection{Dark matter}

Dark matter (DM) is a massive matter component in the universe that does not interact electromagnetically, which makes DM detection extremely challenging, as DM's interactions with standard model particles through other (possibly unknown) forces are expected to be extremely weak. In this section, we will briefly cover some pieces of evidence for the existence of DM, referring the reader to e.g. \cite{Bertone:2016nfn,sanders2010dark} for more elaborate reviews. 

\subsubsection{Rotation curves}
\begin{figure}
    \centering
    \includegraphics[width=0.7\linewidth]{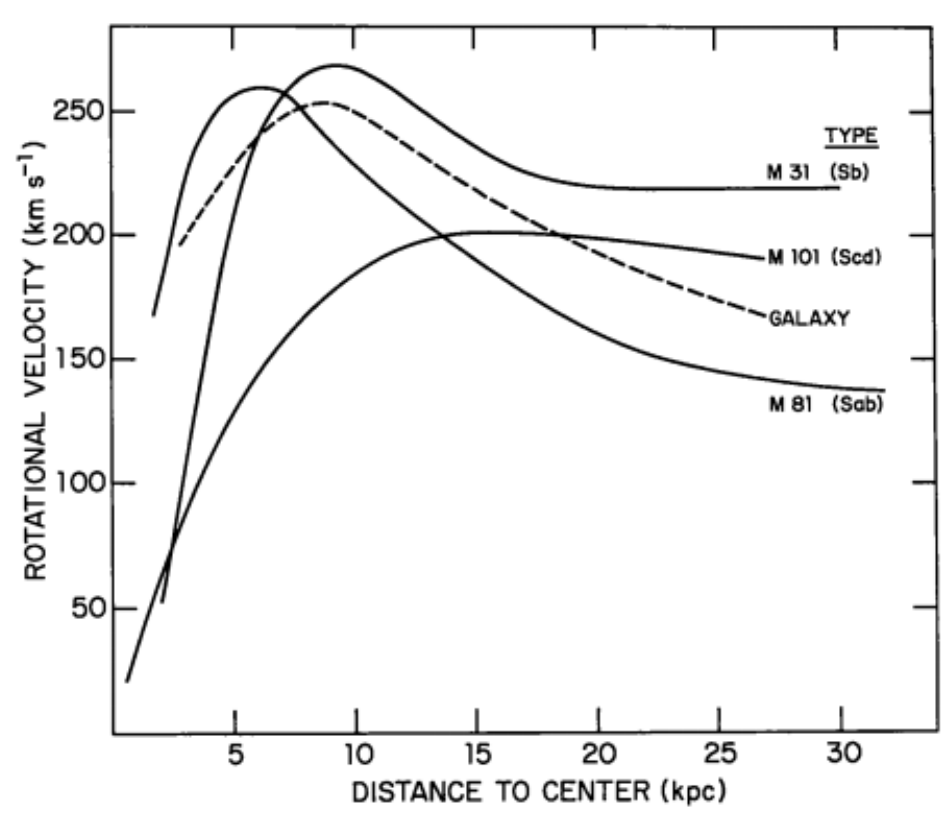}
    \caption{\textbf{Rotation curve data} for galaxies M31, M81 and M101. Figure taken from \cite{Bertone:2016nfn} and originally published in \cite{Roberts:1973}, in which the authors conlude that the spiral galaxies must be larger than photometric measurements indicate.} 
    \label{fig::rotation_curve}
\end{figure}

One of the simplest arguments for the existence of some non-luminous type of matter in galaxies is given by galaxy rotation curves, e.g. that of the Andromeda galaxy M31. Assuming Newtonian dynamics and a circular trajectory of stars around the galaxy's centre, the stars' velocities should satisfy 
\begin{equation}
    v(r) = \sqrt{\frac{GM(r)}{r}},
\end{equation}
where $M(r)$ is the mass integrated over the volume of a sphere of radius $r$. Given that the luminous matter in this galaxy is contained within a finite radius, one expects a $1/r$ drop at large distances. By measuring the circular velocities of stars and gas in this galaxy, various authors found that these velocities do not decrease like $1/r$, but the curve shows a flat behaviour at large distances. The constant behaviour seen in Fig. \ref{fig::rotation_curve} requires that the mass scales like $r$ or equivalently, the energy density like $1/r^2$. 

\subsubsection{Large scale structure}
Dark matter is commonly used to explain large scale structure formation, i.e. the formation of galaxies and galaxy clusters. These structures are thought to stem from small density fluctuations in the early universe, which are enhanced gravitationally until structure forms. Since we think that dark matter only interacts gravitationally, it can start enhancing these inhomogeneities and forming gravitational wells the second they enter the Hubble horizon, unlike baryons, whose coupling to photons causes their inhomogeneities to not grow in amplitude until later. The argument is that if the dark matter perturbations do not start growing early on, there is not enough time for the baryon perturbations to grow into galaxies \cite{Dodelson_2003}. 

\subsubsection{CMB}
Lastly, one can probe the relative dark matter densities using the cosmic microwave background (CMB). The CMB consists of photons emitted at the epoch of last scattering. The spectrum of these photons is a perfect black body spectrum up to small fluctuations, whose power spectrum has peaks at certain positions and with certain amplitudes. These tell us about the DM density today, which is found to be \cite{Planck_2020}
\begin{equation}
    \Omega_\textrm{DM} h^2 = 0.120 \pm 0.001,
\end{equation}
where the relative DM density $\Omega_\textrm{DM} = \rho_\textrm{DM}/\rho_\textrm{c}$, where $\rho_\textrm{c} = 3H_0^3/8\pi G$ is the critical density. $h$ is today's Hubble parameter $H_0$ in units of $100 km s^{-1}Mpc^{-1}$ and is measured as $h = 67.4 \pm 0.5$, meaning that $\Omega_\textrm{DM}$ corresponds to roughly 26\%, compared to roughly 5\% for baryons. 

\section{Primordial black holes}\label{sect::PBHs}

One DM candidate that has received particular attention lately are PBHs, which generically refers to BHs that form in the early universe, through mechanisms other than conventional stellar collapse. The concept of PBHs was introduced first by Zeldovich and Novikov and Hawking \cite{Hawking:1971,Zeldovich:1967} and it was realised soon after that PBHs could e.g. make up (part of) dark matter \cite{Carr:1974, Carr:1975,CHAPLINE1975}, could provide the early universe inhomogeneities necessary for structure formation \cite{Meszaros:1975} and that they might provide seeds for supermassive BHs \cite{Carr:1984cr}. One particularly neat aspect of PBH dark matter is that no new particles or forces are required, in contrast with models for particle dark matter, which usually introduce new types of fields and corresponding particles. However, one does need to allow for large density perturbations in the early universe, to make sure that these cause BH formation. 

The community's interest in PBHs was sparked by two sets of measurements in particular. Firstly, the MACHO collaboration published results on Large Magellanic Cloud microlensing events, which suggested a large number of subsolar massive objects in our galaxy if these lensing events could be explained by PBHs \cite{Aubourg_1993,Alcock_1997}. However, this theory was subsequently disproven by EROS and OGLE results \cite{Tisserand_2007,Wyrzykowski_2010,Wyrzykowski_2011,Wyrzykowski_2011a,Calchi_Novati_2013}, which put stricter constraints of the abundance of PBHs in this mass region. 

Secondly, the recent GW measurements have sparked debate as to whether any of the BHs involved in the observed events can be primordial, and various groups have argued that the detections are consistent with PBHs \cite{Bird_2016,Clesse_2017,Sasaki_2016}. 

Many formation scenarios for PBHs have been proposed and studied, but the standard case remains the collapse of overdensities in the early universe energy density that re-enter the Hubble horizon as it grows post-inflation. This scenario will be discussed in more detail below, and we list several other scenarios in section \ref{sect:intro}.

\begin{figure}
    \centering
    \includegraphics[width=0.7\linewidth]{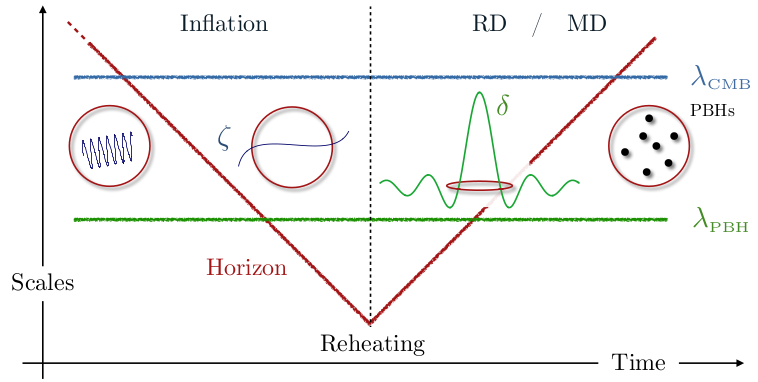}
    \caption{\textbf{Overdensity generation and collapse in the early universe} leading to the formation of PBHs, schematically represented. The comoving Hubble horizon is denoted in red and shrinks with respect to comoving scales during inflation, grows after reheating. The green horizontal line denotes the length scale at which one finds overdensities in the early universe. The blue horizontal line denotes CMB scales, which must be larger than $\lambda_\textrm{PBH}$ so that the overdensities are not imprinted on the CMB. Figure taken from \cite{Franciolini:2021nvv}.} 
    \label{fig::perturbs_inflation}
\end{figure}

The standard formation scenario can be explained as follows. One assumes the early universe energy density has enhanced perturbations with a certain small typical length scale $\lambda_\textrm{PBH}$. This typical length scale will be larger than the Hubble horizon at the end of inflation, so that these perturbations are frozen out and therefore kept from collapsing gravitationally. After inflation, the Hubble horizon will again grow compared to the universe's comoving scales until its size becomes of order $\lambda_\textrm{PBH}$. Gravitational forces become relevant for the perturbation at this point and if the overdensities are large enough, gravitational collapse can proceed and PBHs form. This process is summarized in Fig. \ref{fig::perturbs_inflation}.

In the standard picture, PBHs form in the early radiation-dominated era. In this case, the typical mass of the PBH is proportional to the Hubble mass at the time of formation \cite{Carr:1975}. This picture changes when the equation of state of the universe at the time of PBH formation is softer than in the radiation case, or when the pressure vanishes completely. Another way in which this picture can be altered is in the case of critical collapse \cite{Choptuik:1993,Niemeyer_1998:nj,Yokoyama:1999,Niemeyer_1999:nj,Shibata1999BlackHF,gundlach2000critical,Gundlach_2003,Musco:2004ak,Polnarev_2007,Musco:2008hv,Musco:2012au}.

Today's PBH abundance is quite heavily constrainted, which we will elaborate on below. However, a few mass windows remain in which PBHs could still have an appreciable abundance in terms of the DM density. Firstly, there is the asteroid mass window $10^{17}g<M_\textrm{PBH}<10^{22}g$, but the lack of constraints here may just reflect that it is hard to detect compact objects that are so light \cite{Green:2020jor}. Secondly, there is the sublunar range $10^{20}g<M_\textrm{PBH}<10^{26}g$ and thirdly the intermediate mass range $10\msun<M_\textrm{PBH}<10^{3}\msun$, which is of particular interest because this is the range in which recent GW detections were done. There is also the stupendously large black hole mass region $M_\textrm{PBH} > 10^{11}\msun$, although these BHs are too heavy to be part of a galaxy and could therefore not explain the DM in galactic halos. The final and possibly most exotic mass region is the one below $10^{15}g$, which could be populated by stable Planck mass relics of PBH evaporation, but it has been argued that this window is untestable because the relics would be too small to detect non-gravitationally \cite{Carr:2020gox}.

In conventional GR, any BH is completely characterized by its mass and its angular momentum \footnote{This is when one neglects a BH's electric charge, which is not expected to form in PBH scenarios.}. Therefore, to connect e.g. current GW observations to PBH scenarios, it is important to have an understanding of the expected spins of PBHs. Studies into PBH spin have only been done fairly recently, e.g. the spin probability distribution of PBHs is investigated in \cite{DeLuca:2019buf}, in which the authors find that PBHs formed in a radiation-dominated universes are expected to have small dimensionless spins at the percent level. Similar values for the spin are predicted in \cite{Mirbabayi:2019uph}. The dependence of the formation threshold $\delta_c$ is studied in \cite{He:2019cdb}, in which the authors find that its value increases when the PBH dimensionless spin is non-vanishing, in proportion to its square. Small spins are also predicted in \cite{Harada:2020pzb}, who also note that expected spins may be higher for lighter PBHs, e.g. ones that form in critical collapse scenarios. This motivated the authors of \cite{Chongchitnan:2021ehn} to quantify how rare high-spin PBHs are statistically, and they find that only one in a million PBHs forms with a dimensionless spin parameter larger than $0.8$. 

Literature on PBH spin in a matter-dominated universe is not available much. The authors of \cite{Harada:2017fjm} find that in a pressureless early universe, PBHs are actually rapidly rotating at their formation epoch. The authors of \cite{saito2023spins} study PBH formation with equations of state between completely pressureless and radiation, finding that expected spins increase when the equation of state parameter decreases.

In this work, we will mainly be concerned with initially spherically symmetric overdensities, although we break this symmetry in chapter \ref{Chapter4} to introduce angular momentum into the system. It is known from peak theory that high overdensities are most likely to be spherically symmetric \cite{Bardeen:1986,Heavens:1988}. This makes the assumption of spherical symmetry well-suited for PBH formation scenarios in a radiation-dominated universe, since the collapse threshold is generally large and high density peaks are therefore needed for PBH formation to proceed. This is not the case in a matter-dominated universe, in which the formation threshold is much lower due to the absence of pressure and small overdensities can efficiently accrete. The research in this thesis is focused on comparing to other studies that deal with spherically symmetric scenarios and makes a start at generalizing to setups with less symmetry, and this is certainly an interesting direction for future study. 

Depending on the details of a physical setup, initial deviations from spherical symmetry in an overdensity are expected to grow more or less spherically symmetric over the course of the overdensity's collapse. Deviations from spherical symmetry can be damped e.g. when there is a type of tension present, provided for example by the mass or other self-interactions of a scalar field. As we will see, chapters \ref{Chapter3} and \ref{Chapter4} deal with an overdensity sourced by the gradients of a massless scalar field and we would therefore not expect the spherically symmetric case to act as an attractor, but this should be put to further numerical tests. 

\subsection{Formation threshold} \label{section::PBH_threshold}

One can define an overdensity threshold $\delta_c = \rho/\rho_\textrm{avg}$, above which an overdensity indeed collapses to a PBH. Carr obtained the first value of $\delta_c \sim c_s^2$ in \cite{Carr:1975}, using a Jeans length argument that we will briefly review here, following the discussion in \cite{yoo2022basics}. For a curved FLRW universe with metric 
\begin{equation}
    ds^2 = -dt^2 + a^2(t)\left[\frac{dr^2}{1-Kr^2} + r^2d\Omega^2\right],
\end{equation}
which is a generalisation of \eqn{eqn:FLRWflat}, where we use spherical coordinates, $K$ is the extrinsic curvature and $d\Omega$ the line element of the unit two-sphere, the first Friedmann equation is given by 
\begin{equation}
    H^2 \equiv \left(\frac{\dot{a}}{a}\right)^2 = \frac{8\pi G}{3}\rho - \frac{K}{a^2}.
\end{equation}
On a uniform Hubble time slice, where the Hubble parameter is identical everywhere and equals $H^2 = 8\pi G\rho_b /3$, where $\rho_b$ is the background energy density, we obtain 
\begin{equation}
    \frac{8\pi G}{3}\rho_b = \frac{8\pi G}{3}\rho - \frac{K}{a^2},
\end{equation}
so that we can define a density perturbation as 
\begin{equation}
    \delta \equiv \frac{\rho - \rho_b}{\rho_b} = \frac{K}{a^2H^2}.
\end{equation}
When the perturbation enters the horizon, its physical radius is $R_c \equiv ar_c = 1/H$, so that the value perturbation value at horizon entry is $\delta_c = Kr_c^2$. 

We must take into account pressure gradients when considering whether an overdense region will collapse to a PBH, using the Jeans criterion, which states that if the free-fall timescale of the overdensity is shorter than the sound propagation timescale, collapse will proceed. The soundwave propagation timescale $t_s$ is 
\begin{equation}
    t_s = \frac{R_c}{c_s} = \frac{R_c}{\sqrt{w}},
\end{equation}
where $c_s$ is the sound speed and we use $c_s = \sqrt{w}$. Let $a_\textrm{max}$ be the scale factor when the perturbation reaches its maximum size. The free-fall timescale $t_{ff}$ is 
\begin{equation}
    t_{ff} = \frac{1}{H} = \left(\frac{8\pi G\rho}{3}\right)^\frac{-1}{2} = \frac{a_\textrm{max}}{\sqrt{K}}\left(\frac{a}{a_\textrm{max}}\right)^{3(1+w)/2}.
\end{equation}
Let $a_\textrm{th}$ be the scale factor such that $t_s(a=a_\textrm{th}) = t_{ff}(a=a_\textrm{th})$. If $a_\textrm{max} > t_s(a=a_\textrm{th})$, then $t_s < t_{ff}$ and pressure will prevent gravitational collapse. The condition that must therefore be satisfied for BH formation to occur is 
\begin{equation}
    a_\textrm{max} < t_s(a=a_\textrm{th}) = a_\textrm{max}\left(\frac{w}{Kr_c^2}\right)^{-1/(1+3w)},
\end{equation}
or in terms of $\delta_c$
\begin{equation}
    Kr_c^2 = \delta_c > w = c_s^2.
\end{equation}
It should be noted that in the case that if $w \ll 1$, the Jeans length is much smaller than the Hubble horizon and centrifugal or turbulent effects stemming from any non-sphericity of the perturbation can have a larger effect than the Jeans criterion \cite{Carr:1975,Harada_2013}. 

The above calculation assumes Newtonian gravity and was generalised with GR in \cite{Harada_2013}, obtaining a value of $\delta_c \sim 0.4$ for a radiation dominated universe. Because this computation does not take into account pressure gradients' non-linear effects, this value is just a lower bound. 

The numerical study of BH formation via gravitational collapse was pioneered by several works from the late seventies onwards \cite{Nadezhin:1978,Bicknell:1979,Novikov:1980}. More recently, analytic and numerical studies have shown that this threshold depends on the initial shape of the overdensity, and can range from $\delta_c = 0.4$ to $0.66$, e.g. the authors of \cite{Niemeyer_1998:nj,Niemeyer_1999:nj} obtained $\delta_c \approx 0.7$, after which those of \cite{Shibata1999BlackHF} pointed out that the threshold depends considerably on the shape of the overdensity profile, and their results were shown to be consistent with $0.3\leq \delta_c \leq 0.5$ in \cite{Green:2004}. The work in \cite{Shibata1999BlackHF} also showed that the initial conditions in \cite{Niemeyer_1999:nj} contained nonlinear perturbations, and in \cite{Musco:2004ak} simulations similar to \cite{Niemeyer_1998:nj} but with linear density perturbations were carried out, resulting in a range $0.43 \leq \delta_c \leq 0.47$. The authors of \cite{Hawke_2002} also confirmed that an overdensity could collapse if it met a certain threshold. The authors of \cite{Polnarev_2007} quote $\delta_c \sim 0.45$ and in \cite{Harada:2013epa} the authors obtain an analytic solution $\delta_c \sim 0.41$ by using a relativistic Jeans argument and taking into account the gravitational effect of pressure. The threshold dependence on the shape of the overdensity is studied in e.g. \cite{Musco:2019,Escriva:2019phb}. Other studies into the PBH formation threshold include \cite{Musco:2013, Musco:2020jjb,Escriva:2020tak}. Determining the threshold accurately is vital to sound predictions of the number of PBHs formed (and therefore the PBH density today) from a given density perturbation spectrum. 

\subsection{PBH abundance constraints}

\begin{figure}[t!]
    \centering
    \includegraphics[width=\linewidth]{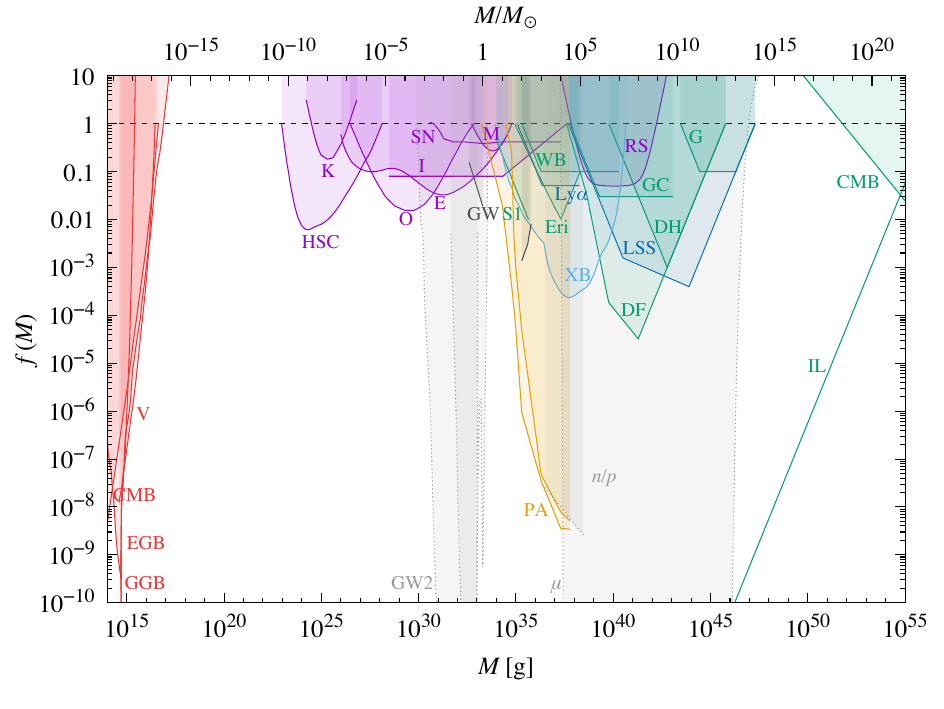}
    \caption{\textbf{PBH abundance constraints today} as a function of mass, assuming a monochromatic mass distribution. The various constraints are colour-coded by type of constraints, i.e. the figure shows constraints from evaporation (red), lensing (magenta), dynamical effects (green), accretion (light blue), CMB distortions (orange), large-scale structure (dark blue) and background effects (grey). Furthermore, the constraints are labeled, e.g. the constraints mentioned in this thesis come from the extragalactic gamma-ray background (EGB), the galactic gamma-ray background (GGB), Subary (HSC), GWs from binaries (GW) and dynamical friction (DF). For a full list of the different types of constraints we refer the reader to \cite{Carr:2020gox}, from which this figure was taken.} 
    \label{fig::pbh_constraints}
\end{figure}

The PBH abundance today is usually characterized by $f_\textrm{PBH} = \Omega_\textrm{PBH}/\Omega_\textrm{DM}$, i.e. the average PBH density as a fraction of the average dark matter density, so $f_\textrm{PBH} = 1$ would mean that all the DM is made up out of PBHs. Many different types of measurements have been used to constrain the PBH abundance in various parts of the mass range, and these constraints are summarized in Fig. \ref{fig::pbh_constraints}. To give the reader an idea of the types of measurements used to derive PBH abundance constraints, we will briefly discuss an inexhaustive list of such measurements, referring the reader to e.g. \cite{Carr:2020gox} and \cite{Franciolini:2021nvv} for more complete accounts. 

\subsubsection{PBH evaporation}
Constraints can be derived from the Hawking radiation \cite{Hawking:1974} that PBHs should emit. When PBHs form with a mass of around $10^{-16}\msun$, they are expected to evaporate completely over a time comparable to the age of the universe. The lighter the PBH, the more energy is emitted through Hawking radiation and this may eventually become detectable today. For example, bounds were obtained from measurements of extragalactic gamma rays \cite{Arbey_2020}, positron annihilations in the galactic centre \cite{DeRocco_2019} and gamma ray observations by INTEGRAL \cite{Laha_2020}. 

\subsubsection{Lensing and GWs}
PBHs can cause gravitational lensing of electromagnetic signatures, due to their compactness. For example, microlensing investigations by Subaru HSC constrain PBH abundance in the mass region $10^{-10}\msun<M_\textrm{PBH}<10^{-7}\msun$. Additionally, PBHs may form binary systems, whose inspirals can emit GWs detectable by the LIGO/Virgo collaboration. One can compare the rate of observed binaries to the rate expected for a given value of $f_\textrm{PBH}$ to derive further constraints \cite{Wong_2021,Ali_Ha_moud_2017,Raidal_2019,Vaskonen_2020}. 

\subsubsection{Dynamical friction}
If the DM halo of our galaxy had a large fraction of heavy PBHs, they would experience dynamical friction from stars and lighter BHs around them, i.e. the gravitational interactions with these lighter objects would slow them down and cause them to move into the galactic centre. By comparing the upper limit on the mass in the galactic centre to predictions from this effect one obtains constraints. \cite{Carr_1999,Carr_2018}.

%% file: Chapter2/chapter2.tex

\chapter{Numerical relativity}\label{Chapter2}  

\graphicspath{{Chapter2/Figs/}}

In this chapter, we discuss the decomposition of the EFE into a 3+1 formulation that is suitable for numerical evolution. We present the ADM formalism \cite{Arnowitt_2008}, in which the space and time directions of GR are explicitly split, and we discuss the BSSN formulation \cite{Nakamura:1987grc,Shibata:1995eot, Baumgarte:1998te}, which represents the same system of PDEs but has been shown to be numerically stable, which the ADM formulation is not. Lastly, we cover gauge conditions that can be employed to stabilize simulations further. This chapter will closely follow the treatment of these topics in \cite{alcubierre} and \cite{baumgarte_shapiro}.

Sticking to the conventions used in sections \ref{section::expanding_universe} and \ref{sect::PBHs}, in this chapter we set $c = 1$ and we keep $G$ explicit. 

\section{3+1 decomposition}\label{section::3plus1}

For the rest of this chapter, curvature tensors without explicit superscript will refer to three-dimensional ones, whilst their four-dimensional counterparts will have superscripts, e.g. ${}^{(4)}R^\mu_{\hspace*{2mm}\nu\rho\sigma}$ is the four-dimensional Riemann tensor, whilst $R^\mu_{\hspace*{2mm}\nu\rho\sigma}$ is the three-dimensional one.

\subsection{Foliation}\label{section::foliation}

The formulation of GR in terms of the full $D$-dimensional EFE is an elegant way to describe the theory, and allows for the formulation of complete $D$-dimensional solutions that can describe a spacetime in its entirety. However, these complete solutions often necessarily have a high degree of symmetry, captured by Killing vectors. A Killing vector field $X$ is such that 
\begin{equation}
    \nabla_\mu X_\nu + \nabla_\nu X_\mu = 0,
\end{equation}
and many exact solutions to the EFE have one or more of these Killing vector fields, such as the asymptotically flat Kerr black hole solution
\begin{equation}
\begin{split}
    ds^2 &= -\left(1 - \frac{2GMr}{\Sigma}\right)dt^2 - \frac{4GMar \sin{\theta}^2}{\Sigma}dtd\phi + \frac{\Sigma}{\Delta}dr^2 + \Sigma d\theta^2 \\
    &\quad + \left(r^2 + a^2 + \frac{2GMa^2 r\sin{\theta}^2}{\Sigma}\right)\sin{\theta}^2d\phi^2    
\end{split}
\end{equation}
where $\Sigma = r^2 + a^2\cos{\theta}^2$, $\Delta = r^2 - 2GMr + a^2$, $M$ is the black hole mass and $J = aGM$ is the black hole angular momentum (note that $a$ should not be confused with the FLRW scale factor here). Because none of the metric components have any dependence on $t$ or $\phi$, $\frac{\partial}{\partial t}$ and $\frac{\partial}{\partial\phi}$ are Killing vectors of this spacetime. When $a = 0$, one retrieves the metric for an asymptotically flat Schwarzschild black hole given in \eqn{eqn::schwarzschild_metric}, which has an additional $\frac{\partial}{\partial\theta}$ Killing vector. The main motivation to solve the EFE equations numerically is to be able to find solutions without easily identifiable Killing vectors and to this end, one must formulate the EFE equations as an initial value problem, in which the  space and time directions are explicitly split. 

To this end, we assume that the spacetime $(M, g_{\mu\nu})$ can be foliated into non-intersecting spatial hyperslices, i.e. each of these hyperslices is spanned by three spacelike vectors. We assume each of these hyperslices is identified by a unique, continuous value of some parameter $t$, which we can for our purposes interpret as a global (but not physical, except in special circumstances) time parameter.
\begin{figure}
    \centering
    \includegraphics[width=0.5\linewidth]{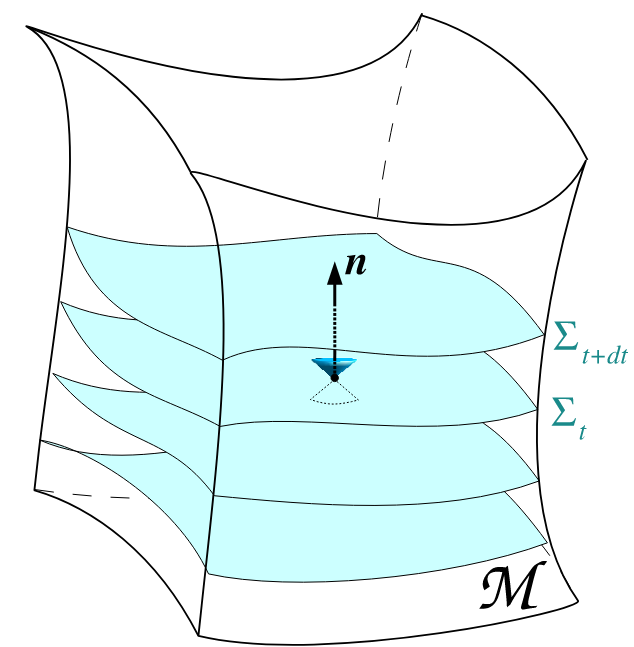}
    \caption{\textbf{Foliation of a spacetime} with two spatial dimensions and one temporal dimension. $\Sigma_t$ corresponds to the hyperslice identified by the parameter $t$ and $n$ is the normal vector to this hyperslice.  Figure taken from \cite{Gourgoulhon:2007ue}.} 
    \label{fig::foliation}
\end{figure}
Using the connection, we can define the 1-forms $\Omega_\mu \equiv \nabla_\mu t$ and $\omega_\mu \equiv \alpha \Omega_\mu$, where $\alpha^2 = -1/g^{\mu\nu}\nabla_\mu t \nabla_\nu t$. The vector field that's normal to the hyperslices and normalized to length one is then given by
\begin{equation}
    n^\mu \equiv -g^{\mu\nu}\omega_\nu.
\end{equation}
The foliation of spacetime and vector normal to a hyperslice are shown schematically in Fig. \ref{fig::foliation}. Using this normal vector, we can define the spatial metric
\begin{equation}
    \gamma_{\mu\nu} \equiv g_{\mu\nu} + n_\mu n_\nu,
\end{equation}
which also serves as a projection operator to project tensors into a spatial slice, by contracting each index of said tensor with $\gamma_{\mu\nu}$, e.g. for some three-index tensor $T_{\mu\nu\sigma}$, one defines its spatial projection as 
\begin{equation}
    T^\textrm{spatial}_{\mu\nu\sigma} = \gamma_\mu^{\hspace*{2mm}\alpha}\gamma_\nu^{\hspace*{2mm}\beta}\gamma_\sigma^{\hspace*{2mm}\delta}T_{\alpha\beta\delta}.
\end{equation}
Similarly, we can define a projection operator along the normal direction as $N^a_{\hspace*{2mm}b} = -n^a n_b$. 

We define a spatial covariant derivative $D_\mu$ by taking the action of the Levi-Civita connection on a tensor and projecting it to the spatial slice, i.e.
\begin{equation}
    D_\mu T_{\rho_1 \ldots \rho_\textrm{n}} \equiv \gamma_\mu^{\hspace{2mm}\nu}\gamma_{\rho_1}^{\hspace{2mm}\sigma_1}\ldots\gamma_{\rho_\textrm{n}}^{\hspace{2mm}\sigma_\textrm{n}}\nabla_\nu T_{\sigma_1 \ldots \sigma_\textrm{n}},
\end{equation}
where $T$ is a tensor of rank $n$. The action on a function $f$ can be found by simply interpreting $f$ as a tensor of rank zero. The components of this connection in terms of the spatial metric take on the same shape as their four-dimensional equivalent
\begin{equation}
    \Gamma^{\mu}_{\rho\nu} = \frac{1}{2}\gamma^{\mu\sigma}\big(\gamma_{\sigma\nu, \rho} + \gamma_{\sigma\rho, \nu} - \gamma_{\nu\rho, \sigma}\big). \label{eqn::chris_symbols_3d},
\end{equation}
and the same holds for the three dimensional Riemann tensor
\begin{equation}\label{eq::riemtens_3d}
    R^\alpha_{\hspace*{2mm}\beta\mu\nu}=
    \partial_\mu\Gamma^\alpha_{\beta\nu}-
    \partial_\nu\Gamma^\alpha_{\mu\beta}+
    \Gamma^\alpha_{\lambda\mu}\Gamma^\lambda_{\beta\nu}-
    \Gamma^\alpha_{\lambda\nu}\Gamma^\lambda_{\beta\mu}.
\end{equation}
It is clear that we cannot capture all the degrees of freedom of the four-dimensional Riemann tensor in its three-dimensional counterpart, which is purely spatial and therefore intrinsic to the hyperslice by design. The missing degrees of freedom are captured by the extrinsic curvature $K_{\mu\nu}$, which is a projection of the gradient of the normal vector $n^\mu$, i.e.
\begin{equation}
    K_{\mu\nu} \equiv -\gamma_\mu^{\hspace*{2mm}\rho}\gamma_\nu^{\hspace*{2mm}\sigma}\nabla_\rho n_\sigma.
\end{equation}
\begin{figure}
    \centering
    \includegraphics[width=0.7\linewidth]{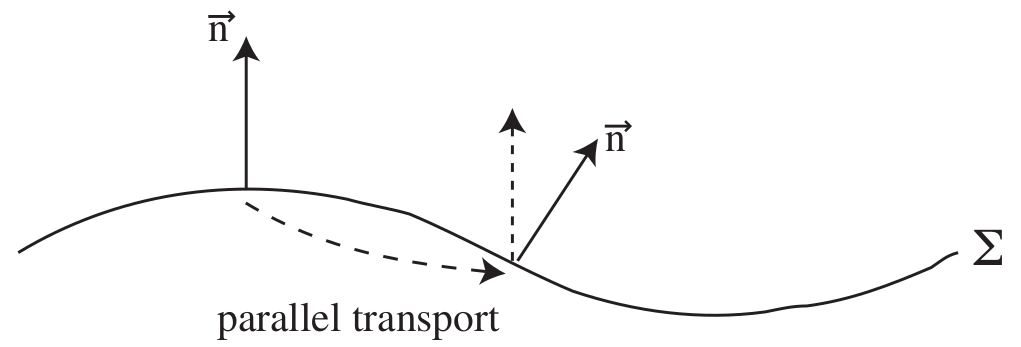}
    \caption{\textbf{Extrinsic curvature} is defined as the projected change of the normal vector as it is parallelly transported along a hypersurface, which is depicted here for a curved line as a hypersurface of two-dimensional flat Euclidean space. The curve is one-dimensional and therefore necessarily intrinsically flat, since the only component of the Riemann tensor is $R^0_{\hspace*{2mm}000}$, which vanishes due to \eqn{eqn::riem_symm_0}. The fact that it looks curved is mathematically represented by the non-vanishing extrinsic curvature. Figure taken from \cite{Aurrekoetxea:2022jux}.} 
    \label{fig::extrinsic_curvature}
\end{figure}
The extrinsic curvature contains information about the embedding of a spatial hyperslice in the full four-dimensional spacetime. This can be nicely illustrated in the case of a one dimensional line as a hypersurface of two-dimensional flat Euclidean space in Fig. \ref{fig::extrinsic_curvature}. One can also think about a two-dimensional torus as a hypersurface of $\mathbb{R}^3$ with a flat Euclidean metric, as even though the torus is intrinsically flat, i.e. initially parallel geodesics stay forever parallel, it is clear that this surface is extrinsically curved. 

Additionally, we can write the Lie derivative (see appendix \ref{app::lie_derivative}) of the spatial metric along the normal vector $n^\mu$ in terms of the extrinsic curvature 
\begin{equation}
    \mathcal{L}_n \gamma_{\mu\nu} = -2K_{\mu\nu},
\end{equation}
and therefore its trace obeys
\begin{equation}
    K = g^{\mu\nu}K_{\mu\nu} = -\mathcal{L}_n \gamma^{\frac{1}{2}}.
\end{equation}
Since we can interpret the determinant $\gamma$ in the above expression as a measure of spatial volume, $K$ is a measure of the change in spatial volume as one moves along a curve with tangent $n^\mu$. 

\subsection{Projecting the EFE}\label{section::EFEprojection}

To relate the three-dimensional curvature tensors to their four-dimensional counterparts, one may project the four-dimensional Riemann tensor onto the spatial hyperslices and their normals. Due to the symmetries of the Riemann tensor, this can be done in three different ways. We will just list the results of these projections, but the interested reader can find the computations in e.g. \cite{Aurrekoetxea:2022jux}. Firstly, projecting all indices spatially yields Gauss' equation 
\begin{equation}
    R_{\mu\nu\rho\sigma} + K_{\mu\rho}K_{\nu\sigma} - K_{\mu\sigma}K_{\nu\rho} = \gamma_{\mu}^{\hspace*{2mm}\kappa}\gamma_{\nu}^{\hspace*{2mm}\lambda}\gamma_{\rho}^{\hspace*{2mm}\tau}\gamma_{\sigma}^{\hspace*{2mm}\upsilon}\hspace*{2mm}{}^{(4)}R_{\kappa\lambda\tau\upsilon}.
\end{equation}
When one index is projected in the normal direction, one obtains the Codazzi equation 
\begin{equation}
    D_\nu K_{\mu\rho} - D_\mu K_{\nu\rho} = \gamma_{\mu}^{\hspace*{2mm}\kappa}\gamma_{\nu}^{\hspace*{2mm}\lambda}\gamma_{\rho}^{\hspace*{2mm}\tau}n^{\upsilon}\hspace*{2mm}{}^{(4)}R_{\kappa\lambda\tau\upsilon}
\end{equation}
and finally, one can project two indices in the normal direction to obtain Ricci's equation 
\begin{equation}
    \mathcal{L}_n K_{\mu\nu} = n^\sigma n^\rho \gamma_\mu^{\hspace*{2mm}\kappa}\gamma_\nu^{\hspace*{2mm}\lambda}\hspace*{2mm}{}^{(4)}R_{\sigma\lambda\rho\kappa}.
\end{equation}
With these relations in hand, we can project the EFE to obtain a 3+1 formulation. The EFE are 
\begin{equation}
    {}^{(4)}R_{\mu\nu} - \frac{1}{2}{}^{(4)}Rg_{\mu\nu} = 8\pi GT_{\mu\nu},
\end{equation}
where we set the cosmological constant $\Lambda$ to zero for simplicity. Two contractions of Gauss' equation combined with the EFE yield the Hamiltonian constraint 
\begin{equation}
    R + K^2 - K_{\mu\nu}K^{\mu\nu} = 16\pi G\rho,
\end{equation}
where $\rho \equiv n_\mu n_\nu T^{\mu\nu}$. Contracting the Codazzi equation and substituting the EFE yields the momentum constraints 
\begin{equation}
    D_\nu K^{\nu}_{\hspace*{2mm}\mu} - D_\mu K = 8\pi GS_\mu,
\end{equation}
where $S_\mu \equiv -\gamma_a^{\hspace*{2mm}\nu}n^\rho T_{\nu\rho}$.

\begin{figure}
    \centering
    \includegraphics[width=0.7\linewidth]{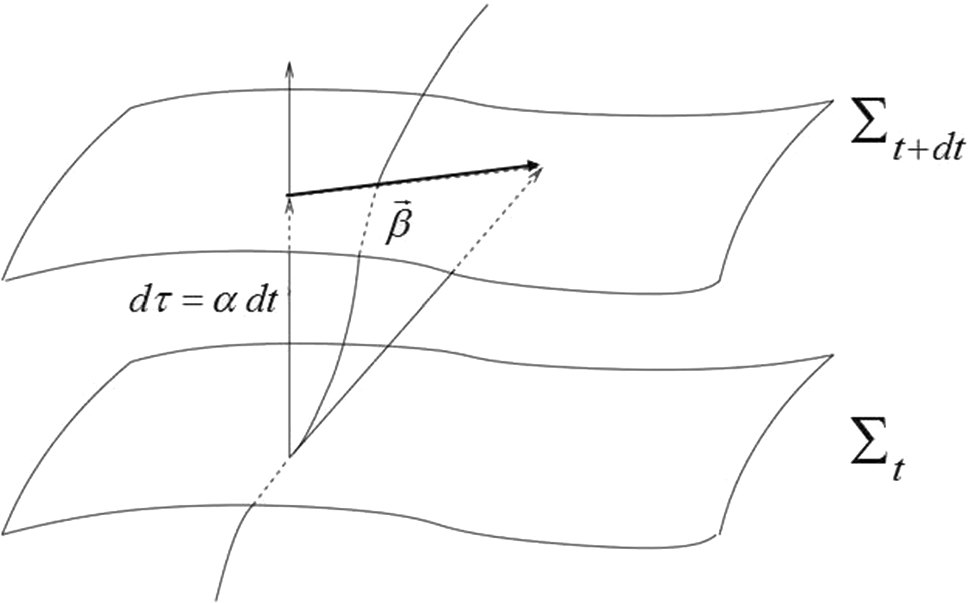}
    \caption{\textbf{The lapse and shift illustrated} between two hyperslices. The lapse $\alpha$ simply sets the length of the vector $d\tau$, while the shift $\beta^i$ encodes the shift in spatial coordinates when moving from one hyperslice to the next along the normal direction. Figure taken from \cite{Santos-Pereira2020}.} 
    \label{fig::lapse_shift}
\end{figure}

To find evolution equations for the spatial metric and the extrinsic curvature, we must think about the curves along which we want to evolve these spatial quantities. There are several reasons to not just use $n^\mu$ for this purpose. Firstly, its dot product with $\Omega_\mu$ is not unity, but this can be easily remedied by multiplying $n^\mu$ by $\alpha$, so that the integral curves of $\alpha n^\mu$ are parametrized by the parameter $t$ \footnote{It must be kept in mind that despite the confusing notation on which the literature seems to agree, the parameter $t$ and vector $t^\mu$ (defined in \eqn{eqn::vector_t}) are distinct objects.}. This is useful, because it means that all vectors $\alpha n^\mu$ originating on one hyperslice end on the same following hyperslice. Secondly, performing time evolution along the integral curves of $n^\mu$ implies that the spatial grid is also propagated along these integral curves, which is not always desirable, for instance in black hole spacetimes (more on this in section \ref{section::gauge}). We therefore allow for a spatial shift to the time integration vector, yielding 
\begin{equation} \label{eqn::vector_t}
    t^\mu \equiv \alpha n^\mu + \beta^\mu,
\end{equation}
where $\alpha$ is referred to as the lapse and $\beta$ is a purely spatial vector, referred to as the shift. $t^\mu$'s dot product with $\Omega_\mu$ is still unity and it is worth noting that $t^\mu$ does not have to be timelike at all, and can be made null or even spacelike with a suitable choice of the shift. Whilst this may seem strange, the only requirement is really that $t^\mu$ is not tangent to the hypersurfaces. The shift does not relate to anything physical, it is merely the change in spatial coordinates when moving from one hyperslice to the next following the normal direction, as illustrated in Fig. \ref{fig::lapse_shift}, i.e.
\begin{equation}
    x^i_{t+dt} = x^i_{t} - \beta^i dt.
\end{equation}

Meanwhile, the lapse measures how much proper time elapses between one hyperslice and the next along the normal direction. Both the lapse and the shift are free parameters and can be set to any value, but in practice have their own evolution equations, which may depend on lapse, shift, spatial metric and/or extrinsic curvature, on which we elaborate in section \ref{section::gauge}. 

By rewriting the Lie derivative of the extrinsic curvature and combining with the Ricci equation, Gauss' equation and the EFE, one can express the Lie derivative of the extrinsic curvature as
\begin{equation}
\begin{split}
    \mathcal{L}_t K_{\mu\nu} &= -D_\mu D_\nu \alpha + \alpha(R_{\mu\nu} - 2K_{\mu\rho}K^{\rho}_{\hspace*{2mm}\nu} + KK_{\mu\nu})\\
    &\quad - 8\pi G\alpha(S_{\mu\nu} - \frac{1}{2}\gamma_{\mu\nu}(S-\rho)) + \mathcal{L}_\beta K_{\mu\nu},    
\end{split}
\end{equation}
whilst the Lie derivative of the spatial metric becomes 
\begin{equation}
    \mathcal{L}_t \gamma_{\mu\nu} = -2\alpha K_{\mu\nu} + \mathcal{L}_\beta \gamma_{\mu\nu}.
\end{equation}
Finally, we pick an adapted set of basis vectors to simplify these expressions. In this basis, the timelike basis vector is $t^\mu = (1, 0, 0, 0)$, so that the Lie derivative along $t^\mu$ becomes $\mathcal{L}_t = \partial_t$. In this basis, $n_i = 0$, where the index $i = 1, 2, 3$, i.e. it runs just over spatial indices. This implies that any tensor with an upper index equal to zero vanishes, e.g. $\beta^\mu = (0, \beta^i)$. Furthermore, the normal vector becomes 
\begin{equation}
    n^\mu =
    \begin{pmatrix}
        \frac{1}{\alpha}, -\frac{1}{\alpha}\beta^i
    \end{pmatrix}
\end{equation}
and the metric can be written as 
\begin{equation}
    g_{\mu\nu} = 
    \begin{pmatrix}
        -\alpha^2 + \beta_l \beta^l& \beta_i\\
        \beta_j& \gamma_{ij}
    \end{pmatrix},
\end{equation}
meaning its corresponding line element becomes 
\begin{equation}
    ds^2 = -\alpha^2 dt^2 + \gamma_{ij}(dx^i + \beta^i dt)(dx^j + \beta^j dt).
\end{equation}
This yields the above constraint and evolution equations in a form that is commonly referred to as the ADM formalism
\begin{subequations}\label{eqn::ADM}
\begin{align}
    R+K^2 - K_{ij}K^{ij} &= 16\pi G\rho,\label{eqn::ADMHam}\\
    D_j(K^{ij} - \gamma^{ij}K) &= 8\pi GS^i,\label{eqn::ADMMom}\\
    \partial_t \gamma_{ij} &= -2\alpha K_{ij} + D_i\beta_j + D_j\beta_i,\label{eqn::ADMgamma}\\
    \partial_t K_{ij} &= -D_i D_j \alpha + \alpha(R_{ij} - 2K_{ik}K^{k}_{\hspace*{2mm}j} + KK_{ij})\label{eqn::ADMK}\\
    &\quad - 8\pi G\alpha(S_{ij} - \frac{1}{2}\gamma_{ij}(S-\rho)) \nonumber\\
    &\quad + \beta^k\partial_k K_{ij} + K_{ik}\partial_j\beta^k + K_{kj}\partial_i\beta^k,\nonumber
\end{align}
\end{subequations}
where the stress tensor's components are given by
\begin{subequations}
\begin{align}
    \rho &= n_\mu n_\nu T^{\mu\nu}\\
    S^i &= -\gamma^{ij}n^a T_{aj}\\
    S_{ij}&= \gamma_{ia}\gamma_{jb}T^{ab}\\
    S &= \gamma^{ij}S_{ij}.
\end{align}
\end{subequations}

\section{Solving the initial constraints}\label{sect:constraints}

\eqn{eqn::ADMHam} and \eqn{eqn::ADMMom} are different from \eqn{eqn::ADMgamma} and \eqn{eqn::ADMK} in the sense that they depend on spatial quantities only, and therefore constrain the allowed metric and matter configuration on any given spatial hyperslice. Because a solution to the full system in \eqn{eqn::ADM} is only a valid solution to the EFE if these constraints are satisfied, it is paramount to find metric and matter configurations that solve them. The evolution equations for $\gamma_{ij}$ and $K_{ij}$ preserve the constraints, which reduces the constraint problem to solving them on the initial hyperslice, after which the contraints are satisfied on subsequent hyperslices, provided that the evolution algorithm is stable (this should always be checked with e.g. convergence tests, which are provided for the research in this thesis in sections \ref{appendix::2} and \ref{Sapp::convergence_testing}).

One popular method to solve the constraints is the conformal transverse traceless (CTT) decomposition, which we will describe in some detail below, referring the reader to reviews \cite{alcubierre,baumgarte_shapiro,Gourgoulhon:2007ue} for more elaborate treatments. 

One may do a conformal transformation of the spatial metric 
\begin{equation}
    \gamma_{ij} = \psi^4\bar{\gamma}_{ij},
\end{equation}
where $\psi$ is a conformal factor and $\bar{\gamma}_{ij}$ is referred to as the conformal metric. In 3 dimensions, choosing $\gamma = \psi^{12}$, where $\gamma$ is the determinant of $\gamma_{ij}$, is convenient since it makes $\bar{\gamma}$, the determinant of $\bar{\gamma}_{ij}$, equal to 1. One can then write the scalar curvature as 
\begin{equation}
    R = \psi^{-4}\bar{R} - 8\psi\bar{D}^2\psi,
\end{equation}
where $\bar{R}$ and $\bar{D}$ are the Ricci scalar and spatial covariant derivative associated to $\bar{\gamma}_{ij}$ respectively. \eqn{eqn::ADMHam} then becomes 
\begin{equation}
    8\bar{D}^2 \psi - \psi\bar{R} - \psi^5K^2 + \psi^5K_{ij}K^{ij} = -16\pi G\psi^5\rho,
\end{equation}
and $K_{ij}$ still needs to satisfy \eqn{eqn::ADMMom}. One continues the conformal decomposition by splitting the extrinsic curvature into a trace and traceless part, i.e. 
\begin{equation}
    K_{ij} = A_{ij} + \frac{1}{3}\gamma_{ij}K.
\end{equation}
These quantities are conformally rescaled separately, namely 
\begin{subequations}
\begin{align}
    A^{ij} &= \psi^{-10}\bar{A}^{ij},\\
    K &= \bar{K},
\end{align}
\end{subequations}
i.e. $K$ is treated as a conformal invariant and we will not use the notation $\bar{K}$ in what follows. One can then rewrite the constraints as 
\begin{subequations}
\begin{align}
    8\bar{D}^2 \psi - \psi\bar{R} - \frac{2}{3}\psi^5K^2 + \psi^{-7}\bar{A}_{ij}\bar{A}^{ij} &= -16\pi G\psi^5\rho,\label{eqn::CTTHam}\\
    \bar{D}_j\bar{A}^{ij} - \frac{2}{3}\psi^6 \bar{\gamma}^{ij}\bar{D}_j K &= 8\pi G\psi^{10}S^i.
\end{align}
\end{subequations}
Finally, one decomposes $\bar{A}^{ij}$ into a divergenceless transverse-traceless part and a longitudinal part, i.e.
\begin{equation}
    \bar{A}^{ij} = \bar{A}^{ij}_{TT} + \bar{A}^{ij}_L,
\end{equation}
where 
\begin{equation}
    \bar{D}_j\bar{A}^{ij}_{TT} = 0
\end{equation}
and 
\begin{equation}
    \bar{A}^{ij}_L = \bar{D}^iW^j + \bar{D}^jW^i - \frac{2}{3}\bar{\gamma}^{ij}\bar{D}_kW^k \equiv (\bar{L}W)^{ij},
\end{equation}
i.e. $\bar{A}^{ij}_L$ is the symmetric traceless gradient of a vector $W^i$. The divergence of $\bar{A}^{ij}$ then becomes 
\begin{equation}
    \bar{D}_j\bar{A}^{ij} = \bar{D}_j\bar{A}^{ij}_L = \bar{D}_j(\bar{L}W)^{ij} \equiv (\bar{\Delta}_LW)^i,
\end{equation}
which defines the vector Laplacian $\bar{\Delta}_L$. The momentum constraints then become
\begin{equation}\label{eqn::CTTMom}
    (\bar{\Delta}_LW)^i - \frac{2}{3}\psi^6 \bar{\gamma}^{ij}\bar{D}_jK = 8\pi G\psi^{10}S^i,
\end{equation}
which together with \eqn{eqn::CTTHam} makes up the CTT expressions of the constraint equations. 

One approach is to solve the Hamiltonian constraint equation for $\psi$ after specifying a profile for $K$. E.g. by choosing a constant mean value $K = \bar{K}$, the second term on the LHS of \eqn{eqn::CTTMom} vanishes and the four constraint equations become a coupled system of elliptic equations. 

There are a number of pitfalls related to finding solutions to this system related to uniqueness and existence of solutions, discussed by e.g. the authors of \cite{Aurrekoetxea:2022mpw}, who propose an alternative approach. Instead of solving the Hamiltonian constraint for $\psi$, it can be solved for $K$ after specifying an initial configuration for $\psi = \psi_0$, reducing it to an algebraic equation for $K$ 
\begin{equation}\label{eqn:CTTKK}
    K^2 = 12\psi_0^{-5}\partial^j\partial_j\psi_0 + \frac{3}{2}\psi_0^{-12}\bar{A}_{ij}\bar{A}^{ij} + 24\pi G\rho.
\end{equation}
This algebraic equation is much more straightforward to solve than its elliptic counterpart \eqn{eqn::CTTHam} and this method is referred to as the CTTK approach. It should be noted that in the original CTT approach, when $K$ is taken to be constant and the momentum densities vanish, \eqn{eqn::CTTMom} is solved trivially. In CTTK, this is no longer the case, since $K$ will generally vary spatially, but the momentum constraint can be linearised to 
\begin{equation}
    (\bar{\Delta}_LW)_i = \frac{2}{3}\psi_0^6\partial_iK + 8\pi G\psi_0^6S_i,
\end{equation}
which can be solved straightforwardly numerically. For instance, by assuming conformal flatness, writing $W_i = V_i + \partial_i U$ for a vector field $V_i$ and a scalar field $U$ and choosing $U$ such that 
\begin{equation}\label{eqn::cttk4}
    \partial^j \partial_j U = -\frac{1}{4}\partial^j V_j,
\end{equation}
one can write the momentum constraints as three Poisson equations in flat space, 
\begin{equation}\label{eqn::cttk5}
    \partial^j\partial_j V_i = \frac{2}{3}\psi^6\partial_i K + 8\pi G\psi^6S_i,
\end{equation}
so that linearising and solving the four coupled equations \eqn{eqn::cttk4} and \eqn{eqn::cttk5} amounts to solving the constraints. 

An alternative method, suggested in \cite{Aurrekoetxea:2022mpw}, is solving \eqn{eqn::cttk4} by choosing 
\begin{equation}\label{eqn::cttk6}
    \partial_i U = -V_i/4,
\end{equation}
in which case the RHS of \eqn{eqn::cttk5} becomes a pure gradient. This can impose restrictions on the initial matter field configuration described by $S_i$, e.g. when $\psi$ is constant, $S_i$ must be a pure gradient as well or equivalently, its curl must vanish. For matter configurations for which the curl of $S_i$ does not vanish, one is then not allowed to use \eqn{eqn::cttk6} and one should instead solve \eqn{eqn::cttk4} and \eqn{eqn::cttk5} as four coupled equations. We will refer back to this point in chapter \ref{Chapter4}.

\section{Well-posedness and hyperbolicity}\label{section::stable_evolution}

Even though the ADM formalism outlined in section \ref{section::EFEprojection} has all the ingredients for succesful numerical evolution, it turns out that it allows for the development of large instabilities over the course of the simulation, because it is ill-posed as opposed to well-posed. To understand this notion better, we may consider a general system of PDEs in the form  
\begin{equation}\label{eqn::PDEsystem}
    \partial_t u = P(D) u,
\end{equation}
where $u$ is a vector function with components dependent on time and space and $P(D)$ is a matrix with components that consist of spatial derivative operators. One can set up an initial value problem, or Cauchy problem, for \eqn{eqn::PDEsystem} by specifying initial data $u(t=0,x)$, i.e. at time $t = 0$ and everywhere in space. Solving this problem is considered finding a solution $u(t,x)$ from the initial data. The problem is considered well-posed if such a solution exists, if the solution is unique and if the solution depends continuously on the initial data or equivalently, if small changes in the initial data correspond to small changes in the solution. The solution should depend continuously on any boundary data, as well. This is captured and quantified by requiring that one is able to define a norm, here denoted by $\Vert\ldots\Vert$, such that
\begin{equation}
    \Vert u(t, \cdot)\Vert \leq C_1 e^{C_2t} \Vert u(0, \cdot)\Vert,
\end{equation}
where $u(t, \cdot)$ denotes the set of evolution variables at a given time and the constants $C_1, C_2$ are independent of the initial data.

We will now introduce the concept of hyperbolicity for a system of first-order PDEs of the form 
\begin{equation}\label{eqn::system_pdes}
    \partial_t u + M^i \partial_i u = 0,
\end{equation}
where the index $i$ runs over the spatial indices, the $M^i$ are constant matrices and we have simplified by setting the RHS to zero. By picking an arbitrary unit vector $\hat{s}$ one can define the principal symbol of this system of equations 
\begin{equation}
    P(s^i) \equiv M^i s_i.
\end{equation}
The properties of the principal symbol determine the hyperbolicity of the system from \eqn{eqn::system_pdes}, i.e. the system is strongly hyperbolic if the principal symbol has real eigenvalues and a complete set of eigenvectors for all $s^i$. On the other hand, if the principal symbol has real eigenvalues for all $s^i$ but not a complete set of eigenvectors, the system is weakly hyperbolic.

Importantly, it can be shown that the initial value problem for the PDE system in \eqn{eqn::system_pdes} is well-posed if and only if the system is strongly hyperbolic \cite{Hilditch:2013sba}. For a more elaborate discussion on well-posedness in the context of Lovelock theories of gravity (such as GR), see e.g. \cite{Papallo:2017qvl} and references therein. 

The hyperbolicity discussion above is valid for the case in which the matrices $M^i$ are constant, i.e. in the case of a linear system of PDEs. When one considers a nonlinear PDE system in which the matrices have dependencies $M^i = M^i(t, x, u)$, one can consider the local form of the matrices $M^i$ by linearising around a background solution. Then, it can be shown that strong hyperbolicity implies well-posedness, as well. For more details, we refer the reader to \cite{alcubierre,Hilditch:2013sba}. 

The discussion above focuses on first-order systems of PDEs. The systems we are ultimately interested in are second-order, e.g. the ADM formulation contains second-order space derivatives of the metric through the Ricci tensor. One can circumvent this by treating first-order derivatives as independent quantities, e.g. one may define $d_{ijk} = \partial_i \gamma_{jk}$ and replace second-order derivatives of $\gamma_{ij}$ by first-order derivatives of $d_{ijk}$ to apply an analysis as described above. We refer the reader to \cite{alcubierre} for a more elaborate discussion.

It can be shown that the ADM formalism would be strongly hyperbolic if it is guaranteed that the momentum constraints are precisely satisfied and if an evolution equation of the Bona-Masso type \cite{Bona:1994dr} is used for the lapse \cite{alcubierre}, i.e.
\begin{equation}
    \partial_t \alpha = \beta^i\partial_i\alpha -\alpha^2 f(\alpha)K,
\end{equation}
where $f(\alpha)$ is an arbitrary positive function of $\alpha$. For computational purposes, since it impossible to guarantee that the first condition will be satisfied over the course of a numerical evolution, one concludes that the ADM formalism is only weakly hyperbolic. 

An alternative 3+1D formulation of the EFE is the Baumgarte-Shapiro-Shibata-Nakamura (BSSN) formalism \cite{Nakamura:1987grc,Shibata:1995eot, Baumgarte:1998te}, which is closely related to the ADM formalism. This formalism is implemented in $\grchombo$ but because it is not primarily used for the research presented in this thesis, we omit it from the main text and refer the reader to appendix \ref{app::BSSN} for further details. 

\section{CCZ4 formalism}\label{app:ccz4}

For the simulations presented in this chapters \ref{Chapter3} and \ref{Chapter4}, we use the CCZ4 formulation \cite{Alic_2012} of the Einstein equations, based on the Z4 system with the inclusion of damping terms \cite{Bona_2003,Bona:2003qn,Gundlach:2005eh}. 

In the Z4 system \cite{Bona_2003,Bona:2003qn}, the four-dimensional EFE are modified to 
\begin{equation}
    R_{\mu\nu} + \nabla_\mu Z_\nu + \nabla_\nu Z_\mu = 8\pi G\left(T_{\mu\nu} - \frac{1}{2}g_{\mu\nu}T\right),
\end{equation}
where the new four-vector $Z^\mu$ vanishes for physical solutions. This gives rise to the following 3+1D decomposition for a vacuum spacetime:
\begin{subequations}
\begin{align}    
    \partial_t \gamma_{ij} &= \beta^i\partial_i \gamma_{ij} - 2\alpha K\\
    \partial_t K_{ij} &= \beta^i\partial_i K_{ij} - D_i\alpha_j + \alpha\big[R_{ij} + D_iZ_j + D_jZ_i \\
    &\quad - 2K_{im}K^m_{\hspace*{2mm}j} + (K - 2\Theta)K_{ij}\big]\nonumber\\
    \partial_t \Theta &= \beta^i\partial_i \Theta + \frac{\alpha}{2}\big[R + (K - 2\Theta)K - K_{mn}K^{mn}\\
    &\quad + 2D_mZ^m - 2Z^m\partial_m\ln{\alpha}\big]\nonumber\\
    \partial_t Z^i &= \beta^i\partial_i Z^i + \alpha\big[D_mK^m_{\hspace*{2mm}i} - D_iK + \partial_i\Theta\\
    &\quad - 2K^m_{\hspace*{2mm}i}Z_m - \Theta \partial_i\ln{\alpha}\big]\nonumber,
\end{align}
\end{subequations}
where $\Theta = N_\mu Z^\mu$. The elliptic Hamiltonian and momentum constraints have been replace by extra evolution equations, and instead one now imposes the four constraints 
\begin{equation}\label{eqn::ccz4constraints}
    \Theta = 0, \quad Z^i = 0.
\end{equation}
Note that the last two evolution equations reduce to the Hamiltonian and momentum constraints when \eqn{eqn::ccz4constraints} is satisfied. It can now be shown that this system is strongly hyperbolic when a generalized Bona-Masso type lapse evolution equation 
\begin{equation}
    \partial_t \alpha = \beta^i\partial_i\alpha -\alpha^2 f(\alpha)\big(K - m\Theta\big)
\end{equation}
is used, for a positive function $f(\alpha)$. Because one must choose $m=2$ when $f(\alpha) = 1$, it is most convenient to set $m=2$ generally \cite{alcubierre}. 

Finally, it is possible to add damping terms to the Z4 system \cite{Gundlach:2005eh}, so that the true physical solutions to the EFE (i.e. the ones for which \eqn{eqn::ccz4constraints} is satisfied) effectively become an attractor of the complete set of solutions to the Z4 system. This damped formalism in four-dimensional covariant form is 
\begin{equation}
    R_{\mu\nu} + 2\nabla_{(\mu}Z_{\nu)}-2\kappa_1 n_{(\mu}Z_{\nu)} + \kappa_1 (1+\kappa_2)g_{\mu\nu}n_\alpha Z^\alpha = 8\pi G\left(T_{\mu\nu} - \frac{1}{2}g_{\mu\nu}T\right),
\end{equation}
in which we have set the cosmological constant $\Lambda$ to zero $n^\mu$ is the unit normal to the hyperslice foliation and $\kappa_1$ and $\kappa_2$ are constant damping parameters.

The conformal and traceless decomposition of the damped Z4 system is known as the CCZ4 formulation \cite{Alic_2012}
\begin{subequations}
    \begin{align}
        \partial_t\chi=&\beta^k\partial_k\chi + \frac{2}{3}\chi (\alpha K-\partial_k\beta^k)\,, \label{eq:ccz4chi}\\
        \partial_t\bar{\gamma}_{ij}
        =&\beta^k\partial_k
        \bar{\gamma}_{ij}+-2\alpha\tilde{A}_{ij}+\bar{\gamma}_{ik}
        \partial_j\beta^k+\tilde{\gamma}_{jk}
        \partial_i\beta^k-\frac{2}{3}
        \bar{\gamma}_{ij}\partial_k\beta^k
        \,, \label{eq:ccz4gamma}\\
        \partial_t\tilde{A}_{ij}=&\beta^k\partial_k\tilde{A}_{ij}+\chi\left\lbrace-D_iD_j\alpha
        +\alpha\left(\hat{R}_{ij}-8\pi G\alpha S_{ij}\right)\right\rbrace^{TF}\label{eq:ccz4A}\\
        &\quad+\alpha\left(\tilde{A}_{ij}\left(K - 2\Theta\right)-2\tilde{A}_{ik}
        \tilde{A}^k_j\right)+\tilde{A}_{ik}\partial_j\beta^k
        +\tilde{A}_{jk}\partial_i\beta^k-\frac{2}{3}\tilde{A}_{ij}\partial_k\beta^k
        \,, \nonumber\\
        \partial_t K=& \beta^k\partial_k K + \alpha \left(\hat{R} + K\left(K - 2\Theta\right)\right) - 3\alpha\kappa_1\left(1 + \kappa_2\right)\Theta\label{eq:ccz4K}\\
        &\quad - \gamma^{jk}D_\alpha D_j k + 4\pi G\alpha \left(S-3\rho\right)
        \,,\nonumber\\
        \partial_t \Theta =& \beta^k \partial_k \Theta + \frac{1}{2}\alpha \left(\hat{R} - \bar{A}_{jk}\bar{A}^{jk} + \frac{2}{3}K^2 - 2\Theta K\right) - \alpha\kappa_1\Theta\left(2 + \kappa_2\right)\\
        &\quad - \Theta^k\partial_k \alpha - 8\pi G\alpha\rho,\nonumber\\
        \partial_t\hat{\Gamma}^i=&\beta^k\partial_k
        \hat{\Gamma}^i+\bar{\gamma}^{jk}+\bar{\gamma}^{jk}\partial_j\partial_k\beta^i+\frac{1}{3}\bar{\gamma}^{ij}
        \partial_j\partial_k\beta^k-\hat{\Gamma}^j\partial_j\beta^i\label{eq:ccz4conf}\\
        &\quad +\frac{2}{3}\left[\partial_j\beta^j\left(\hat{\Gamma}^i + 2\kappa_3\frac{\Theta^i}{\chi}\right) - 2\alpha K\frac{\Theta^i}{\chi}\right] - 2\alpha K\frac{\Theta^i}{\chi} + 2\bar{\gamma}^{ik}\left(\alpha\partial_k\Theta - \Theta\partial_k\alpha\right) \nonumber\\
        &\quad -2\tilde{A}^{ij}\partial_j\alpha+2\alpha\left(
        \hat{\Gamma}^i_{jk}
        \tilde{A}^{jk}-\frac{3}{2}\tilde{A}^{ij}\frac{\partial_j\chi}{\chi}-\frac{2}{3}
        \bar{\gamma}^{ij}\partial_jK-8\pi G\chi S^i
        \right),
    \end{align}
    \end{subequations}
where $\hat{R}_{ij}$ is the modified Ricci tensor 
\begin{equation}
    \hat{R}_{ij} \equiv R_{ij} + 2D_{(i}\Theta_{j)}
\end{equation}
and 
\begin{equation}
    \hat{\Gamma}^i = -\partial_j\bar{\gamma}^{ij} + 2\frac{\gamma^i_{\hspace*{2mm}\alpha}Z^\alpha}{\chi}.
\end{equation}

\section{Gauge conditions}\label{section::gauge}

Gauge conditions were already mentioned above, and it is instructive to look at specific examples of gauge conditions and at how the gauge conditions can help a simulation progress, or crash it. For instance, because the physical time between hyperslices is proportional to the lapse, a high lapse value can cause the effective Courant factor to become too high for the simulation to run succesfully. Furthermore, in some scenarios black holes may be present, which implies the presence of singularities somewhere in the spacetime, and clever choices for lapse and shift can steer the grid points away from these singularities. 

\subsection{Geodesic slicing}\label{section::geodesic_slicing}

Arguably the most straightforward gauge choice is a geodesic slicing, in which $\alpha = 1$ and $\beta^i = 0$. Whilst this is extremely easy to implement, it has obvious drawbacks, one of which can be seen by defining the proper acceleration of the unit normal vector field as 
\begin{equation}\label{eqn::normal_acc}
    a_i \equiv n^j\nabla_jn_i = D_i \ln{\alpha},
\end{equation}
which vanishes in the case of geodesic slicing, and because the shift vanishes the coordinates are free-falling and follow geodesics, which can be seen by comparing \eqn{eqn::normal_acc} to \eqn{eqn::geodesic_2}. This would clearly be an issue in spacetimes with a concentrated region of increased energy density, e.g. a Schwarzschild spacetime, because over time the coordinates would cover a smaller and smaller part of the spacetime and eventually all fall into the black hole. This effect could be countered by setting non-vanishing shift values but other issues remain, such as grid points eventually reaching the singularity. 

\subsection{Hyperbolic slicing}\label{section::hyperb_slicing}

Dynamical gauges that have proven very succesful for spacetimes with black holes are hyperbolic formulations, such as 
\begin{equation}\label{eqn::moving_puncture_gauge}
    \partial_t \alpha = -\mu_{\alpha_1}\alpha^{\mu_{\alpha_2}}K + \mu_{\alpha_3}\beta^i\partial_i\alpha,
\end{equation}
which for specific choices of the constants ($\mu_{\alpha_1} = 2, \mu_{\alpha_2} = \mu_{\alpha_3} = 1$) is known as 1+log slicing \cite{Bona:1994dr}, although the optimal choice of constants depends on the exact physics one is simulating and sometimes on the observables one is interested in. This gauge choice is singularity avoiding, in the sense that the lapse is reduced in regions of high curvature, slowing the effective speed of coordinate observers. This can be understood by setting $\beta^i =0$ in \eqn{eqn::moving_puncture_gauge}, so that $\partial_t\alpha = -2\alpha K$ with 1+log constants, which gives $\alpha(t) = \alpha_0 e^{-2Kt}$, where $\alpha_0$ is the initial lapse value, often set to one. This makes clear that the lapse will go to zero exponentially fast in regions where $K>0$, i.e. collapsing regions of spacetime. 
\begin{figure}
    \centering
    \includegraphics[width=0.825\linewidth]{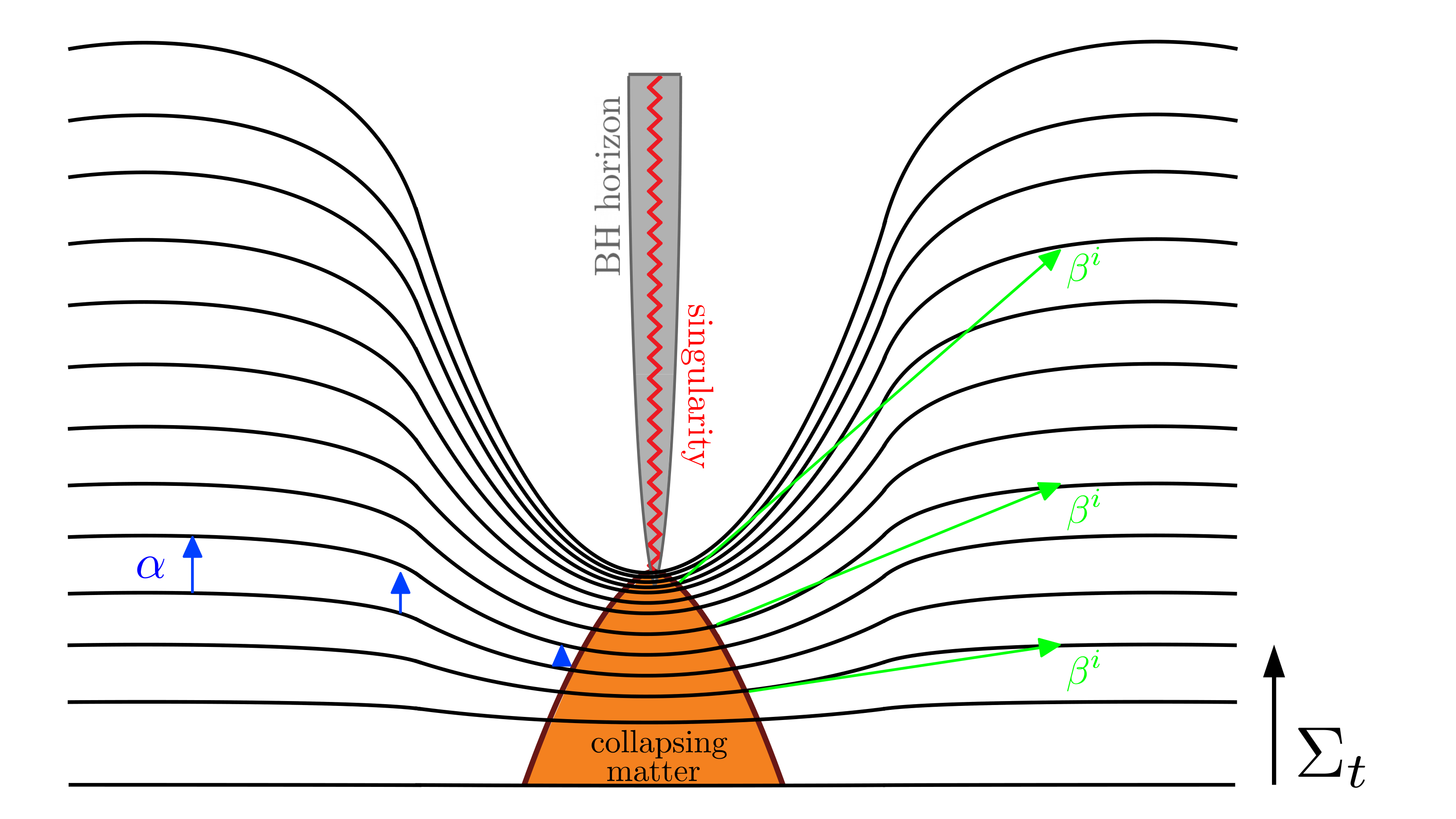}
    \caption{\textbf{Hyperslice stretching} due to a dynamical hyperbolic gauge condition for the lapse. As the passage of physical time effectively comes to a halt near the singularity, but not away from it, hyperslices become progressively more stretched. Figure taken from \cite{Aurrekoetxea:2022jux}.} 
    \label{fig::stretching}
\end{figure}
Another advantage of this method is that lapse evolution equation only depends on local quantities, i.e. evolution variables and derivatives thereof. Other singularity avoiding lapse choices exist, such as maximal slicing, but these require solving elliptic equations numerically, which is computationally costly especially on three dimensional grids.

Slowing down the effective passage of time in some regions but not in others has a major disadvantage, which is that it creates slice stretching, as illustrated in Fig. \ref{fig::stretching}. Due to this stretching, large gradients can develop, not just in the lapse profile near an overdensity but also in other evolution variables, in particular the extrinsic curvature. This problem can be at least partially mitigated by letting the shift vector dynamically evolve, too. We discussed in section \ref{section::EFEprojection} how the shift moves coordinates away from the normal trajectories and by moving the coordinates away from the overdensity when the lapse around the overdensity decreases, one can counteract the effects of slice stretching. The gamma-driver condition \cite{Alcubierre:2002kk} for the shift vector is 
\begin{subequations}
\begin{align}
    \partial_t\beta^i &= \eta_{\beta_1}B^i,\\
    \partial_t B^i &= \mu_{\beta_1}\alpha^{\mu_{\beta_2}}\partial_t\hat{\Gamma}^i - \eta_{\beta_2}B^i,
\end{align}
\end{subequations}
with $\eta_{\beta_1} = 3/4$, $\eta_{\beta_2}=\mu_{\beta_1}=1$ and $\mu_{\beta_2} = 0$ is a common choice. When one combines the 1+log slicing for the lapse and this gamma driver condition for the shift one speaks of the moving puncture method or moving puncture gauge \cite{Baker:2005vv, Campanelli:2005dd, vanMeter:2006vi}. These conditions maintain the well-posedness of both the BSSN and CCZ4 formalisms. In general, it is important to use gauge conditions that keep the formalism well-posed as a whole - a detailed discussion of suitable gauge conditions for different formalisms is outside the scope of this thesis, but we refer the reader to e.g. \cite{vanMeter:2006vi,Beyer:2004sv,Gundlach:2006tw,Palenzuela:2020}.

\section{GRChombo}

All research presented in this thesis was done using the NR code $\grchombo$ \cite{Clough:2015sqa,Andrade2021}, widely used to study strong-gravity phenomena \cite{Figueras:2015hkb,Clough:2016jmh,Clough:2016ymm,
Helfer:2016ljl,Tunyasuvunakool:2017wdi,
Figueras:2017zwa,Clough:2017ixw,Clough:2017efm,
Helfer:2018vtq,Alexandre:2018crg,Widdicombe:2018oeo,
Dietrich:2018bvi,Clough:2018exo,Helfer:2018qgv,
Kunesch:2018jeq,Clough:2019jpm,Muia:2019coe,
Bantilan:2019bvf,Widdicombe:2019woy,
Aurrekoetxea:2019fhr,Drew:2019mzc,Widdicombe:2020kjo,Helfer:2020gui,
Aurrekoetxea:2020tuw,Figueras:2020dzx,Nazari:2020fmk,
Andrade:2020dgc,Bamber:2020bpu,Joana:2020rxm,Radia:2021hjs,Clough:2021qlv, Traykova:2021dua,
Drew:2021ckb,deJong:2021bbo,Radia:2021smk,Figueras:2021abd,Wang:2022hra,Joana:2022uwc,Aurrekoetxea:2022ika,Croft:2022bxq,Aurrekoetxea:2022mpw,Clough:2022ygm,Croft:2022gks,Cheung:2022rbm,AresteSalo:2022hua,Bamber:2022pbs,Figueras:2022zkg,Drew:2022iqz,Evstafyeva:2022bpr,Evstafyeva:2022rve,Joana:2022pzo,Chung-Jukko:2023cow,Aurrekoetxea:2023jwd,Traykova:2023qyv,Cayuso:2023aht, deJong:2023gsx,AresteSalo:2023mmd,Doneva:2023oww,Aurrekoetxea:2023fhl,Franca:2023bed}. For a particularly detailed account of the code's capabilities and technical challenges involved in optimizing its performance, we refer the reader to \cite{Radia:2021smk}. $\grchombo$ is an open-source code and inherits the foundation of its functionality from the PDE solver software Chombo \cite{Adams:2015kgr}. As mentioned above, $\grchombo$ evolves the CCZ4 equations and does so using a fourth-order Runge-Kutta method. Spatial derivatives are computed using fourth-order stencils normally, but it is possible to switch to sixth-order stencils.  $\grchombo$ is specifically well-suited to simulate physical scenarios in which an adaptable grid structure is important and part of its philosophy is to keep the code easily modifiable and adaptable. This is achieved through heavy use of object oriented programming and templating concepts in C++. 

$\grchombo$ has full Adaptive Mesh Refinement (AMR), meaning that the full simulation box is covered by a uniform coarse grid, on top of which a hierarchy of Cartesian grids of increasing resolution is stacked. These finer grid levels can be added in one or in several regions of the simulation domain at a time. If the coarsest grid has level number $l = 0$, then the grid spacing at level $l$ is
\begin{equation}
    \Delta x_l = 2^{l_\textrm{max} - l}\Delta x_{l_\textrm{max}},
\end{equation}
i.e. the resolution increases by a factor of 2 each time a new level is added. The Courant factor $\Delta t/\Delta x$ is adjusted accordingly, i.e. grid level $l + 1$ is progressed twice as slowly as grid level $l$, so that the Courant factor for all grid levels is equal. 

Using AMR means that the grid shape adjusts to the physical system that is simulated and can do so dynamically and independently. The AMR algorithm does require user-specified tagging criteria, based on which the code decides whether or not to add or eliminate particular grid levels. Such tagging criteria can be based on truncation errors, degree of constraint violation or other measures, such as energy density or the gradients of an evolution variable. When regridding takes place, the code evaluates the tagging criteria at each grid point and adds finer levels at and around the tagged cells. 

$\grchombo$ employs both the BSSN and CCZ4 formalisms and uses Kreiss-Oliger dissipation \cite{kreiss1973methods} to control high-frequency noise, which typically originates from truncation errors and from interpolation errors that stem from the addition or deletion of grid levels. It is currently able to implement periodic, Sommerfeld, extrapolating and reflective boundary conditions. The public version of the code includes examples for scalar matter, Kerr black holes and black hole binaries, with (semi-)analytic constraint-satisfying initial data. 

$\grchombo$ simulations are typically run on high-performance computing clusters, and the code is therefore efficiently parallelised, using a hybrid of the MPI and OpenMP implementations. For instance, each AMR level is split into boxes, which are then distributed as equally as possible over the MPI processes available. Within an MPI process, OpenMP is used to thread the loops over grid points to e.g. compute spatial derivatives. 

Several diagnostics tools are available in the public code. An apparent horizon finder is provided to locate black holes and deduct their masses and angular momenta. Furthermore, it is possible to extract gravitational waves, energy densities and momentum fluxes.

%% file: Chapter3/chapter3.tex
\chapter{Primordial black hole formation with full numerical relativity}\label{Chapter3}

\graphicspath{{Chapter3/Figs/}}

This chapter contains the article \textit{Primordial black hole formation with full numerical relativity} \cite{deJong:2021bbo}, published in the \textit{Journal of Cosmology and Astroparticle Physics (JCAP)}. 

In this chapter and the next, we use Planck units $\hbar=c=1$ such that $G=\mpl^{-2}$, where $\mpl$ is the non-reduced Planck mass. 

\section{Introduction} \label{sect:intro}

Primordial black holes (PBHs) form in the early stages of the universe, and their idea was first conceived in the late sixties and early seventies \cite{Zeldovich:1967,Hawking:1971,Carr:1974}. It is notable that it was the potential existence of small black holes from primordial origin that led Hawking to theorize black hole evaporation \cite{Hawking:1974}. It was realised shortly after that PBHs could constitute a significant part of cold dark matter \cite{CHAPLINE1975}, and interest in PBHs has spiked in the recent past as a result. Evaporating PBHs have been suggested as explanations for galactic and extra-galactic $\gamma$-ray backgrounds, short $\gamma$-ray bursts and anti-matter in cosmic rays \cite{Page:1976wx,Carr:1976zz,Wright:1995bi,Lehoucq:2009ge,Kiraly1981,MacGibbon:1991vc,Cline:1996zg} and PBHs could provide seeds for the formation of supermassive black holes and large-scale structure \cite{Carr:1984cr,Bean:2002kx}. Moreover, PBHs could be responsible for certain lensing events \cite{Hawkins:1993,Hawkins:2020}, with recent analysis suggesting that the population of black holes (BHs) detected by the LIGO/Virgo/KAGRA (LVK) observatories \cite{LIGOScientific:2020ibl} may be primordial \cite{LIGOScientific:2020ufj,Franciolini:2021tla}. Additionally, work is underway to use next generation gravitational wave experiments to detect PBH formation and mergers \cite{kozaczuk2021signals,ng2021singleeventbased}. Results obtained by the NANOGrav Collaboration \cite{Arzoumanian_2020} have been associated to PBHs, as well \cite{DeLuca:2020agl,Vaskonen:2020lbd,Kohri:2020qqd,Domenech:2020ers}.

Various formation mechanisms could be relevant for PBHs \cite{Carr:2020gox, Green:2020jor}. These mechanisms include the formation of PBHs during inflation \cite{Clesse_2015,Inomata_2017,Garc_a_Bellido_2017,Ezquiaga:2017fvi}, the collision of bubbles that result from first order phase transitions \cite{Crawford1982,Hawking:1982,Kodama:1982,Leach:2000ea,Moss:1994,Kitajima:2020kig,Khlopov:1998nm,Konoplich:1999qq,Khlopov:1999ys,Khlopov:2000js,Kawana:2021tde,Jung:2021mku}, the collapse of cosmic strings \cite{Kibble:1976sj,Hogan:1984zb,Hawking:1987bn,Polnarev:1991,Garriga:1993gj,Caldwell:1995fu,
MacGibbon:1997pu,Wichoski:1998ev,Hansen:1999su,Nagasawa2005,Carr:2009jm,
Bramberger:2015kua,Helfer:2019,Bertone:2019irm,James-Turner:2019ssu,Aurrekoetxea:2020tuw,Jenkins:2020ctp,Blanco-Pillado:2021klh}, the collapse of domain walls produced during a second order phase transition \cite{Dokuchaev:2004kr,Rubin:2000dq,Rubin:2001yw,Garriga:2015fdk,Deng:2016vzb,Liu:2019lul}, the collapse of a scalar condensate in the early universe \cite{Cotner:2016cvr,Cotner:2017tir,Cotner:2018vug,Cotner:2019ykd} and specific baryogenesis scenarios \cite{Dolgov:1992pu,Dolhov:2020hjk,Green_2016,Dolgov_2009,Kannike_2017}. However, the mechanism that is most relevant for this work is the collapse of overdense regions that are present in the early universe \cite{Carr:1975,Nadezhin:1978,Bicknell:1979, Choptuik:1993,Evans:1994,Niemeyer_1998:nj,Green_1999:gl,Musco:2012au,Yoo:2020lmg}, which may originate from e.g. pre-inflation quantum fluctuations \cite{Carr:1993cl,Carr_1994:cgl,Hodges:1990hb,Ivanov:1994inn,Garcia:1996glw,Randall:1996rsg,Taruya_1999,Bassett_2001}.

In the standard picture, these fluctuations collapse post inflation, while the universe is dominated by radiation energy. The nonzero radiation pressure resists collapse, meaning that the inhomogeneities must be fairly large for PBHs to form. 

It was suggested early on, by using a Jeans length approximation, that an overdensity $\delta$ must be larger than a critical value $\delta_c$ equal to $\frac{p}{\rho} = 1/3$ if PBHs are to form \cite{Carr:1975} - further analytic and numerical studies into this threshold are mentioned in section \ref{sect::PBHs}.

PBH formation in matter-dominated epochs has also been extensively studied analytically and semi-analytically. In various non-standard universe histories, inflation is followed by a period of matter-domination \cite{Khlopov:1985kmz,Carr_2018:cdow,Martin_2020,Allahverdi_2021}. PBH formation in such an early epoch of matter-domination was considered early on \cite{Khlopov:1980mg}. More recently, a threshold amplitude for the collapse of a scalar field overdensity was found \cite{Hidalgo:2017dfp}, the effects of non-sphericity \cite{Harada:2016mhb} and inhomogeneity \cite{Kokubu:2018fxy} on the collapse were investigated, the resulting spin of the PBHs was studied \cite{Harada:2017fjm}, the duration of an early epoch of matter-domination was constrained by considering the PBH abundance \cite{carrion2021complex} and constraints on the amplitude and spectral index of the collapsing scalar field were obtained \cite{Carr:2017edp}.

\begin{figure}[t!]
    \includegraphics[width=\linewidth]{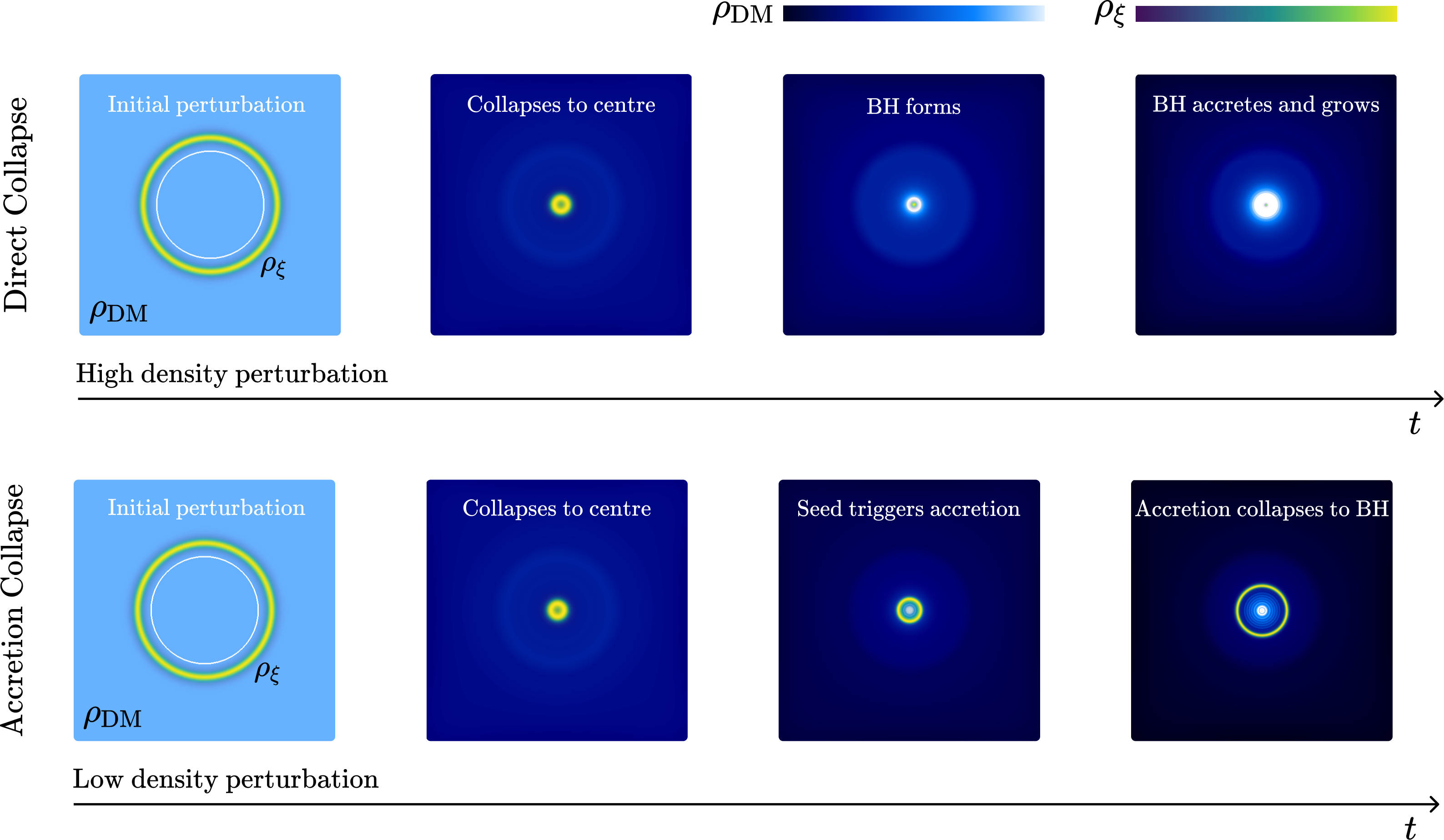}
    \caption{\textbf{Direct collapse and accretion driven mechanisms:} The figure summarizes the two distinct processes of PBH formation studied in this paper. 
    Top panel shows the \textit{direct collapse} mechanism, where a high density initial superhorizon perturbation collapses and forms a black hole as soon as the perturbation reaches the centre.
    Bottom panel depicts the \textit{accretion driven collapse} mechanism, where a lower density initial perturbation collapses and acts as a seed to trigger the accretion of the massive scalar field, which subsequently collapses to form a black hole. Both start from the same initial radius $R_0$, but with different initial amplitudes $\Delta \xi$.   In the leftmost figures, we show the initial size of the Hubble horizon (white solid line), which will grow as time evolves. In the other figures, the Hubble horizon has grown larger than the box size. Colourbars are shown in the top right, with lighter (darker) colours signifying higher (lower) energy densities, and scales fixed per mechanism. Video comparisons of these mechanisms can be found \href{https://youtu.be/fqEBlCybF8I}{here} \cite{Movie1} and \href{https://youtu.be/4N5e2RnUkmU}{here} \cite{Movie2}.} 
    \label{fig:panel_summary}
\end{figure}

In this work, we use full 3+1D numerical relativity simulations to investigate the collapse of subhorizon and superhorizon non-linear perturbations in an expanding universe that is dominated by matter. We model the expanding background and the collapsing perturbation using an oscillating massive scalar field and massless scalar field respectively. The massless scalar field's initial energy is thusly contained purely in its gradients. We will show that there are two broad mechanisms of black hole formation -- via \emph{direct collapse} for the case where the overdensity is sufficiently large that it will form a black hole, and via \emph{post-collapse accretion} for the case where the overdensity is smaller. In both cases, the process is rapid and its duration is around a single Hubble time, forming PBHs with initial masses of $\mbh H\sim 10^{-2}\mpl^2$. We illustrate these mechanisms in Fig. \ref{fig:panel_summary}.

Our choice of fundamental scalar fields as matter component, instead of a pressureless cold fluid (as suggested by \cite{East:2019chx}), is prompted by our focus on an early time (i.e. pre-BBN) matter-dominated phase instead of the present late time matter-dominated phase. Such early era matter-domination is often driven by a non-thermal fundamental scalar or moduli dynamics \cite{Acharya:2008bk}  instead of the more familiar cold pressureless fluid such as thermal WIMP dark matter. Furthermore, early matter phases will eventually transition into a radiation-domination epoch, such that the standard Hot Big Bang cosmological evolution can proceed. Such a phase transition from matter-domination into radiation-domination can then be achieved through the decay of the scalar field into either standard model particles or intermediaries.

This paper is organised as follows. In section \ref{sect:setup} we explain the numerical setup we use to evolve a scalar perturbation in a matter-dominated background. In section \ref{appendix:initialdata}, we cover our numerical setup. In section \ref{sect:bhformation} we introduce the two aforementioned formation mechanisms and their characteristics, and we study the properties of the black holes that are formed post collapse. In section \ref{sect:mass_distribution} we comment on the post formation growth of the PBHs, and we conclude in section \ref{sect:conclusions}.

\section{Early matter-domination epoch with scalar fields} \label{sect:setup}

The action we will consider is 
\begin{equation}
S = \int ~d^4 x \sqrt{-g}\Big[\frac{\mpl^2}{16\pi}R - \mathcal{L}_\phi - \mathcal{L}_\xi\Big]~,
\end{equation}
involving a massive scalar field $\phi$ with mass $m$ that models the ambient matter, and a massless scalar field $\xi$ that sources the initial perturbation. They are both minimally coupled to gravity but not otherwise coupled to one another, i.e.
\begin{align}
\mathcal{L}_\phi &=\frac{1}{2}\nabla_{\mu}\phi\nabla^\mu\phi + \frac{m^2 \phi^2}{2}~,~\mathrm{and}~ \\
\mathcal{L}_\xi &= \frac{1}{2}\nabla_{\mu}\xi\nabla^\mu\xi ~.
\end{align}
Since the field $\xi$ has no potential, it will only influence dynamics via its gradients. Furthermore, it will dilute much more rapidly than $\phi$, and hence it does not affect the long term dynamics of the system once its initial job of sourcing a perturbation is done. 

In principle, we could use a single massive scalar $\phi$, which would be particularly suitable if $\phi$ is the inflaton oscillating at the bottom of its own potential after slow-roll. However, if $\phi$ models a more general early-universe scalar field with a high mass that is not necessarily the inflaton, other (potentially massless) scalar fields may also be present and the setup presented here is more realistic. Such a massive scalar field would temporarily be the dominant driver of universe expansion, before it necessarily decays into radiation before BBN. Moreover, in practice, we find that large perturbations of the massive scalar introduce a large infusion of potential energy into the dynamics of the background resulting in non-matter-dominated evolution, at least initially. Whilst such effects certainly provide an interesting research direction, we are interested mainly in processes where the universe expands in a matter-like fashion until the perturbation has completely collapsed, so that the setup with two scalar fields is more suitable.

When the gradients in $\xi$ are negligible, the spacetime dynamics are dominated by the behaviour of the background scalar field $\phi$. When $\phi$ is additionally homogeneous on a given spatial hyperslice, the metric of the spacetime is well described by the Friedman-Lema\^{i}tre-Robertson-Walker (FLRW) line element
\begin{equation}
ds^2 = -dt^2 + a(t)^2(dr^2+r^2d\Omega^2_2)~,
\end{equation}
where $d\Omega^2_2 =d\theta^2 + \sin^2\theta d\phi^2$. The scale factor $a(t)$ evolves according to the Friedmann equation $H^2 = 8\pi\rho/3\mpl^2$, where $H(t)\equiv \dot{a}/a$ is the Hubble parameter\footnote{Dotted variables are derivatives with respect to cosmic time $t$.}. The equation of motion for $\phi$ reduces to the Klein-Gordon equation
\begin{equation} \label{KGphi}
  \ddot{\phi} + 3H\dot{\phi} + \frac{dV}{d\phi} = 0~,
\end{equation}
where the Hubble parameter is
\begin{equation}\label{eq:friedmann}
H^2 \equiv \frac{8\pi}{3\mpl^2}\left(\frac{1}{2}\dot{\phi}^2 + V(\phi)\right)~,
\end{equation}
and the corresponding pressure is given by
\begin{equation}\label{eos}
  p_{\mathrm{DM}} = \frac{1}{2}\dot{\phi}^2 - V(\phi)\,.
\end{equation}
If the oscillation of $\phi$ is sufficiently undamped, which is the case if $2m \gg 3H$ \footnote{This condition is obtained by interpreting \eqn{KGphi} as the equation of motion for a damped oscillator of the form $m\frac{dx^2}{dt^2} + c\frac{dx}{dt} + kx = 0$. The condition for undamped oscillation then becomes $c^2 - 4mk < 0$ or $9H^2 - 4m^2 < 0$. }, the friction term in \eqn{KGphi} can be neglected. The dynamics of $\phi$ are then approximately given by a simple harmonic oscillator $\phi(t) = \phi_0\cos{\big(mt\big)}$, whose pressure is

\begin{equation}
    p_{\mathrm{DM}} = \frac{\phi_0^2 m^2}{2}\Big(\sin^2{\big(mt\big)} - \cos^2{\big(mt\big)}\Big)~.
\end{equation}
As long as the oscillation period $T$ is sufficiently smaller than one Hubble time, this averages to zero over one Hubble time, i.e. $\langle p_{\mathrm{DM}} \rangle =0$, resulting in a matter-dominated expansion, which can be interpreted as a model for pressureless dust \cite{gu2007oscillating} at large scales.

Meanwhile, the massless scalar field $\xi$ provides the energy density perturbation that will trigger BH formation. In this paper, we exclusively consider initially static spherically symmetric perturbations and we leave the generalisation to fewer degrees of symmetry for future work. We choose the initial configuration of $\xi$ to be space dependent as
\begin{equation} \label{xi_profile}
    \xi(t=0,r) = \Delta\xi ~ \tanh{\Big[\frac{r - R_0}{\sigma}\Big]},
\end{equation}
where $\Delta\xi$, $\sigma$ and $R_0$ are the amplitude, width and the initial size of the perturbation respectively. We comment further on this perturbation shape in appendix \ref{section:initialdata}. The mass of the initial perturbation scales roughly as $R_0^2$.  We emphasise that this perturbation is non-linear, despite its moniker. Nevertheless, its massless nature means that it will propagate very close to the speed of light. 

We work with an overdensity shaped as a spherical shell to make the overdensity's initial length scale very obvious, i.e. it is natural to assign an initial size $R_0$ to the overdensity specified by \eqn{xi_profile},making it also straightforward to label overdensities as initially sub- or superhorizon. It is common in PBH formation research to work with overdensities that are initially peaked at the origin instead. However, this is more difficult to set up here, because the field providing the overdensity is massless and therefore only provides energy through its gradients. We expect that our results would hold up, at least qualitatively, if one were to use an overdensity peaked at the origin, although this should be checked numerically in future numerical studies. 

The initial staticity of the perturbation reflects the limited dynamics of a superhorizon perturbation that is frozen out. However, frozen out does not imply motionless. Nevertheless, the $\xi$ field is massless and accelerates rapidly to velocities around $c$, regardless of its initial velocity. Thus, combined with the fact that a static configuration simplifies the constraints the initial data must satisfy (which is discussed more detail in appendix \ref{section:initialdata}), we deem this an acceptable simplification.

Given the initial static configuration, we expect to see the perturbation split into an infalling mode, which drives the PBH formation, and an outgoing mode, which rapidly disperses. Our simulation box has periodic boundary conditions -- we ensure that the simulation domain is sufficiently large that the outgoing mode does not reach the boundary before PBH formation takes place. The background scalar field $\phi$ starts from rest, so that $\dot{\phi}=0$ and the initial Hubble parameter in the absence of inhomogeneities is $H_0^2 = 8\pi \mpl^{-2} V(\phi_0)/3$ via \eqn{eq:friedmann}.

Since the configuration of $\xi$  breaks the homogeneity of the initial spatial hyperslice, to set up the correct initial conditions for the metric, we will solve the Hamiltonian constraint. We choose a conformally flat ansatz for the 3-metric $\gamma_{ij}$,
\begin{equation}
dl^2 = \psi^{4}(dx^2+dy^2+dz^2)\,.
\end{equation}
Then, the Hamiltonian constraint reduces to an equation for the conformal factor $\psi$
\begin{equation} \label{eq:hamconstraint}
    \mathcal{H} = \Delta\psi - \frac{\psi^5}{12}K^2 + 2\pi\mpl^{-2}\psi^5\rho = 0\,,
\end{equation} 
where 
\begin{equation}
\rho = \rho_\xi + \rho_{\mathrm{DM}} = \frac{\psi^{-4}}{2}\left(\partial_i\xi\right)^2 + V(\phi_0) ~.
\end{equation}
Here the local expansion $K$ is the trace of the extrinsic curvature, $K = \mathrm{Tr} K_{ij}$.  \eqn{eq:hamconstraint} then becomes 
\begin{equation} 
\Delta\psi - \frac{\psi^5}{12}\left(K^2 - 9H_0^2\right) + \pi\mpl^{-2}\psi\left(\partial_i\xi\right)^2 = 0\,.
\end{equation}
We choose an initially expanding spacetime with $K = -3H_0$, so that the periodic integrability condition is satisfied\footnote{The initial energy density of the system is completely dominated by the scalar potential of the homogeneous matter field $\phi$, which allows us to neglect the perturbation field $\xi$. Then, the main contribution to the initial energy density is given by the (homogeneous) value of the potential and thus $K^2=24\pi V(\phi_0)$.} \cite{Bentivegna:2013xna,Yoo:2018pda,Clough:2016ymm}. This condition requires that the following integral over the simulation domain vanishes:
\begin{equation}
    \mathcal{I}_\mathcal{H} \equiv \int_D ~dV \left(\frac{2}{3}\psi^5 K^2 - 16\pi\psi^5\rho\right) = 0.
\end{equation}
There are similar conditions related to the three momentum constraints, see e.g. \cite{Aurrekoetxea:2022mpw}, which vanish trivially in our setup because we set $K$ to a constant value and all linear momentum densities vanish (the momentum constraints themselves vanish trivially for the same reason). The eventual Hamiltonian constraint only depends on the radial coordinate $r$ due to the spherical symmetry of the setup, and we solve for the conformal factor $\psi$ numerically
\begin{equation} \label{eq:ham_constraint}
\frac{\partial^2\psi}{\partial r^2} + \frac{2}{r}\frac{\partial\psi}{\partial r} - \frac{\psi^5}{12}\left(K^2 - 9H_0^2\right) + \frac{\pi\psi}{\mpl^{2}}\left(\frac{\partial\xi}{\partial r}\right)^2 = 0\,.
\end{equation}
We note that the solution of \eqn{eq:ham_constraint} is not strictly compatible with periodic boundary conditions, which causes discontinuities near the boundaries of the simulation domain. We make sure that these discontinuities are small enough to not affect the stability of the evolution and we make the simulation domain large enough so that they never propagate to the collapsing region before the simulation ends. Furthermore, it is worth specifying that the spherically symmetric scenarios described in this chapter can be studied in a computationally more efficient manner using a dimensionally reduced 1+1D code. However, we choose a 3+1 setup to make subsequent generalisation to scenarios with fewer degrees of symmetry more straightforward in terms of implementation. 

\section{Numerical methodology} \label{appendix:initialdata}

\subsection{Evolution equations}

\begin{figure}
    \centering
    \includegraphics[width=0.6\linewidth]{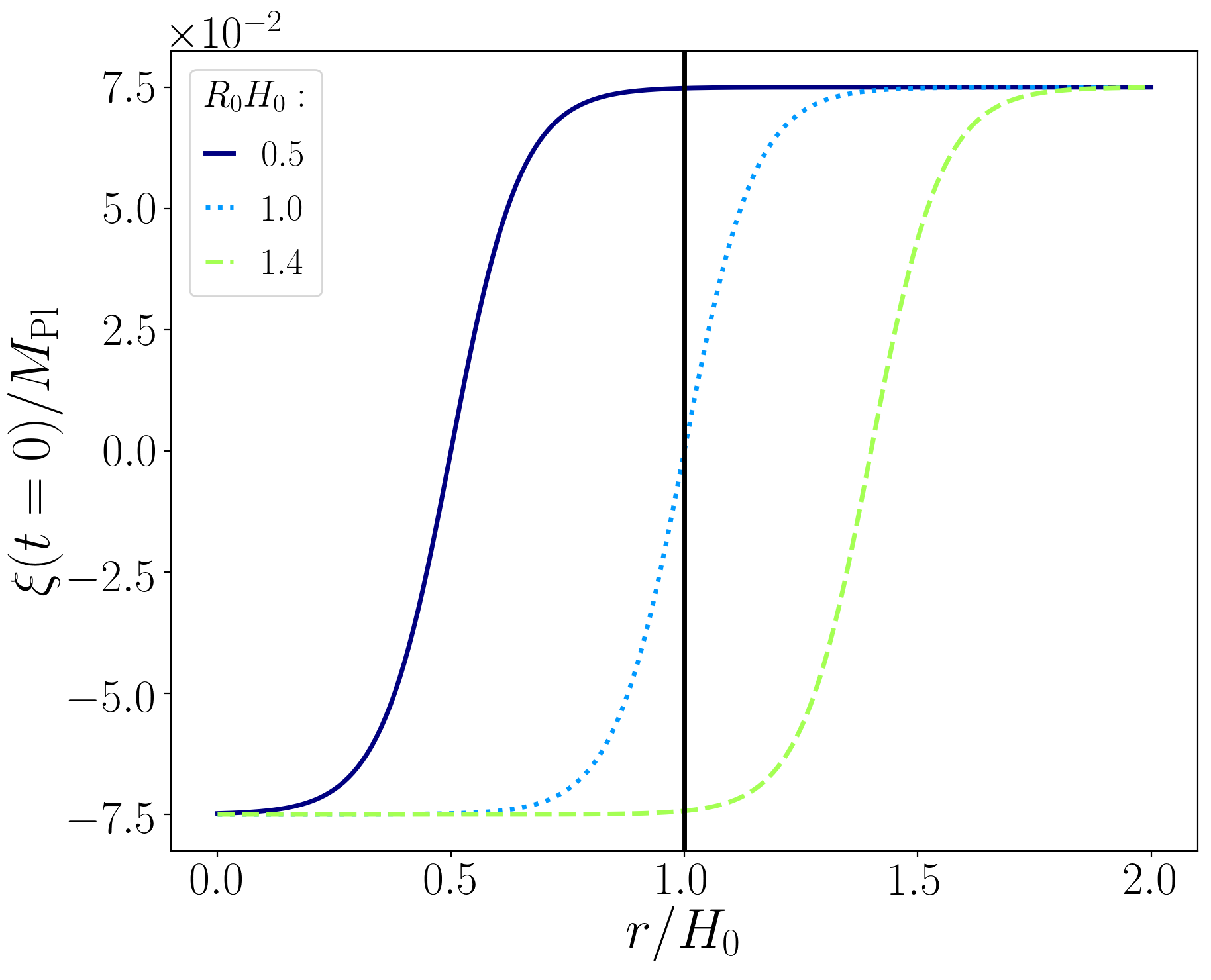}
    \caption{\textbf{Initial profiles} of the massless $\xi$ field, as a function of radial distance away from the center. Depicted are initially subhorizon, horizon and superhorizon sized perturbations, with same amplitude and thus same energy densities.} \label{fig::initial_profile_xi}
\end{figure}

This work was written based on simulations run using $\grchombo$ \cite{Clough:2015sqa}, with the CCZ4 formulation of the Einstein equations \cite{Alic_2012}. This formulation relies on the 4-dimensional spacetime being foliated into 3-dimensional non-intersecting hyperslices, whose intrinsic curvature is described by a spatial metric $\gamma_{ij}$, whilst their embedding in the 4-dimensional spacetime is encoded in the extrinsic curvature $K_{ij}$. The line element is decomposed as
\be 
    ds^{2} = -\alpha^2 dt^2 + \gamma_{ij}\big(dx^i + \beta^i dt\big)\big(dx^j + \beta^j dt\big) \,,
\ee
where the lapse $\alpha$ and the shift vector $\beta^i$ are user-specified gauge functions. The spatial metric $\gamma_{ij}$ is often written as a product of a  \textit{conformal factor} $\psi$ and a \textit{background} (or \textit{conformal}) metric $\bar{\gamma}_{ij}$, so that the determinant of the conformal metric equals one,
\be 
    \gamma_{ij} = \psi^4\bar{\gamma}_{ij}\,, \q \det{\bar{\gamma}_{ij}} = 1\,, \q \psi = \frac{1}{(\det{\gamma_{ij}})^{1/12}} \,.
\ee

Time evolution proceeds along the vector $t^a = \alpha n^a + \beta^a$, where $n^a$ is the unit normal vector to the hyperslice that is being evolved. The gauge functions $\alpha$ and $\beta^i$ are specified on the initial hyperslice, and then evolved using evolution equations suitable for long-term stable numerical simulations. The choice in this work is
\be
\begin{split}
    \partial_t\alpha &= -\mu_{\alpha_1}\alpha K + \beta^i \partial_i \alpha\,,\\
    \partial_t\beta^i &= \eta_{\beta_1} B^i\,,\\
    \partial_t B^i &= \beta^j\partial_j\bar{\Gamma}^i + \partial_t \bar{\Gamma}^i - \eta_{\beta_2}B^i\,,
\end{split}
\ee
where $\bar{\Gamma}^i\equiv \bar{\gamma}^{jk}\bar{\Gamma}^i_{jk}$, and $\bar{\Gamma}^i_{jk}$ are the conformal Christoffel symbols associated to the  conformal metric via the usual definition
\be 
    \bar{\Gamma}^{i}_{jk} = \frac{1}{2}\bar{\gamma}^{il}\big(\bar{\gamma}_{lk, j} + \bar{\gamma}_{lj, k} - \bar{\gamma}_{jk, l}\big)\,.
\ee
This choice is known as the \textit{moving-puncture gauge} \cite{Bona:1994dr,Baker:2005vv,Campanelli:2005dd,vanMeter:2006vi}. We choose $\mu_{\alpha_1} = 0.5,\, \eta_{\beta_1} = 0.75,\, \eta_{\beta_2} = 10^{-4}$, which helps us control constraint violation growth and evolve black hole spacetimes.

\subsection{Initial data} \label{section:initialdata}

\begin{figure}
    \centering
    \includegraphics[width=.6\linewidth]{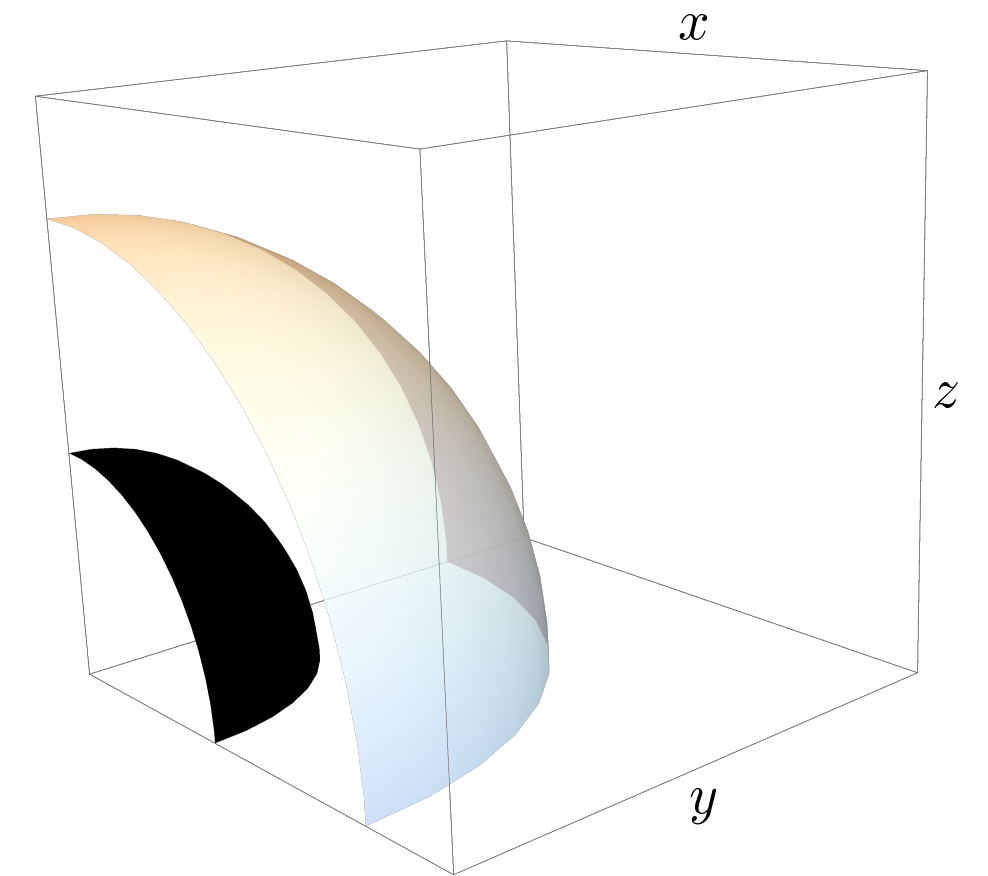}
    \caption{\textbf{Initial setup} of collapsing superhorizon overdensity (transparent shell) and inner initial Hubble horizon depicted in black. Using symmetric (reflective) boundary conditions we simulate an eighth of the system.} \label{fig::initial_setup}
\end{figure}

The matter content of this work is comprised by a massive $\phi$ field that dominates the background dynamics, and an inhomogeneous massless $\xi$ field which provides the local overdensity through its gradients. We choose potentials
\be 
    V_\phi(\phi) = \frac{1}{2}m^2\phi^2, \qquad V_\xi(\xi) = 0\,,
\ee
and spherically symmetric initial field configurations
\be 
\begin{split}
    \phi(t = 0, x^i) &= \phi_0
    \,,\\
    \xi(t = 0, x^i) &= \Delta\xi\tanh{\Big[\frac{r - R_0}{\sigma_0}\Big]}
    \,,\\
    \frac{\partial\phi(t = 0, x^i)}{\partial t} &= \frac{\partial\xi(t = 0, x^i)}{\partial t} = 0\,.
\end{split}
\ee
Examples of initial field profiles for the massless $\xi$ field are given in Fig. \ref{fig::initial_profile_xi}. The hyperbolic tangents allow us to localize the gradients in the $\xi$ field, and thereby its energy density, as the field is without potential. In our simulations, $mH_0^{-1} = 62.6$. We also choose a conformally flat metric $\bar{\gamma}_{ij}=\delta_{ij}$, so the energy density on the initial hyperslice is given by
\begin{equation}
    \rho(t = 0, x^i) = \frac{\psi^{-4}}{2}\delta^{ij}\partial_i\xi\partial_j\xi  
    +\frac{1}{2}m^2\phi_0^2\,,
\end{equation} 
which corresponds to a shell-like overdensity in a dark matter environment.

\begin{figure}
    \centering
    \includegraphics[width=0.6\linewidth]{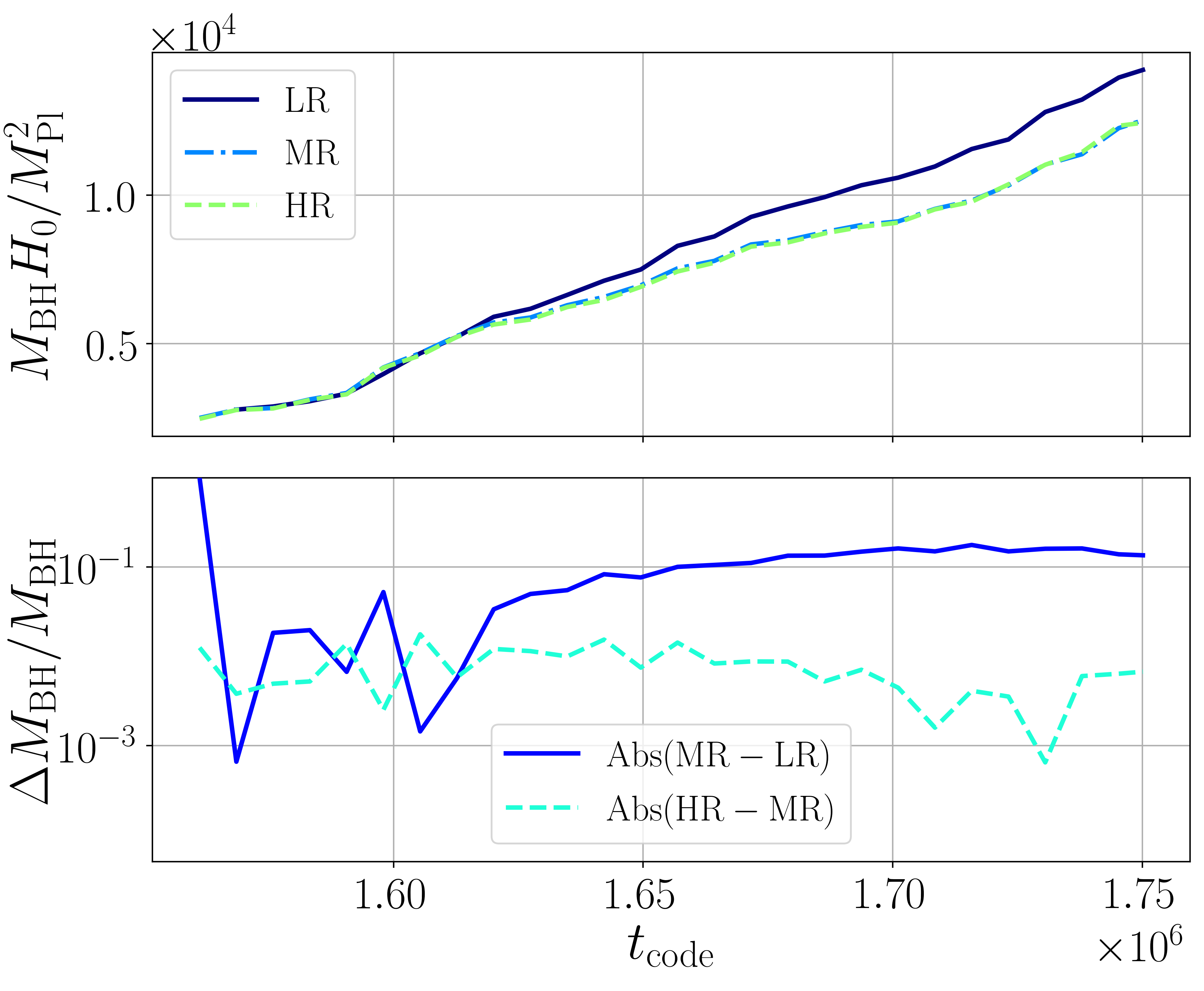}
    \caption{\textbf{Convergence test} of the black hole mass formed from an initial perturbation of $R_0=1.6H_0^{-1}$ and $\Delta\xi\mpl^{-1} = 0.075$. Top panel shows the estimated mass for three base grid resolutions $N_\mathrm{LR}=80$, $N_\mathrm{MR}=96$, $N_\mathrm{HR}=128$. Bottom panel shows errors in mass measurements between high-middle and middle-low resolutions showing convergence to $1\%$.} \label{fig::convergenceR1.6}
\end{figure}

This setup is spherically symmetric and we reduce the computational cost of evolution by simulating one eighth of the system using symmetric boundary conditions. A schematic depiction of the initial setup is given in Fig. \ref{fig::initial_setup}. 

The equation of motion for a massive homogeneous field $\phi$ is given by the Klein-Gordon equation
\be \label{KGeom}
    \ddot{\phi} + 3H\dot{\phi} + \frac{\partial V(\phi)}{\partial\phi} = 0\,,
\ee
where $H\equiv \dot{a}/a$ is the Hubble function defined with the scale factor of the universe $a(t)$.

For the universe to stay matter-dominated for an extended period of time, we require the solution of \eqn{KGeom} to be an undamped oscillation. This constrains the initial value of $\phi$ to
\be 
    \phi_0 < \frac{1}{\sqrt{3\pi}}\mpl\,,
\ee
and we use $\phi_0 = 7.8 \times 10^{-3}\mpl$ for all our simulations.

\subsection{Grid details and convergence testing} \label{appendix::2}
The length of the simulation box is $L_\textrm{box} = 2/H_0$ ($L_\textrm{box} = 2.5/H_0$) for initially subhorizon (superhorizon) perturbations. $\grchombo$ uses AMR to dynamically add resolution in regions of the box that are physically interesting. The grid size of the coarsest level, which covers the entire simulation box, is $48^3$ and the refinement factor between a given AMR level and the next is 2. The Courant factor $\Delta t/\Delta x = 0.0384$ and is equal for each AMR level, since level $l + 1$ is evolved using timesteps twice smaller than for level $l$. The maximum number of levels achieved for these simulations is 4, including the base level. Grid points are tagged for refinement when either $\rho_\xi$ or the gradients of $K$ pass a threshold value. 

We test the robustness of our numerical results by finding the mass of the black hole formed from an initial perturbation of radius $R_0=1.6 H_0^{-1}$ and $\Delta\xi= 0.15H_0^{-1}$, using three different base grid resolutions, namely $N_\mathrm{LR}=80$, $N_\mathrm{MR}=96$, $N_\mathrm{HR}=128$. Fig. \ref{fig::convergenceR1.6} shows the mass obtained with an apparent horizon finder \cite{Thornburg:2003sf} for these three runs, indicating that convergence is achieved.

Additionally, we made sure the code reproduces the FLRW limit for appropriate initial conditions. If one uses the same initial setup as described in appendix \ref{section:initialdata}, but gives the field $\xi$ a uniform value throughout the simulation box, evolution should proceed in an FLRW manner, as the simulated universe is now completely homogeneous. We checked this by averaging the values of the energy density $\rho$ and the scale factor $a$ over the box, and tracking these throughout the evolution. We satisfactorily find that the scaling between these two quantities is then $\rho \sim a^{-3}$, as expected for a matter-dominated FLRW universe.

\section{Primordial black hole formation} \label{sect:bhformation}

Our main scale of reference will be the initial size of the unperturbed  Hubble horizon $H_0$, which is fixed for all simulations by choosing the initial value of the scalar field $\phi$ to be $\phi_0=7.8\times 10^{-3}\mpl$, with $m = 62.6 H_0$. For larger values of $\phi_0$, the $\phi$-oscillation amplitude decays rapidly during the first Hubble time, which makes it hard to track its evolution accurately. At the same time, value for $m$ makes sure that $\phi$ performs around 10 oscillations during the first Hubble time, which we find is sufficiently rapid to cause a matter-like universe expansion. At the same time, it is not so fast that we need to reduce the size of the Courant factor to a value so small that it would make the simulation unnecessarily computationally heavy. In the following, we will vary the initial size of the perturbation from subhorizon to superhorizon, $R_0 H_0 \in \left[0.575,~1.6\right]$. We will also vary the perturbation amplitude within the range $\Delta\xi\mpl^{-1} \in \left[0.075,~0.12\right]$, whilst keeping the initial width fixed to $\sigma_0=0.15 H_0^{-1}$, such that the ratio between the maximum gradient energy density to matter energy density is $\rho_{\xi}/\rho_\phi\sim 1$. As the perturbation mass is small compared to the Hubble mass in all scenarios we consider, the background expansion is not significantly influenced by the perturbation's presence.

Our simulations reach just into the superhorizon regime and at the moment, we are limited from exploring this regime further numerically by the computational costs of such simulations. Nonetheless, we will argue in section \ref{sect:accretion} that our simulations already capture some superhorizon dynamics. Moreover, we argue that it may not be necessary to probe the superhorizon regime much further, as our model is effectively pressureless. The corresponding speed of sound is therefore small, and physics already propagate very slowly on scales just past the horizon size.

We find that, for both subhorizon and superhorizon perturbations, PBH formation occurs via two possible mechanisms -- a \emph{direct collapse} mechanism whereby the PBH is formed by the initial perturbation of $\xi$ itself, and a \emph{post-collapse accretion} mechanism whereby the initial perturbation of $\xi$ sources a gravitational potential that then accretes the background matter  $\phi$ until a PBH forms. What determines the type of PBH formation process depends (unsurprisingly) on both the geometry and mass of the initial perturbation shell, as well as the expansion rate of the background cosmology. We will discuss these two mechanisms below \footnote{In this paper, we have used the background energy density, or equivalently the background scale factor, as time. This corresponds to the cosmic time infinitely far away from the centre of the PBH. However, in numerical relativity simulations, the foliation of spatial hyperslices is dynamically driven by the so-called puncture gauge, which is required to enforce numerical stability in the presence of future singularities. In that context, we have assumed that the mass of a PBH is identified with its foliation, when in principle one should identify it via null geodesics from the black hole horizon to infinity.  In other words, what the cosmological observer (with their own clocks tuned to cosmic time) far away from the black hole deduces as the properties of the black hole, e.g. its mass, should be information that is emitted (by light or GW) from the black hole and then propagated to this observer. Our approximation assumes this ``time lag'' between the local time (i.e. foliation time) and the cosmic time is negligble. This inaccuracy should be minor and would not affect the main conclusions of the paper. } We expect that .

\subsection{Direct collapse} \label{sect:direct_collapse}

\begin{figure}[t!]
    \centering
    \includegraphics[width=0.65\linewidth]{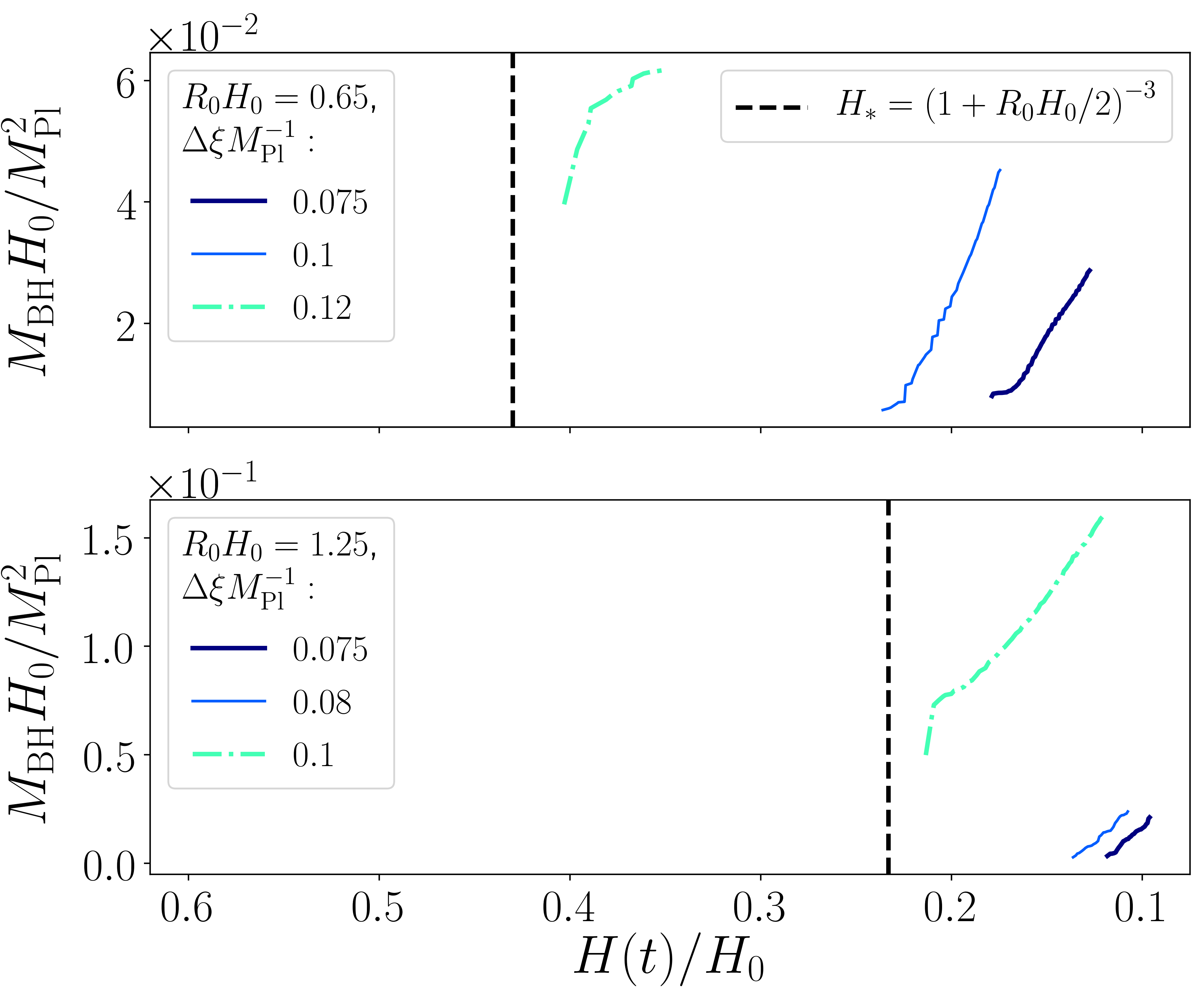}
        \caption{\textbf{Black hole formation for different perturbation amplitudes.} The top (bottom) panel shows mass of the formed BHs as a function of the Hubble parameter $H(t)$ at infinity, for subhorizon (superhorizon) collapse respectively. Vertical dashed black lines correspond to the time at which the perturbation reaches the centre according to \eqn{eqn:astar}. BHs formed through direct (accretion) collapse are shown in dash-dotted (solid) lines. For accretion collapse BHs, increasing the amplitude $\Delta\xi$ makes that the BH forms earlier with a smaller initial mass. Our simulations are in good agreement with the hoop conjecture prediction that the threshold is $\Delta\xi\mpl^{-1}\approx 0.1$ for $R_0H_0 =0.65$ and $\Delta\xi\mpl^{-1}\approx 0.07$ for $R_0H_0 =1.25$. In direct collapse, part of the collapsing perturbation ends up within the black hole, corresponding to a larger initial mass.}
    \label{fig:figure_A1}
\end{figure}

In the direct collapse scenario, the perturbation collapses towards its geometric centre (to which we will henceforth simply refer as the centre) and forms a black hole directly on its own, without significant accretion of the background DM density. We will now estimate the time a shell takes to undergo direct collapse. The perturbation field $\xi$ is massless, and hence if we ignore the backreaction of the shell on the background geometry, it propagates along null-like geodesics \footnote{If the perturbation field $\xi$ instead has mass $m_{\xi} \sim m_{\phi}$, then the collapse time is roughly the free fall time $\tau_{\mathrm{ff}} \sim \sqrt{\mpl/m_{\xi}\xi}$ which is also roughly one Hubble time.}. In an FLRW background, the scale factor $a$ is given by the null element $dt^2 =a^2(t)dr^2$. Solving this kinematic equation, the co-moving radius of the shell is then
\begin{equation}
\rsh = R_0a_0^{-1}-2H_0^{-1}a_0^{-1}\left[\left(\frac{a}{a_0}\right)^{1/2}-1\right]~,\label{eqn:shell_trajectory}
\end{equation}
where we set $a_0\equiv 1$ to be the initial scale factor at the initial time. The value of the scale factor at the moment the shell collapses to the centre $a_*$ is then the solution to the equation $\rsh(a_*)=0$, namely
\begin{equation}
a_* = a_0\left[1+ \frac{R_0H_0}{2}\right]^2~,\label{eqn:astar}
\end{equation}
which happens after roughly a Hubble time. 
Notice that $a_*$ is independent of the initial mass and depends only on $R_0$. We show in Fig. \ref{fig:figure_A1} that this analytic estimate is in good agreement with our numerical results.

To determine whether or not a given initial perturbation shell will undergo direct collapse into a black hole, we consider the \emph{width} of the shell at the time when the shell reaches the centre $\sigma_* = \sigma(a_*)$. Ignoring backreaction again, since the field $\xi$ is massless, the width of the shell as it collapses towards the centre scales as the expansion rate, i.e. 
\begin{equation}
\sigma(a) = \sigma_0 a ~. \label{eqn:width_eom}
\end{equation}
Thus the width of the shell when it reaches the centre is simply $\sigma(a_*) = \sigma_0a_*$. At this moment, applying the hoop conjecture\footnote{The presence of an expanding background modifies the hoop conjecture somewhat in general \cite{Saini:2017tsz}, but we checked that the effects are negligible in our analytic estimates. } suggests that if the condition
\begin{equation}
\sigma(a_*) < 2GM_{\mathrm{infall}}~,\label{eqn:hoop}
\end{equation}
is satisfied,  where $M_{\mathrm{infall}}$ is half\footnote{The infalling mass is half the initial mass, since the other half will radiate outwards to infinity, so the shell's initial (vanishing) momentum is conserved.}
the initial mass of the shell obtained by integrating the gradient energy of the profile $\xi(r)$ roughly given by  \eqn{xi_profile} in flat space
\begin{equation}
M_{\mathrm{infall}} \approx \frac{1}{2}\int dr~4\pi r^2 \frac{1}{2}\left(\frac{\partial \xi}{\partial r}\right)^2~,\label{eqn:infall_mass}
\end{equation}
then a black hole will form through direct collapse. This result is again in good agreement with our numerical results, as shown in Fig. \ref{fig:figure_A1}.

We note that because we do not simulate a perfect fluid with vanishing equation of state, we cannot assume that perturbations with arbitrary small amplitudes eventually lead to BH formation, even through accretion collapse. However, all perturbations whose collapse is presented in this work collapse to BHs, and we have not determined a minimum amplitude below which a BH does not form.

\begin{figure}[t!]
    \begin{subfigure}[t]{0.45\linewidth}
      \centering
      \includegraphics[width=1.\linewidth]{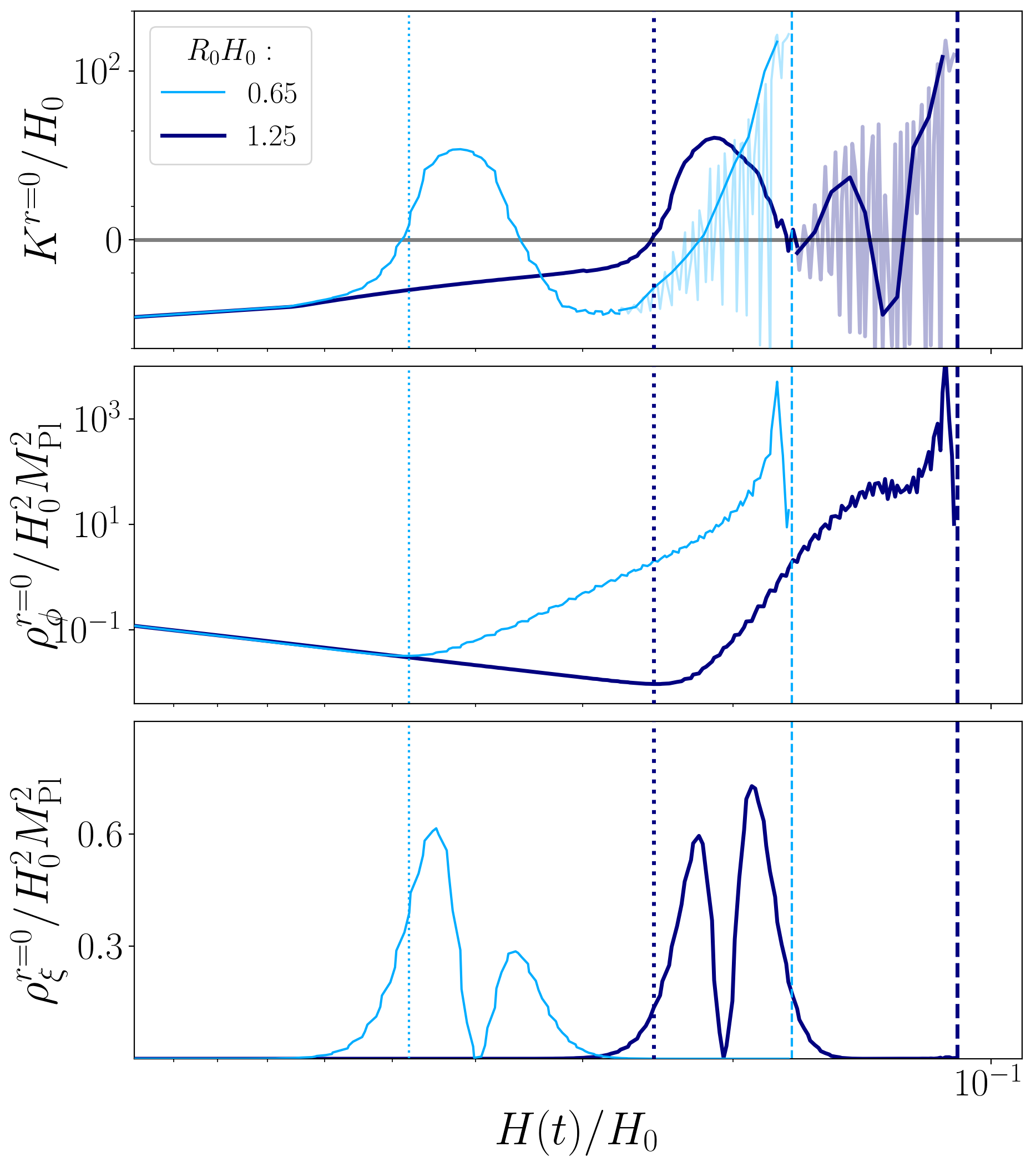}
      \caption{In this subfigure, a representative subhorizon (superhorizon) is shown in thin light blue (thick dark blue) in the {\bf accretion collapse case}. Note that here, $\rho_\xi$ disperses completely before the BH forms, and BH formation happens late compared to the direct collapse case. }
      \label{fig:rhos_vs_r_accretion}
    \end{subfigure}\hfill
    \begin{subfigure}[t]{0.45\linewidth}
      \centering
      \includegraphics[width=1.\linewidth]{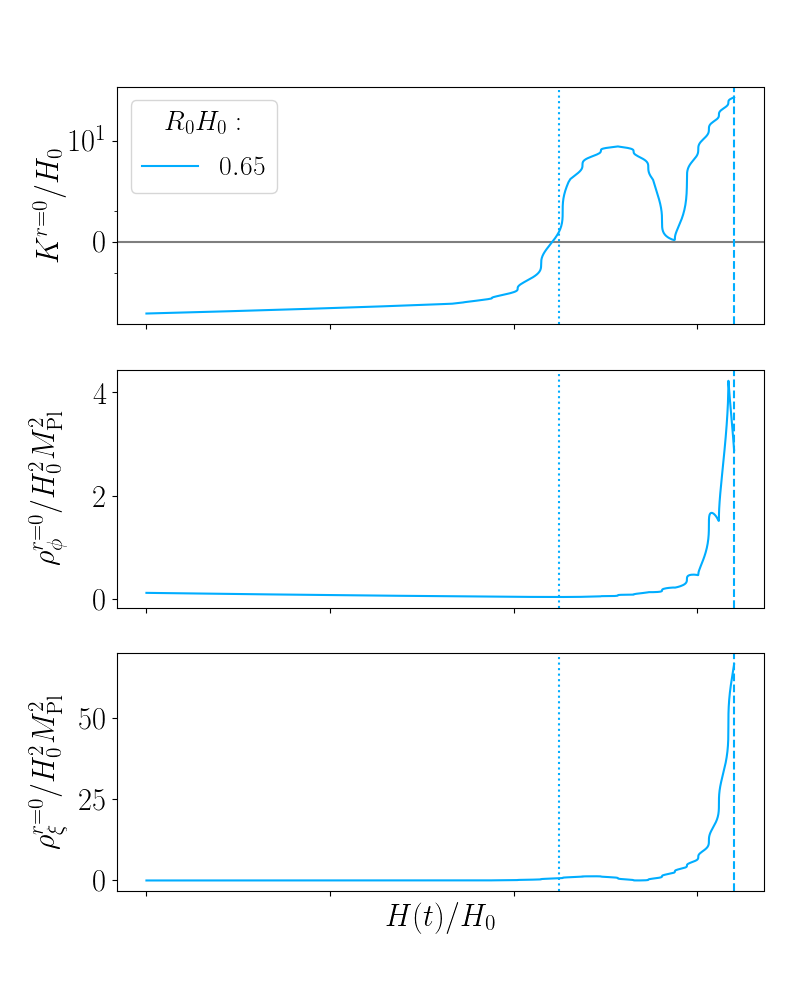}
      \caption{In this subfigure, a representative subhorizon is shown in thin light blue in the {\bf direct collapse case}. Here, $\rho_\xi$ increases at the center, decreases and increases again as the shell rebounds back into the center. $\rho_\xi$ is still high when the BH forms.}
      \label{fig::rhos_vs_r_direct}
    \end{subfigure}\hfill
    \caption{\textbf{Evolution of the local expansion $K$ and energy densities $\rho_{\phi}$ and $\rho_\xi$} at the centre of the collapse $r=0$, as a function of the Hubble parameter $H(t)$ at infinity -- recall that $K>0$ corresponds to locally collapsing spacetime. The top, middle and bottom panels show the evolution of the expansion, the background energy density and gradient energy density respectively. Initially, the background energy density decays as $\rho_\mathrm{DM} \sim a(t)^{-3}$. When the perturbation reaches the centre (dotted vertical lines) and disperses, gravitational effects decouple the system and stop the local expansion, acting as a seed for the accretion of the background matter $\rho_\mathrm{DM}$. The accretion of the background matter continues until and after a black hole forms (dashed vertical lines). We attribute the oscillatory behaviour of $K$ at late times to gauge effects in regions of high energy density.}
    \label{fig::rho_vs_r}
\end{figure}

The fact that such simple estimates agree with our numerical results suggests that the backreaction of the perturbation on the background dynamics is not very important, at least at the level of determining when and how a black hole will form, even if the shell density is large and locally $\rho_\xi > \rDM$. This is backed up by our numerical simulations, where we see that the $\rDM$ profile is not strongly affected by the presence of $\rho_\xi$, at least initially, as can be seen in a video of the numerical evolution of the energy densities \href{https://youtu.be/4N5e2RnUkmU}{here} \cite{Movie2}.

\subsection{Accretion collapse} \label{sect:accretion}

\begin{figure}
    \centering
    \includegraphics[width=0.65\linewidth]{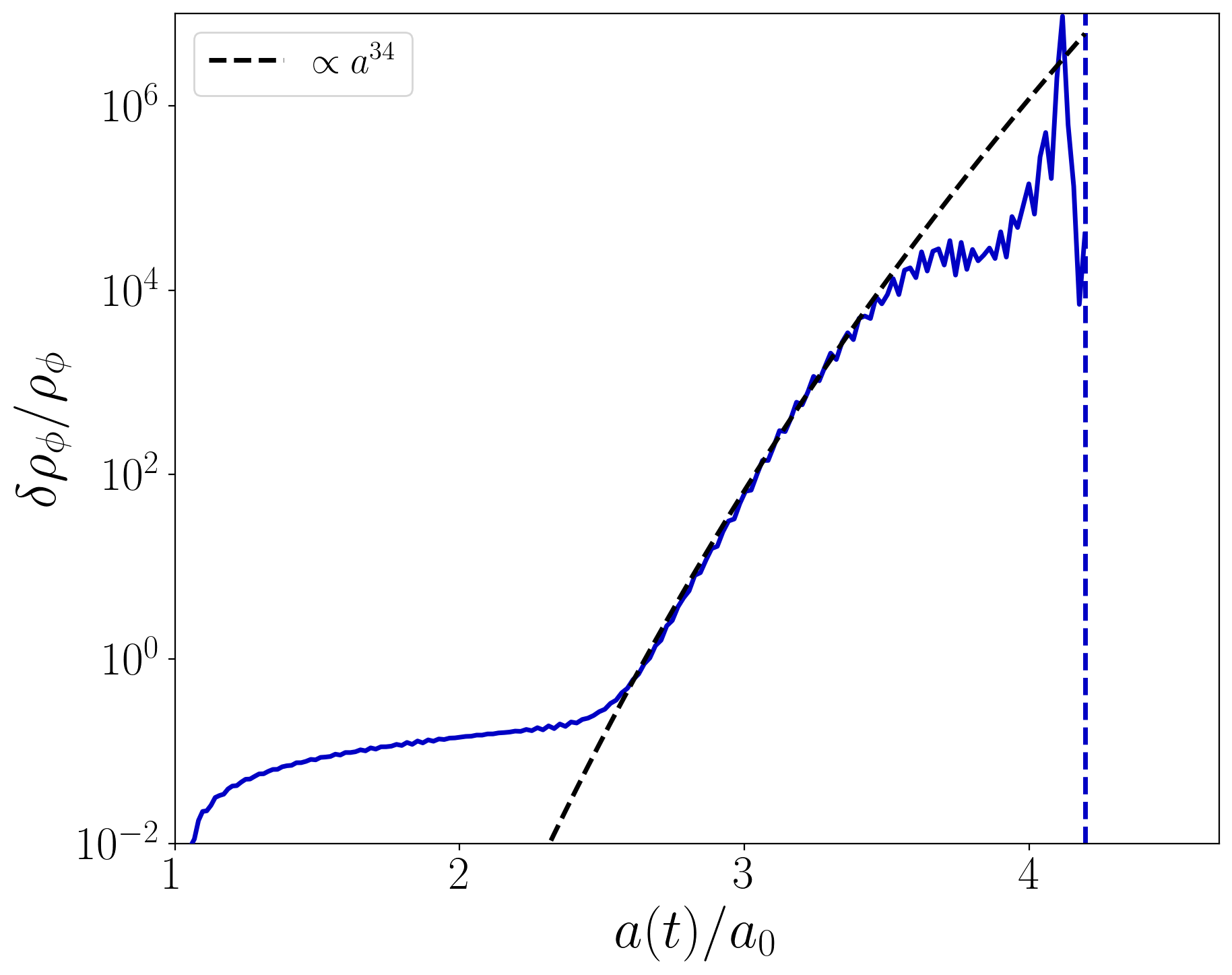}
        \caption{\textbf{Rate of growth for the local matter overdensity} $\delta \rho_{\phi}/\rho_{\phi}$ at the centre of the collapse is well beyond the linear approximation as $\delta\rho_\phi/\rho_\phi\propto a^{34}$. Near black hole formation (vertical dashed blue line), the accretion rate tapers off, although some of this tapering effect is due to our gauge condition.}  
    \label{fig:deltarho_pert}
\end{figure}

On the other hand, if $\sigma(a_*) > 2GM_{\mathrm{infall}}$, a black hole does not form directly. In this case, the energy density of the perturbation $\rho_{\xi}$ disperses after reaching the centre and becomes locally sub-dominant to the background energy density $\rdm$. Nevertheless, the presence of $\xi$ generates a gravitational potential well in the centre, which seeds accretion of the background DM and eventually causes a collapse into a black hole. We illustrate this process in Fig. \ref{fig:rhos_vs_r_accretion}, providing an equivalent graph for the direct collapse mechanism in Fig. \ref{fig::rhos_vs_r_direct}.

In this phase, the initially homogeneous and expanding background spacetime is made to locally collapse by the perturbation, with the expansion $K$ locally changing sign from negative (expanding) to positive (contracting), decoupling the region near the centre from the rest of the expanding background. The local matter begins to accrete at an extremely high rate $\delta \rdm/\rdm \propto a^{34}$, as shown in Fig. \ref{fig:deltarho_pert}. Once sufficient DM mass has accumulated, a PBH forms. This process takes around an e-fold to complete.  This rapid accretion rate is much higher than that predicted from linear theory, which is $\delta \rdm/\rdm  \propto a$ \cite{Gunn:1972,Shapiro_1999}, indicating that the process is highly non-linear.

\begin{figure}
\centering
\includegraphics[width=0.65\linewidth]{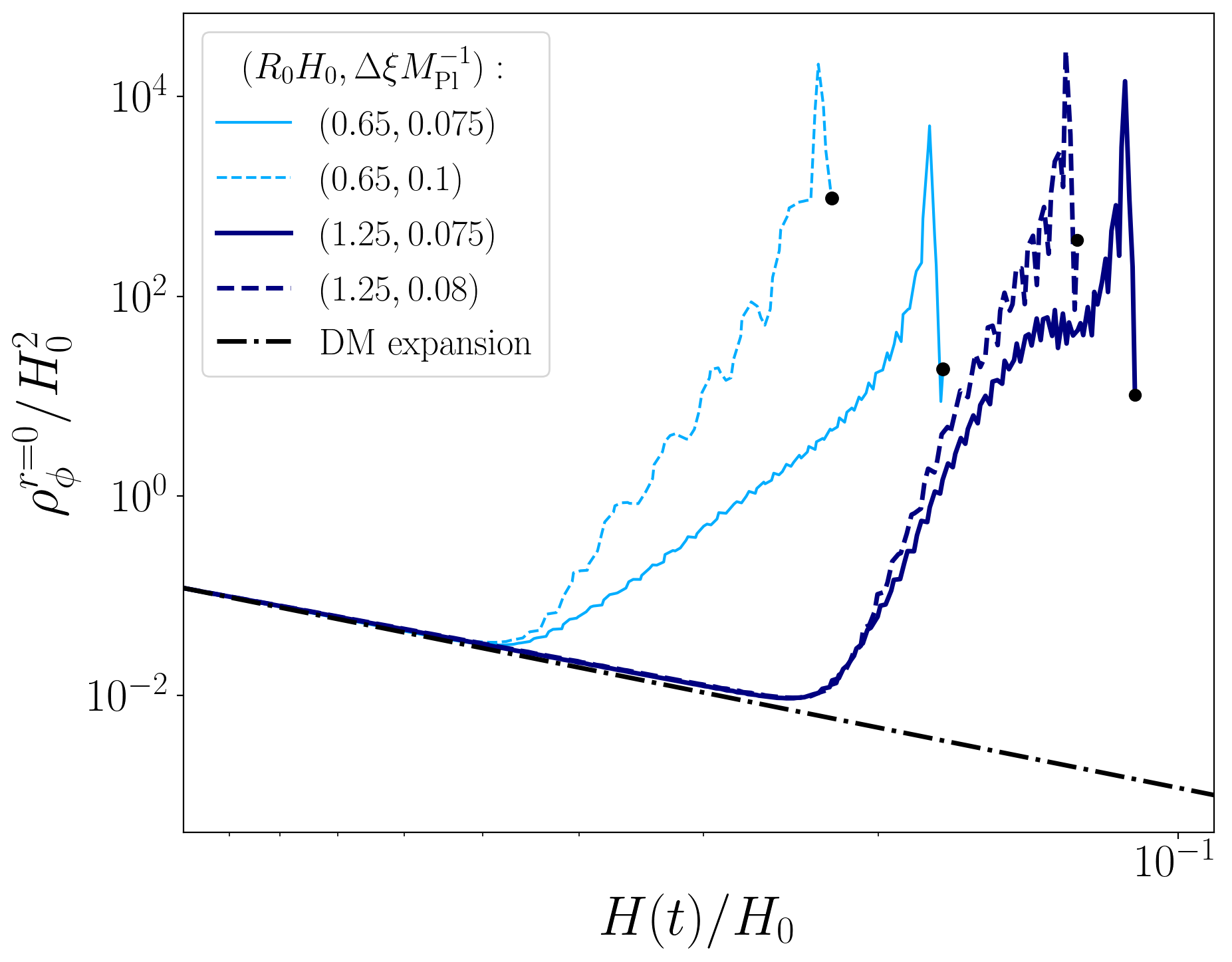}
    \caption{\textbf{Evolution of the matter energy density} at the centre of the collapse for a set of initial radii $R_0$ and amplitudes $\Delta\xi$. For same radii perturbations, accretion begins at the same time. However, the accretion rate is larger for larger amplitudes, which results in the formation of a black hole at an earlier time.} 
\label{fig:rhoDM_amp}
\end{figure}

From our simulations, we note two salient points. Firstly, if we consider shells that undergo accretion collapse, for fixed initial amplitude $\Delta\xi$, the smaller the initial $R_0$ (and therefore the smaller the mass) of the initial perturbation, the more massive the initial mass of the PBH. This somewhat counter-intuitive result is due to the fact that the PBH forms via accreting DM, thus a more massive seed will generate a \emph{steeper} potential well, and hence the Schwarzschild radius is reached earlier and at a smaller value for the PBH mass. To confirm this, we checked that keeping $R_0$ fixed but increasing $\Delta \xi$ also yields a less massive initial PBH -- this is true for both subhorizon and superhorizon cases, as can be seen in Fig. \ref{fig:figure_A1}. In Fig. \ref{fig:rhoDM_amp} we plot the matter energy density for two different values for $R_0$ and $\Delta\xi$. We confirm that larger amplitudes (and thus more massive seeds) result in a faster accretion rate. 

Secondly, as $R_0$ approaches $H_0^{-1}$, the expansion rate of the universe begin to exert a competing effect. For shells with larger $R_0$, it takes \emph{longer} for the shell to reach the centre, and thus a smaller $\rho_{\xi}$ and less steep potential when accretion begins. This leads to an increase in the initial mass of the PBH following our argument above -- resulting in the ``bump'' in the initial mass of the black holes (e.g. the black dots in Fig. \ref{fig:mass_vs_rho}).

We note that the distinction between direct and accretion collapse depends mostly on how deep the gravitational well is that is formed at the moment that the overdensity reaches its own centre. When the overdensity is relatively light (heavy), the immediate gravitational well is shallow (deep) and we expect a BH to form via accretion(direct) collapse. We show that this distinction is manifest for the case of two scalar fields, but we expect that the same distinction will be present for other number of fields, including when there is only one matter field, as long as the universe's expansion is suitably matter-like. 

\section{PBH growth and final Mass} \label{sect:mass_distribution}

\begin{figure}[t!]
    \begin{subfigure}[t]{0.48\linewidth}
      \centering
      \includegraphics[width=1.\linewidth]{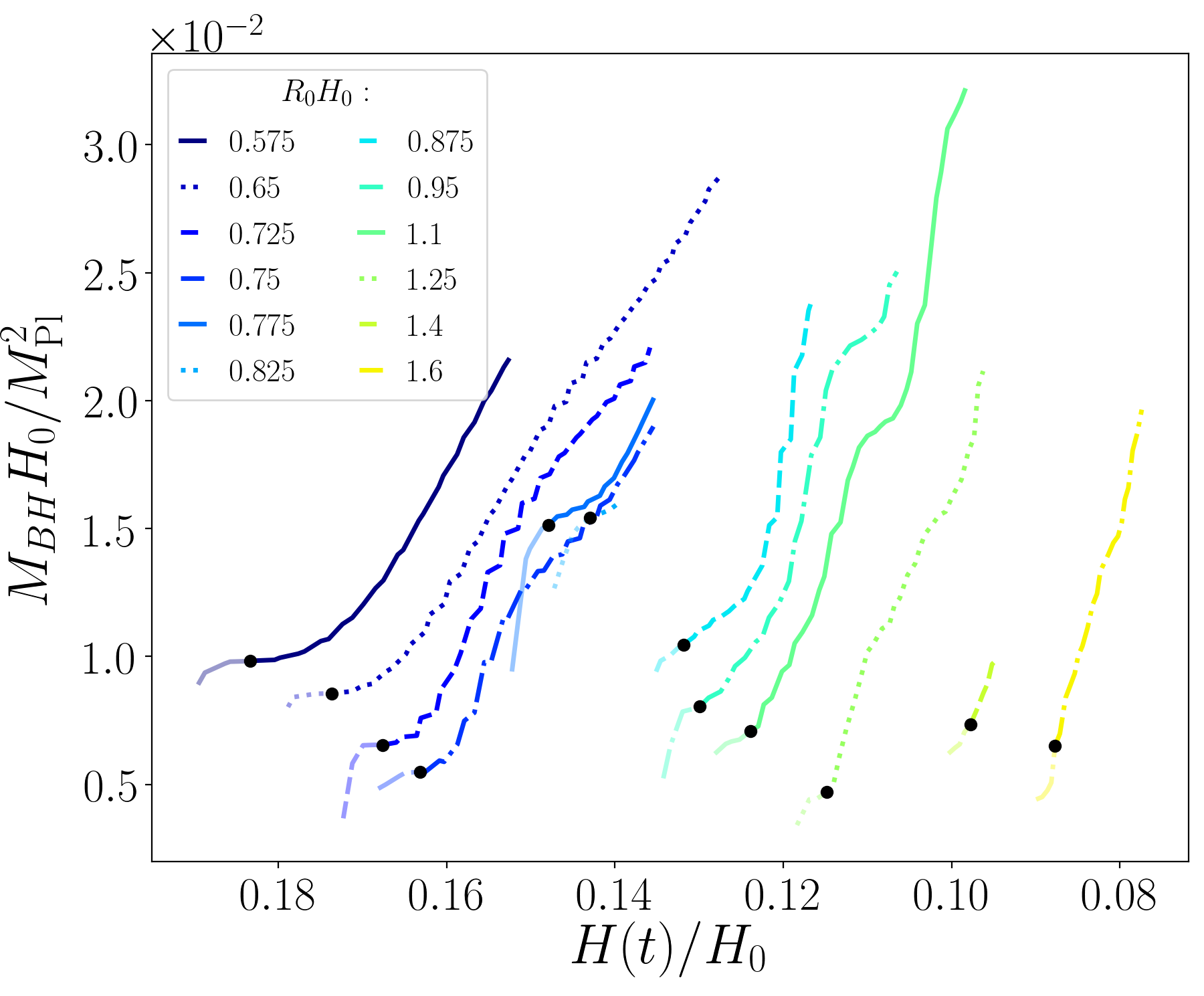}
      \caption{\textbf{Summary of simulations} showing the black hole $M_\mathrm{BH}$ as a function of the Hubble parameter $H(t)$ at infinity, for various initial radii $R_0$, for $\Delta\xi\mpl^{-1} = 0.075$. The growth rate of the black hole mass is larger for larger shells, because they source a larger gravitational potential. Black dots correspond to the initial black hole masses at formation, identified using an apparent horizon finder. Shaded parts of the curves before the black dots where the apparent horizon finder results are inaccurate, e.g. because the apparent horizon is not covered by a sufficient number of grid points.}
      \label{fig:mass_vs_rho}
    \end{subfigure}\hfill
    \begin{subfigure}[t]{0.48\linewidth}
      \centering
      \includegraphics[width=1.\linewidth]{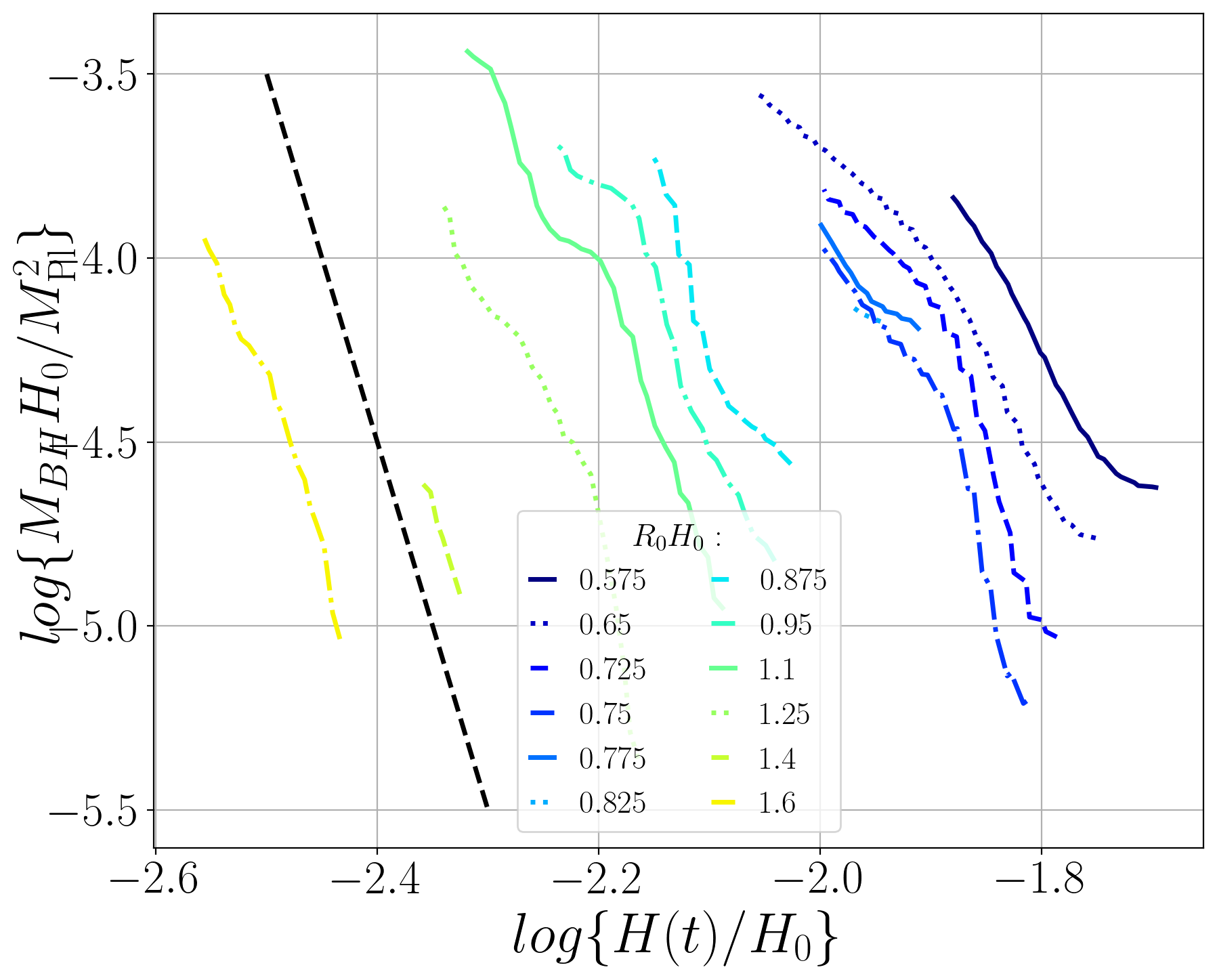}
      \caption{The same data as in figure \ref{fig:mass_vs_rho} on a log-log plot. Time runs from right to left on the $x$-axis. The dashed black line on the left hand side shows the relation $M_\textrm{BH}\propto H^{-10}$ to make clear that for all perturbations shown, there is a large negative power law relation between these two quantities, at least in the early stages of the accretion process.  }
      \label{fig::mass_vs_rho_log}
    \end{subfigure}\hfill
    \caption{\textbf{Evolution of the local expansion $K$ and energy densities $\rho_{\phi}$ and $\rho_\xi$} at the centre of the collapse $r=0$, as a function of the Hubble parameter $H(t)$ at infinity -- recall that $K>0$ corresponds to locally collapsing spacetime. The top, middle and bottom panels show the evolution of the expansion, the background energy density and gradient energy density respectively. Initially, the background energy density decays as $\rho_\mathrm{DM} \sim a(t)^{-3}$. When the perturbation reaches the centre (dotted vertical lines) and disperses, gravitational effects decouple the system and stop the local expansion, acting as a seed for the accretion of the background matter $\rho_\mathrm{DM}$. The accretion of the background matter continues until and after a black hole forms (dashed vertical lines). We attribute the oscillatory behaviour of $K$ at late times to gauge effects in regions of high energy density.}
\end{figure}

In the cases of both direct and accretion collapse, the initial mass of the PBH formed is small compared to the Hubble horizon, $\mbh H_0\sim 10^{-2}\mpl^2$ -- see Fig. \ref{fig:mass_vs_rho}.  Once the initial PBH has formed, the PBH accretes DM from its surroundings in the growth phase at a rate that depends on the steepness of the potential and the density of the surrounding DM ``scalar cloud'' \cite{Hui:2019aqm,Clough:2019jpm,Bamber:2020bpu,Hui:2021tkt}. In general and regardless of the details of the parameters, we find that the initial accretion rate is much higher than the linear theory prediction of $\delta \rdm/\rdm \propto a$, as mentioned above.  This growth rate is roughly constant, at least initially, and its contribution to the mass of the PBH will rapidly dwarf that of its initial mass.  

In a matter-dominated universe,  naive Newtonian collapse suggests that the maximum mass of the black hole is bounded by $\mbh H\sim \alpha \mpl^2$ \cite{Carr:1974}, where $\alpha\lesssim 1$ is some constant which depends on the exact details of the accretion. This suggests that a BH will not grow more rapidly than self-similarly once its mass is of the same order as the Hubble mass. In references \cite{Harada:2004pf,Harada:2004pe}, it was demonstrated numerically that while the initial growth can be rapid, it will not achieve self-similar growth as accretion is not efficient once the black hole decouples from the background spacetime. However, these works used a massless scalar field equivalent to a fluid with a stiff equation of state $p = \rho$ as ambient matter instead of a massive scalar field, and the latter models an early universe dominant matter component more accurately. From  Fig. \ref{fig:mass_vs_rho}, we find that $M\sim H^{-\beta} $ where $\beta\gg 1$. As $M$ approaches the Hubble horizon, we expect $\beta \leq 1$ although unfortunately, we were unable to track the growth of PBH beyond a few factors of their initial mass, as the numerical cost becomes prohibitive. 

As long as the universe is dominated by DM, the black hole will continually accrete and grow without end. This would be the case if the PBH is formed in the present late time DM dominated epoch -- however such late time PBH has already been ruled out \cite{Carr:2020gox,Green:2020jor}. As we mentioned in the introduction, we consider instead an early phase of DM domination before transitioning into the era of radiation-domination prior to the onset of Big Bang Nucleosynthesis  (BBN), i.e. before the temperature of the universe is around $1~\mathrm{MeV}$. This provides a natural cut-off for the growth of the PBH. 

Nevertheless, if we assume that the rapid growth we observe continues until $\mbh H \sim \mpl^2$, and that the BH grows self-similarly after, it is implied that the final mass of the PBH is independent of when it forms.  This means the final mass of the PBH is given by
\begin{equation}
M_\mathrm{BH} \approx 10^{38} \left(\frac{1~\mathrm{MeV}}{T}\right)^2 g \approx 10^5 \left(\frac{1~\mathrm{MeV}}{T}\right)^2 M_\odot~, 
\end{equation}
where $T$ is the temperature of the universe at the onset of radiation-domination and $M_\odot \approx 10^{33}g$. Taking $T_\mathrm{BBN}=1~\mathrm{MeV}$ as the natural cut-off for the growth of the PBH, the most massive black holes that can be formed via this accretion mechanism are $M_\mathrm{BH} \approx 10^{38} g \approx 10^5 M_\odot~$ \cite{Carr:2004kgc,Green:2014faa}.

On the other hand, if the PBH growth asymptotes to a slower rate than  the self-similar rate, or achieve self-similarity before $\mbh \sim H^{-1} \mpl^2$, then our simulations suggests that $\mbh \gtrsim 10^{-2} H^{-1} \mpl^2$, where $H$ is the Hubble parameter when the PBH forms. This means that PBH formed around $T\sim 5$ MeV, $ \mbh \gtrsim  40M_\odot$ could form the basis of the  population of massive BH that are being detected today by the LVK observatories.

\section{Summary and discussion} \label{sect:conclusions}
In this paper, we demonstrate that superhorizon non-linear perturbations can collapse and form PBHs in a matter-dominated universe, using full numerical relativity. We show that, depending on the mass of the initial perturbation shell, this happens via either the direct collapse or the accretion collapse mechanisms. We provide an analytic criterion  \eqn{eqn:hoop} using the hoop conjecture to determine which mechanism is relevant for a given setting, and compute the timescale of collapse using the same prescription. Despite the ${\cal O}(1)$ non-linearity, we find that the dynamics of collapse can be modeled as a simple superhorizon mass shell collapsing in an expanding background. This suggests that semi-analytic estimates of PBH formation in a matter-dominated era are broadly accurate. 

On the other hand, details matter. We showed that  even in the cases where the perturbation is insufficient on its own to form a PBH in a direct collapse, non-linear accretion rates are far higher than what standard linear theory predicts, causing a rapid collapse into a PBH via accretion of ambient DM.  In both the direct collapse and accretion collapse formation cases, the initial mass of the PBH is roughly $\mbh \sim 10^{-2}H^{-1}\mpl^2$, but formation is followed by an extremely rapid growth $M\propto H^{-\beta}$ where $\beta\gg 1$. Presumably, this growth will asymptote to either the self-similar rate $\beta=1$ or the decoupled rate $\beta <1$ \cite{Harada:2004pf,Harada:2004pe}.

Interestingly, even if the self-similar rate is not achieved, the fact that most of the mass of the PBH is gained through post-formation accretion suggests that there might be a mechanism to generate PBHs with non-trivial spin.  Such non-trivial spin might for example be generated by the collapse of a non-spherically symmetric shell, even if the shell is initially spinless. In that case, the PBH might not form in the centre of the initial mass distribution and thus form with spin, whilst outgoing radiation carries away angular momentum of opposite sign, such that angular momentum is still globally conserved, as suggested by \cite{Harada:2016mhb, Harada:2017fjm}. We will explore this possibilty in a future publication.

It is an interesting question whether our results would change if we simulated a proper dust-dominated universe using fluids, rather than an effective matter-domination using a massive scalar field. The main difference is that in our case, the scalar field's mass introduces a length scale below which the dust approximation breaks down. This means e.g. that we expect that there is a minimum perturbation amplitude below which an overdensity does not form a PBH, because the length scales at which collapse takes place are small enough for there to be significant pressure support. However, we have no reason to believe that we have probed this regime and we therefore do not believe that using fluids would significantly affect our conclusions.

%% file: Chapter4/chapter4.tex
\chapter{Spinning primordial black holes formed during a matter-dominated era}\label{Chapter4}

\graphicspath{{Chapter4/Figs/}}

This chapter contains the article \textit{Spinning primordial black holes formed during a matter-dominated era} \cite{deJong:2023gsx}, published in the \textit{Journal of Cosmology and Astroparticle Physics (JCAP)}. 

In this chapter, like in chapter \ref{Chapter3}, we use Planck units $\hbar=c=1$ such that $G=\mpl^{-2}$, where $\mpl$ is the non-reduced Planck mass. 

\section{Introduction} \label{Ssect:intro}

The realization that cold dark matter may be partly made up out of primordial black holes (PBHs) \cite{Zeldovich:1967,Hawking:1971,Carr:1974,Novikov:1979,Carr:2016ks, Carr:2017rt,Carr:2020xqk} and the opportunities to test and constrain this hypothesis \cite{Page:1976wx,Carr:1976zz,Wright:1995bi,Lehoucq:2009ge,Kiraly1981,MacGibbon:1991vc,Cline:1996zg, Carr:1984cr,Bean:2002kx, Hawkins:1993,Carr:2019kxo,Hawkins:2020, LIGOScientific:2020ufj,Franciolini:2021tla, DeLuca:2020agl,Vaskonen:2020lbd,Kohri:2020qqd,Domenech:2020ers, Arzoumanian_2020,Atal:2022zux,Carr:2023tpt} has renewed interest in PBHs. The mechanism considered standard to form these relics is via the collapse during the radiation era of superhorizon perturbations originating from the growth of quantum fluctuations during inflation \cite{Carr:1975,Nadezhin:1978,Bicknell:1979, Choptuik:1993,Evans:1994,Niemeyer_1998:nj,Green_1999:gl,Musco:2012au,Yoo:2020lmg,Carr:1993cl,Carr_1994:cgl,Hodges:1990hb,Ivanov:1994inn,Garcia:1996glw,Randall:1996rsg,Taruya_1999,Bassett_2001,Clesse_2015,Inomata_2017,Garc_a_Bellido_2017,Ezquiaga:2017fvi,Geller:2022nkr,Qin:2023lgo}, but could also be the result of other early universe dynamical processes or epochs \cite{Crawford1982,Hawking:1982,Kodama:1982,Leach:2000ea,Moss:1994,Kitajima:2020kig,Khlopov:1998nm,Konoplich:1999qq,Khlopov:1999ys,Khlopov:2000js,Kawana:2021tde,Jung:2021mku,Dokuchaev:2004kr,Rubin:2000dq,Rubin:2001yw,Garriga:2015fdk,Deng:2016vzb,Liu:2019lul, Hogan:1984zb,Hawking:1987bn,Polnarev:1991,Garriga:1993gj,Caldwell:1995fu,
MacGibbon:1997pu,Wichoski:1998ev,Hansen:1999su,Nagasawa2005,Carr:2009jm,
Bramberger:2015kua,Helfer:2019,Bertone:2019irm,James-Turner:2019ssu,Aurrekoetxea:2020tuw}. 

There is next to no observational data to constrain the thermal history of our universe before big bang nucleosynthesis \cite{Carroll:2001bv,Hooper:2023brf}. Models of early matter-dominated expansion epochs include overproduction of non-relativistic particles \cite{Khlopov:1980mg,Polnarev:1982a}, the presence of moduli fields ubiquitous in string theory models \cite{Green:1997,Kane:2015jia}, or at the end of inflation during a process known as reheating \cite{Kofman:1994rk, Kofman:1997yn,Albrecht:1982mp,Amin:2014eta,Aurrekoetxea:2023jwd,Carr:2018}, when the energy stored in the inflaton is transferred into standard model particles.

PBHs' properties and abundances, and therefore corresponding detection prospects, depend on the details of the era in which they form. In this work, we investigate the expected angular momentum of PBHs. Angular momentum is relevant to possible PBH signatures, e.g. the amplitude of the stochastic gravitational wave background from spinning PBHs could increase by 50\% \cite{Kuhnel:2019zbc}, and PBHs with spin may avoid some of the abundance bounds related to evaporation due to their lower Hawking temperatures \cite{Arbey:2019jmj}. It is suggested that PBHs that form during a radiation-dominated era generally have small spins \cite{DeLuca:2019buf, Mirbabayi:2019uph, Harada:2020pzb,Chongchitnan:2021ehn}, reflected by the fact that the PBH formation threshold is increased proportional to the square of the angular momentum \cite{He:2019cdb}, although they could develop non-negligible spins through Hawking radiation \cite{Calza:2021czr,Calza:2023rjt}. On the other hand, PBH production is more efficient in a matter-dominated era in the absence of pressure support, even though non-spherical effects that may resist gravitational collapse become important \cite{Harada:2016mhb}. PBH formation in the context of an early matter-dominated epoch has also been studied in e.g. \cite{Green:1997jkl,Cotner:2016cvr,Hidalgo:2017dfp,Georg:2016yxa,Georg:2017mqk,Carr:2017edp,Kokubu:2018fxy,DeLuca:2021pls,Padilla:2021zgm,Harada:2022xjp,Hidalgo:2022yed}. Of particular relevance to this work is the argument by the authors of \cite{Harada:2017fjm} that most PBHs formed in a matter-dominated setting are near-extremal, although this result does not take into account PBH mass accretion.

In this work, we report on fully non-linear 3+1D numerical relativity simulations that aim to shed more light on the role of angular momentum in PBH formation during a matter-dominated era. We simulate the collapse of superhorizon non-linear perturbations sourced by a massless scalar field, on a matter-dominated expanding background driven by an oscillating massive scalar field.

\begin{figure*}[t]
    \href{https://youtu.be/CC4xBLol4aE}{
    \includegraphics[width=\linewidth]{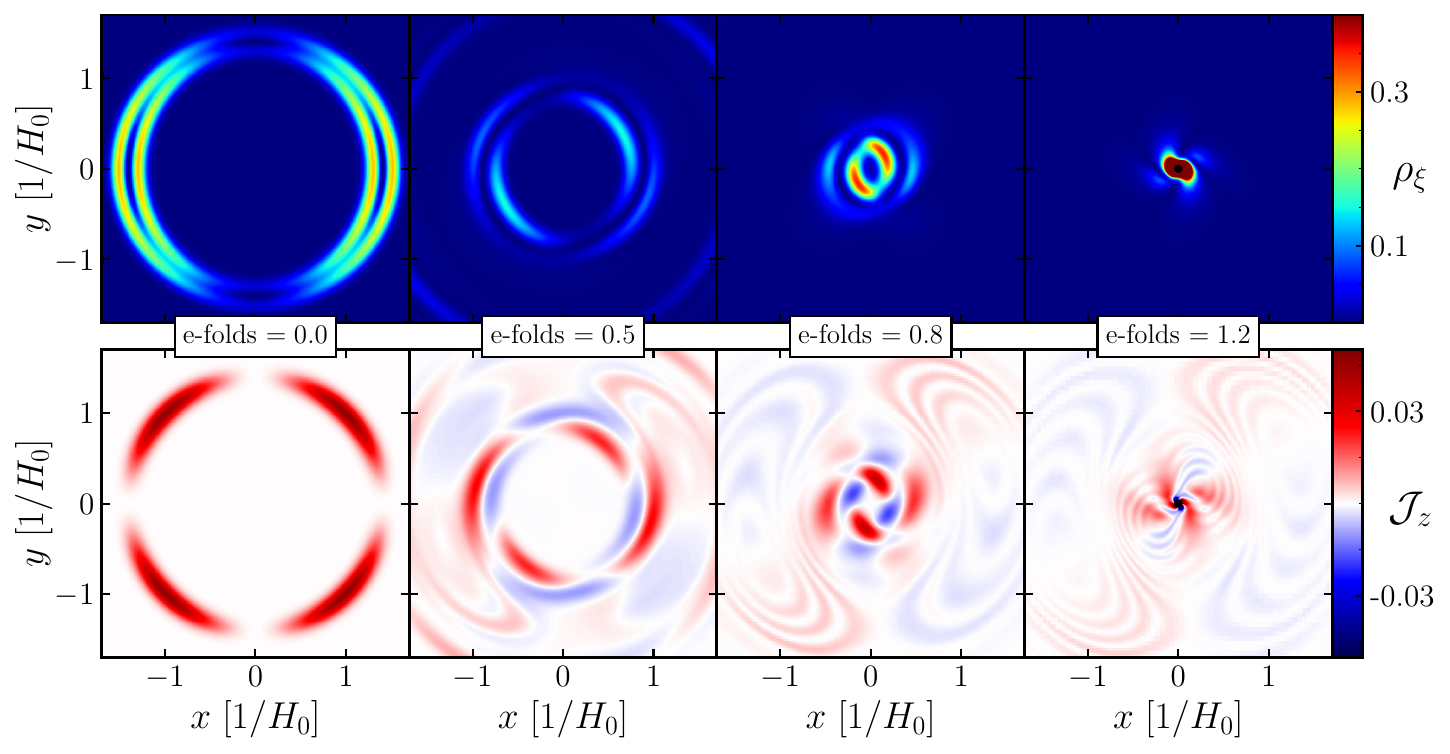}}
    \caption{\textbf{Evolution panel} showing the evolution of the energy density of the field $\xi$ (total angular momentum density around the $z$ axis) in the top (bottom) row, in units of $[\rho_{\xi}]=H_0^2 \mpl^2$ ($[\mathcal{J}_z]=H_0 \mpl^2$), for a perturbation with $R_0 = 1.4 H_0^{-1}$. The leftmost column shows the initial conditions, in which $\rho_\xi$ shows two peaks corresponding to the gradients of the gaussian in \eqn{Seq::xi_r_init}, whilst the angular momentum density takes strictly positive values. In the top row, we see the perturbation collapse and acquire an asymmetric configuration before black hole formation in the third column, signalling the presence of angular momentum in the collapsing region. In the bottom panels, we see $\mathcal{J}_z$ take on both positive and negative values over the course of the evolution, on which we comment in section \ref{Ssect::results_pbhspin}. In the rightmost column the PBH has formed, and the small area already hidden behind the apparent horizon is shown in black. A video of this data can be found \href{https://youtu.be/CC4xBLol4aE}{here} \cite{Movie3}.
    } 
    \label{Sfig:evol_panel}
\end{figure*}

We find that the formation process is quite efficient, i.e. at horizon formation the PBH contains $\mathcal{O}(10\%)$ of the collapsing overdensity's mass and angular momentum. However, the PBH dimensionless spin goes down as it accretes non-rotating background matter, resulting in negligible final spins if the matter-dominated era lasts several e-folds. We illustrate typical collapse behaviour in Fig. \ref{Sfig:evol_panel}. Note that since a radiation-dominated era must  intervene before the  onset of the present matter-dominated era, the background massive scalar must reheat. Any PBHs that were formed during this era will remain as a matter component which presumably can be the dark matter component today.

The paper is organised as follows. In section \ref{Ssect:initialconditions}, we discuss the numerical setup of our simulations, initial conditions and diagnostics. In particular, we provide details on finding black hole apparent horizons in expanding spacetimes. In section \ref{sect:num_meth_4}, we provide details on the numerical methodology in this work. In section \ref{Ssect:results} we report on our main results, after which we conclude in section \ref{Ssect:conclusions}.

\vspace*{\fill} 

\section{Setup} \label{Ssect:initialconditions}

We consider a massive scalar field $\phi$ and a massless scalar field $\xi$, both minimally coupled to gravity\footnote{We use the mostly plus $-+++$ signature  and Planck units $\hbar=c=1$, such that $G=\mpl^{-2}$, where $\mpl$ is the non-reduced Planck mass.}, i.e.
\begin{equation}
S = \int ~d^4 x \sqrt{-g}\left[\frac{\mpl^2}{16\pi}R - \mathcal{L}_\phi - \mathcal{L}_\xi\right]~.
\end{equation}
We also assume there is no direct coupling between them, i.e.
\begin{subequations}
\begin{align}
\mathcal{L}_\phi &=\frac{1}{2}\nabla^{\mu}\phi\nabla_\mu\phi + \frac{1}{2}m^2 \phi^2~, \\
\mathcal{L}_\xi &= \frac{1}{2}\nabla^{\mu}\xi\nabla_\mu\xi ~.
\end{align}
\end{subequations}

The scalar field $\phi$ with mass $m$ plays the role of the ambient matter driving the background expansion. In the limit in which $\phi$ is homogeneous and dominates, the spacetime is well-described by the Friedman-Lema\^{i}tre-Robertson-Walker (FLRW) line element
\begin{equation}
    ds^2 = -dt^2 + a(t)^2(dr^2+r^2d\Omega^2_2)\,.
\end{equation}
If $\phi$ oscillates coherently with a period considerably smaller than a Hubble time, $2\pi/m \ll 1/H$, its pressure averages to zero and the scale factor $a(t)$ grows in a matter-like fashion, i.e. $a \propto t^{2/3}$ and $\rho\propto a^{-3}$.
The massless field $\xi$, on the other hand, plays the role of the collapsing superhorizon perturbation, and will thus be the seed triggering black hole formation. 

Going beyond the FLRW limit to study black hole formation requires solving the Einstein field equations \cite{Einstein:1916} for a more general line element, which we decompose in the usual ADM form \cite{Arnowitt_2008}
\begin{equation}
    ds^2 = -\alpha^2dt^2 + \gamma_{ij}(dx^i + \beta^i dt)(dx^j + \beta^j dt),
\end{equation}
where $\gamma_{ij}$ is the three-dimensional spatial metric. The lapse and shift gauge functions $\alpha$ and $\beta^i$ determine the choice of spatial hypersurface and their coordinates, which in numerical relativity are dynamically determined. At each hypersurface, the rate of change of $\gamma_{ij}$ is given by the extrinsinc curvature $K_{ij} \equiv -(1/2){\cal{L} }_{{\bf n}}\gamma_{ij}$, where ${\bf n}$ is the vector normal to the hypersurface. $K_{ij}$ can be decomposed further into its trace $K$ and tracefree components $A_{ij}$, where $K_{ij} = A_{ij} + (1/3)K\gamma_{ij}$. 
The trace $K$ measures the local expansion, where in our sign convention negative (positive) $K$ indicates locally expanding (collapsing) space.
 
We evolve the Einstein field equations using the CCZ4 formulation \cite{Alic_2012} and the moving puncture gauge \cite{Bona:1994dr,Baker:2005vv,Campanelli:2005dd,vanMeter:2006vi}, using the numerical relativity code $\grchombo$ \cite{Clough:2015sqa,Radia:2021smk,Andrade2021}. We list explicit expressions for the matter variable evolution equations in section \ref{Sapp::evolution_equations}.

\subsection{Initial data}\label{Ssect::init_data}

We choose the initial gauge $\alpha=1$ and $\beta^i=0$.
We set the scalar field $\phi$ to a homogeneous value $\phi_0$ starting from rest, i.e. $\partial_t{\phi}=0$. The initial unperturbed Hubble parameter $H_0$ is then 
\begin{equation} \label{Seq::initial_hubble_param}
    H_0^2 = \frac{4\pi m^2}{3\mpl^2}\phi_0^2 \,.
\end{equation}
We break spherical symmetry and inject angular momentum into the system via the elevation and azimutal angle-dependence\footnote{Note the difference between $\varphi$, which denotes the azimuthal angle of a spherical coordinate system, and $\phi$, denoting the massive scalar field.} of the massless field $\xi$ and its conjugate momentum $\Pi_\xi=\left(\partial_t \xi -\beta^i\partial_i\xi\right)/\alpha$
\begin{subequations}\label{eqn::init_time}
\begin{align}
    \label{Seq::xi_init_time} \xi(t,r,\theta,\varphi) &= R(r)\left[1 + \sin{(\theta)}\Phi(t,\varphi)\right],\\
    \label{Seq::pi_init_time} \Pi_\xi (t,r,\theta,\varphi) &=R(r)\sin{(\theta)}\partial_t\Phi(t,\varphi),
\end{align}
\end{subequations}
where
\begin{subequations}
    \begin{align}
        \label{Seq::xi_r_init} R(r) &= A \exp{\left[-\frac{(r-R_0)^2}{\lambda^2}\right]},\\
        \label{Seq::xi_phi_init} \Phi(t,\varphi) &= B \cos{(k\varphi-\omega t)}.
    \end{align}
\end{subequations}
The constants $A,\, R_0,\, \lambda,\, B,\, k,\, \omega$ represent the perturbation shell's initial radial amplitude, radius, width, spin amplitude, spin wavenumber and spin angular velocity, respectively -- Fig. 
\ref{Sfig:init_params} in section \ref{Sapp::initial_data} schematically clarifies the meaning of these constants. The expressions in \eqn{eqn::init_time} are time-dependent, and we obtain initial data by setting $t=0$. 
The angular momentum density around any given axis through the origin is given by 
\begin{equation}
    \mathcal{J}^i = \epsilon^{ijk}x_j S_k
\end{equation}
where $S^i = -\gamma^{ij}n^a T_{aj}$ and $n^a = (1,-\beta^i)/\alpha$ is the normal vector to the hypersurface. Here $\epsilon^{ijk}$ is the antisymmetric Levi-Civita symbol and $i,j,k=1,2,3$ label the spatial Cartesian basis.
We are interested in the $z$-component of the angular momentum density\footnote{The $\theta$-dependence introduces additional angular momentum around the $x$- and $y$-axes, but we choose parameters that makes sure these are negligible compared to the angular momentum around the $z$-axis.}, which in the initial data is only sourced by the massless field $\xi$ and in a non-orthonormal polar basis $(\partial_r, \partial_\varphi,\partial_z)$ is expressed as
\begin{equation}\label{Seqn::angmomdensity_z}
\begin{aligned}
    \mathcal{J}_{\xi,z}^0 &= xS_{\xi,y} - yS_{\xi,x}\\
    &= -\Pi_{\xi}\left(x\partial_y\xi - y\partial_x\xi\right)\\
    &= -\Pi_{\xi} \partial_\varphi\xi,
\end{aligned}
\end{equation}
where we take a derivative with respect to the azimuthal angle $\varphi$ directly in the last line and all quantities on the RHS are evaluated on the initial slice. This expression can be integrated over the volume to find the total initial angular momentum $J_{\xi,z}^0$. 
In our notation, the $\mathcal{J}$ in a calligraphic font will denote an angular momentum density, while $J$ will denote angular momentum, i.e. $J_i = \int ~dV \mathcal{J}_i$. In what follows, if a directional subscript is omitted, a $z$-subscript is implied. 
Finally, we note the important point that the angular momentum is \emph{not} concentrated on the equatorial plane of the shell, but is roughly Gaussian distributed around it with angular width $\Delta \theta \approx 0.5\pi$ (see Fig. \ref{Sfig:theta_plot}). The implications of this distribution is that, as we shall see, matter which is spinning but not along the equatorial plane will still spiral into the center. 

As is well known, all initial data in general relativity must obey a coupled system of Hamiltonian and momentum constraint equations, which we solve using the CTTK method \cite{Aurrekoetxea:2022mpw}. In particular, we choose an initial spatially flat metric $\gamma_{ij}=\delta_{ij}$ and we follow the CTTK prescription described in section \ref{sect:constraints}, choosing $\partial_i U = -V_i / 4$ as in \eqn{eqn::cttk6}. We will refer to this method as the ``original method''. The initial conditions for the results presented in this paper were obtained using this method.

As discussed in section \ref{sect:constraints}, the RHS of \eqn{eqn::cttk5} then becomes a pure gradient, i.e. one must be able to write the momentum density vector $S_i$ as $S_i = \partial_i F$ for a function $F$ (note that $\psi$ is constant in our case). It turns out that the matter configuration described above does not satisfy this requirement. As a result, one must solve \eqn{eqn::cttk4} and \eqn{eqn::cttk5} as four coupled equations, without imposing \eqn{eqn::cttk6}. We will refer to this as the ``corrected method''.

We have checked that we obtain the same initial profiles for $K$ and $A_{ij}$ with the original and corrected methods to within 0.02\% and that our results for final PBH mass and spin are identical in both cases - we provide further details in appendix \ref{app:CTTK_check}. Hence, we are confident that the profiles for the trace $K$ and traceless parts $A_{ij}$ obtained using the original method provide physically reliable results. This profile is equivalent to choosing an initially homogeneous cosmological scale factor, where the rate of local expansion is determined by the matter distribution. Note that in the absence of inhomogeneities, the Hamiltonian constraint reduces to the usual form of the Friedmann equation $H^2 = 8\pi\rho/3\mpl^2$, with $K=-3H$.

\subsection{Diagnostics quantities}\label{Ssect::diagnostics}

The key three diagnostic quantities we will track are the PBH mass $M_\mathrm{BH}$, dimensionful angular momentum $J_\mathrm{BH}$ and dimensionless spin (dimensionless Kerr parameter)
\begin{equation}
    a_\mathrm{BH}\equiv \frac{J_\mathrm{BH}}{G\mbh^2}~.
\end{equation}
We obtain these values by measuring the properties of the black hole \emph{apparent} horizon (AH) --  the outermost marginally trapped surface on which the expansion of outgoing null geodesics vanishes. 
To find the AH we follow the procedure described in \cite{Thornburg:2003sf} -- the next paragraph is rather technical and the reader may skip it. 

If the spatial metric on a given three dimensional spatial hypersurface is $\gamma_{ij}$, a two dimensional surface $S$ with a spacelike unit outward-pointing normal vector $s^a$ induces a two dimensional metric $m_{ab}$ on $S$ equal to
\begin{equation}
    m_{ab} = \gamma_{ab} - s_a s_b.
\end{equation}
The expansion of outgoing null vectors $k_+^{~a} = n^a + s^a$ in this surface is defined as $\Theta_+ \equiv m_{ab}\nabla^a k_+^{~b}$, or equivalently
\begin{equation} \label{Seqn::expansion_outgoing}
    \Theta_+\equiv D_i s^i + K_{ij} s^i s^j - K,
\end{equation}
which vanishes on the AH.
To find this surface during the formation stage, we shoot rays in different directions from a PBH centre \emph{guess} point, which we take to be the location of the maximum energy density, and solve $r^\kappa\Theta_+(r) = 0$ for $r$ along these rays. Here $r$ is the $\theta$- and $\varphi$-dependent coordinate distance along a given ray and $\kappa\geq 0$ is an arbitrary AH expansion radius power that can be chosen to optimise AH searches depending on underlying spacetimes - note that any solution to $r^\kappa\Theta_+(r) = 0$ also solves $\Theta_+(r) = 0$, provided that $r\neq 0$, but performance of an algorithm solving the former equation may be significantly better. We find that in expanding spacetimes, $\kappa = 2$ provides reliable performance. 

To measure the area, spin and mass of the AH, we use \cite{Caudill:2006hw, Ashtekar:2000hw}
\begin{subequations}
\begin{align}
    A_\textrm{BH} &= \oint_{S} d^2 V,\\
    J^{i}_\textrm{BH} &= \frac{1}{8\pi} \oint_S ~ d^2 V \left(\xi^i\right)^l s^j K_{jl},\\
    M_\textrm{BH} &= \sqrt{\frac{A_\textrm{BH}}{16\pi} + \frac{4\pi J_\textrm{BH}^2}{A_\textrm{BH}}},
\end{align}
\end{subequations}
where $d^2V$ is the natural area measure on $S$ constructed from $m_{ab}$ and $\xi^i$ are approximate Killing vectors of $m_{ab}$ approximated well by the flat space rotational Killing vectors, $\left(\xi^i\right)^l=\epsilon^{ijl}x_j$, since the AH surface closely resembles a coordinate sphere. We present a more detailed discussion of the AHs in FLRW spacetimes  in appendix \ref{Sapp::AH}.

In particular, when focusing on the efficiency of the formation process, we will be interested in ratios  of the collapsing seed, i.e. $(\mbh/\Mxi)$, $(J_\mathrm{BH}/J^0_\xi)$ and $(a_\mathrm{BH} / \axi)$, where $\Mxi$ is the initial mass of the collapsing perturbation given by integrating the initial gradient and kinetic energy 
\begin{equation}
    \Mxi = \int dV~ \left( \frac{1}{2}(\partial_i \xi)^2 +\Pi_\xi^2\ \right),.\label{Seq::infall_mass}
\end{equation}
The dimensionful angular momentum $J_\xi^0$ is given by integrating the expression in \eqn{Seqn::angmomdensity_z} over volume on the initial slice and $\axi = J^0_\xi/\left(G\Mxi\right)^{2}$. We note that in the non-static, non-axisymmetric spacetime that we consider there is no local definition of energy of angular momentum density. Similarly, there is no definition of total mass or angular momentum because the spacetime isn't asymptotically flat, and any measures of these quantities are therefore coordinate-dependent.

\section{Numerical methodology}\label{sect:num_meth_4}

\subsection{Initial data} \label{Sapp::initial_data}
\begin{figure*}[t]
    \includegraphics[width=\linewidth]{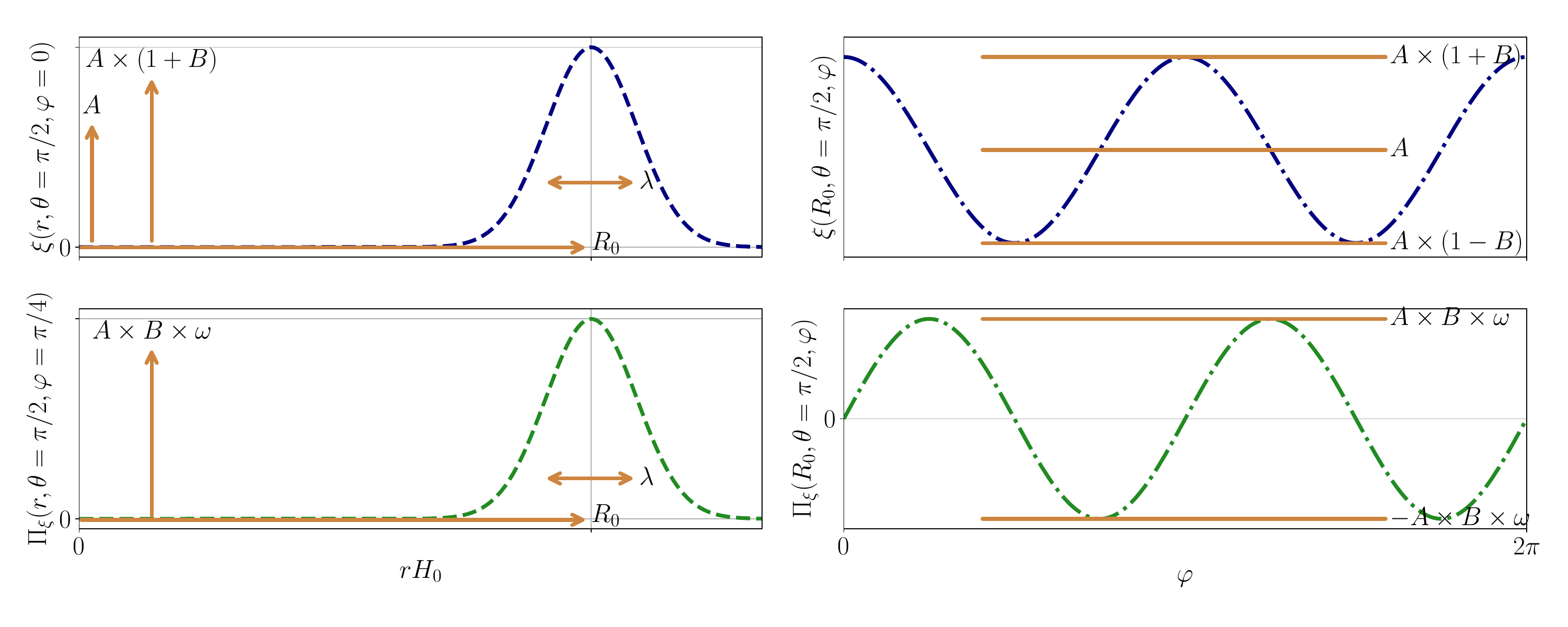}
    \vspace*{-10mm}
    \caption{\textbf{Initial configurations for $\xi$ and $\Pi_\xi$} give by \eqn{Seq::xi_init_time} and \eqn{Seq::pi_init_time}, for parameters $A=0.0825\mpl, R_0 = 1.2H_0^{-1}, \lambda = 0.15H_0^{-1}, B = 0.5, k = 2, \omega = 12H_0^{-1}$. The navy blue (green) dashed lines in the top (bottom) row represent initial profiles for $\xi$ ($\Pi_\xi$). The dashed (dashdotted) lines in the left (right) columns represent initial $\xi$ or $\Pi_\xi$ profiles as a function of coordinate radius $r$ with elevation angle $\theta$ and azimuthal angle $\varphi$ kept constant (as a function of $\varphi$ with $r$ and $\theta$ kept constant). The various solid orange arrow and lines show how the constants $A, R_0, \lambda, B, \omega$ relate to the initial shape of these profiles, whilst the final constant $k$ is set to $2$ here, so that the profiles in the right column oscillate twice per $\varphi$ rotation.} 
    \label{Sfig:init_params}
\end{figure*}

\begin{figure}[t]
    \centering
    \includegraphics[width=.6\linewidth]{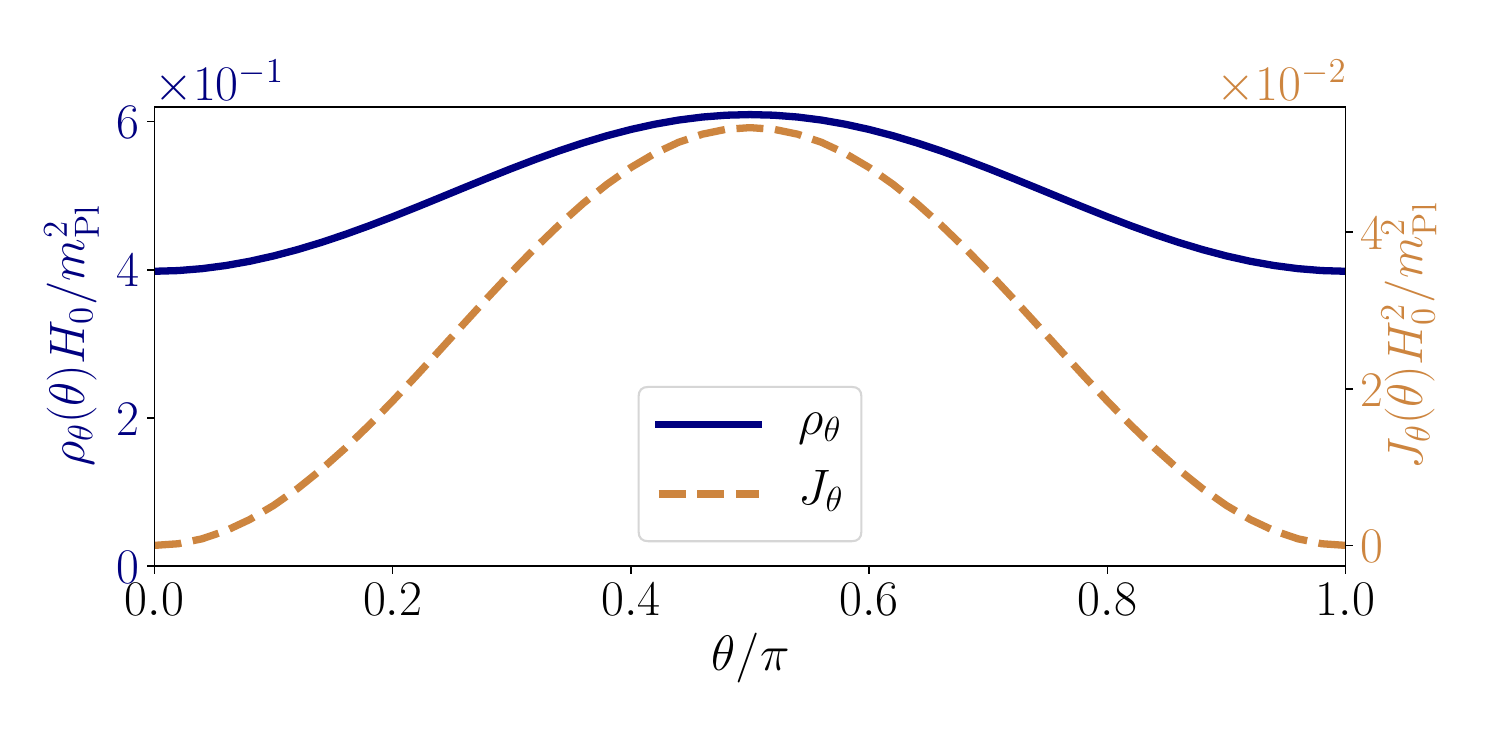}
    \vspace*{-5mm}
    \caption{\textbf{Initial configurations of $\rho_\theta(\theta)$ and $J_\theta(\theta)$,} defined in \eqn{Seqn::rhotheta} and \eqn{Seqn::Jtheta}, as a function of elevation angle $\theta$, for initial parameters $A = 0.0875\mpl$, $R_0 = 1.4H_0^{-1}$, $\lambda = 0.15H_0^{-1}$, $B = 0.5$, $k = 2$, $\omega = 12H_0^{-1}$. The solid blue (dashed light brown) line corresponds to the y-axis on the left (right).} 
    \label{Sfig:theta_plot}
\end{figure}

The initial configurations for $\xi$ and its time derivative $\Pi_\xi$ are described in section \ref{Ssect::init_data} of the main text and illustrated in Fig. \ref{Sfig:init_params}. 

In section \ref{Ssect::results_pbhmass} of the main text, we show that the PBHs' initial masses and accretion rates do not decrease significantly with increasing initial shell angular momentum and we attribute this to the initial shape of the shell. In Fig. \ref{Sfig:theta_plot}, we show typical initial profiles of the quantities

\begin{subequations}
\begin{align}
    \rho_\theta(\theta) &= \iint ~dr d\alpha \hspace{1mm} r^2 \rho(r, \theta, \alpha) \label{Seqn::rhotheta} \\
    J_\theta(\theta) &= \iint ~drd\alpha \hspace{1mm} r^2 \mathcal{J}(r, \theta, \alpha), \label{Seqn::Jtheta}
\end{align}
\end{subequations}
i.e. the energy density and angular momentum density integrated over a conical surface at elevation angle $\theta$. From this figure, we note that the largest part of the shell's mass is located away from the equatorial $z=0$ plane, where it is less susceptible to orbital effects and more likely to spiral into the centre. Additionally, a significant part of the angular momentum is located away from the equatorial plane. We expect that if the mass distribution peaked more sharply around the equatorial plane, the spin effects on the initial PBH mass and accretion rate would be stronger. This is an interesting direction for future research. 

\subsection{Grid details and convergence testing} \label{Sapp::convergence_testing}

We use periodic boundary conditions in the $x$- and $y$-directions and reflective boundary conditions in the $z$-directions, so that we effectively only simulate the top half of the perturbation. The length of the simulation box is $L_\textrm{box} = 4/H_0$ ($L_\textrm{box} = 5/H_0$) for initially subhorizon (superhorizon) perturbations in the $x$- and $y$ directions, and half in the $z$-direction. The grid details for the simulations presented in this chapter are otherwise largely equivalent to the ones from chapter \ref{Chapter3}, but we list them here for completeness. The grid size of the coarsest level, which covers the entire simulation box, is $96$ in the $x$- and $y$ directions, and half in the $z$-direction. The refinement factor between a given AMR level and the next is 2. The Courant factor $\Delta t/\Delta x = 0.0384$ and is equal for each AMR level. The maximum number of levels achieved for these simulations is 4, including the base level. Grid points are tagged for refinement when either $\rho_\xi$ or the gradients of $K$ pass a threshold value. 
\begin{figure*}[t]
    \includegraphics[width=\linewidth]{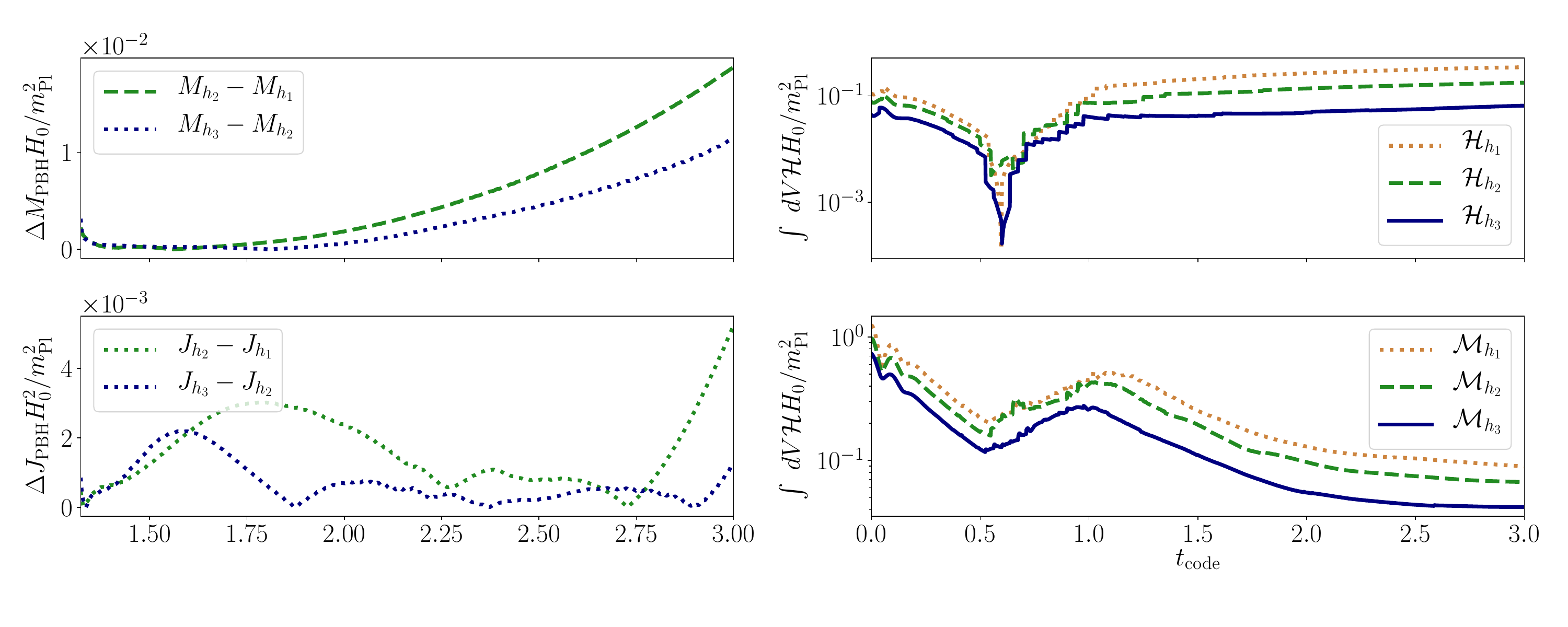}
    \vspace*{-10mm}
    \caption{\textbf{Convergence testing.} We show data for the formation of a PBH formed by an initial perturbation with $R_0 = 1.4$ and $\chi_0^\xi = 6.1\times 10^{-2}$. The top left panel shows the absolute value of the difference between the masses detected by simulations of different resolutions, while the bottom left panel shows the absolute value of the difference between the dimensionless spins. The top right panel shows the Hamiltonian constraint violation integrated over the simulation domain for different resolutions, while the bottom right panel shows the integrated momentum constraint violation.} 
    \label{Sfig:convergence}
\end{figure*}

We perform several convergence tests on the robustness of our numerical results, by finding the mass and spin of a PBH formed by an initial perturbation with $R_0 = 1.4H_0^{-1}$ and $a_\xi^0 = 6.1\times 10^{-2}$. We do so using three different base grid resolutions, $N_1 = 80$, $N_2 = 96$ and $N_3 = 128$, which correspond to base grid spacings $h$ of $h_1 = 6.25\times 10^{-2}H_0^{-1}$, $h_2 = 5.21\times 10^{-2} H_0^{-1}$ and $h_3 = 3.90\times 10^{-2} H_0^{-1}$. We track Hamiltonian and momentum constraint violation, as well. 

Fig. \ref{Sfig:convergence} shows mass and spin difference between simulations of different base resolutions and Hamiltonian and momentum constraint violation for all simulations, indicating that convergence is achieved. We note that we checked the code simulates a homogeneous FLRW universe for appropriate initial conditions in an earlier publication \cite{deJong:2021bbo}.

\subsection{Scalar field evolution equations}\label{Sapp::evolution_equations}

Whilst the metric variable evolution equations can be found in e.g. \cite{Alic_2012}, we list the matter evolution equations here explicitly for completeness. For the generalised case of $N$ minimally coupled scalar fields $\phi_i$, the evolution equations are 
\begin{subequations}
\begin{align}
    \partial_t \phi_i &= \alpha \Pi_i + \beta^j\partial_j \phi_i, \\
    \partial_t \Pi_i &= \beta^j\partial_j \Pi_i + \alpha\partial^j\partial_j \phi_i + \partial^j \phi_i \partial_j \alpha\\ 
    &\quad+ \alpha\left(K\Pi_i - \gamma^{jk}\Gamma^l_{jk}\partial_l\phi_i - \frac{dV(\phi_i)}{d\phi_i}\right)\nonumber,
\end{align}
\end{subequations}
whilst the energy-momentum expressions appearing in the metric variables' evolution equations are
\begin{subequations}
\begin{align}
    \rho &= \frac{1}{2}\sum_{i=1}^N \left[\Pi_i^2 + \partial^j\phi_i\partial_j\phi_i + V(\phi_i)\right],\\
    S_j &= \sum_{i=1}^N -\Pi_i \partial_j\phi_i,\\
    S_{jk} &= \sum_{i=1}^N\Big[\partial_j\phi_i\partial_k\phi_i - \frac{1}{2}\bar{\gamma}_{jk}\big(\bar{\gamma}^{lm}\partial_l\phi_i\partial_m\phi_i\\
    &\quad - \Pi_i^2 - 2V(\phi_i) \big) \Big],\nonumber\\
    S &= \gamma^{jk}S_{jk}.
\end{align}
\end{subequations}
In our work we set $N=2$. 

\section{Results} \label{Ssect:results}

\begin{figure*}[t]
    \includegraphics[width=\linewidth]{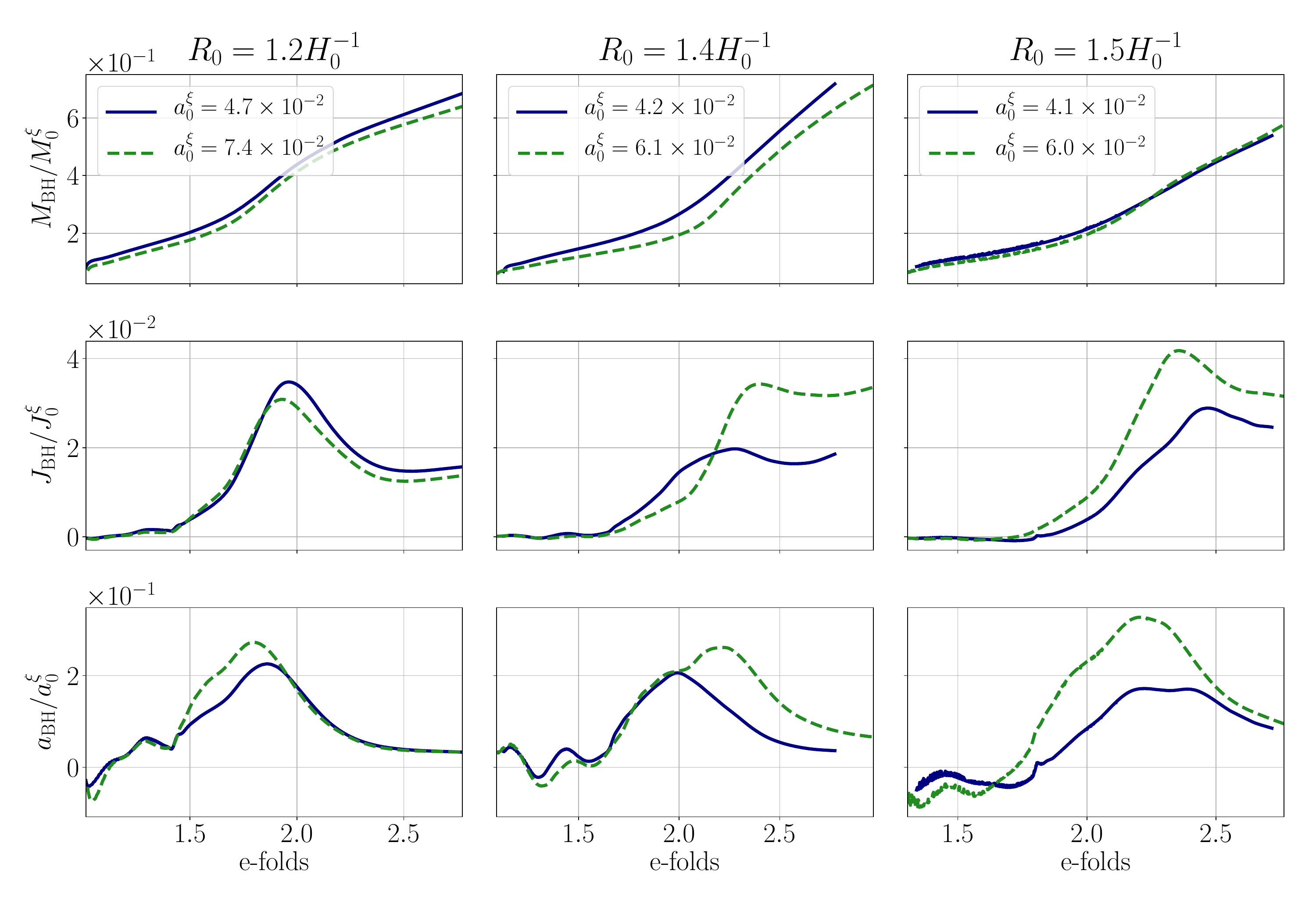}
    \vspace*{-10mm}
    \caption{\textbf{PBH masses, angular momenta and dimensionless spins for perturbations of varying initial size $R_0$.} The top, middle and bottom rows show normalized PBH mass $\mbh$, PBH dimensionful angular momentum $J_{\mathrm{BH}}$ and PBH dimensionless spin $a_{\mathrm{BH}}$ respectively, all as a function of number of e-folds after collapse initiated. The data in different columns is for shells of different initial size, as annotated at the top. The data is labelled by the shell's initial dimensionless spin. The legends in the top plots are valid for their entire respective columns. Our results show that the efficiency of the angular momentum transfer from shell to PBH is ${\cal O}(5)$\%.} 
    \label{Sfig:pbh_panel}
\end{figure*}

The length scale of interest in this work is the unperturbed Hubble horizon $H_0$, set through \eqn{Seq::initial_hubble_param}. For all simulations we choose $\phi_0 = 7.8\times 10^{-3} \mpl$ and the mass $m = 62.6 H_0$, for the same reasons as presented in section \ref{sect:bhformation}. For the initial size of the superhorizon perturbation we use a range of $R_0 \in [1.2, 1.5]H_0^{-1}$ with initial width  $\lambda = 0.15 H_0^{-1}$.  The spin amplitude and wavenumber are fixed to $B = 0.5$ and $k=2$, respectively.

In our previous work \cite{deJong:2021bbo} we found there exist two distinct mechanisms by which PBHs can form in a matter-dominated era depending on the amplitude of the initial perturbation, i.e.
\begin{equation}\label{Seq::direct_threshold}
    \frac{4 G \Mxi}{\lambda \left(2 + R_0 H_0\right)^2} \quad
    \begin{cases}
    < 1\quad \text{accretion collapse,}\\
    > 1\quad \text{direct collapse.}
    \end{cases}
\end{equation}

The two mechanisms are 
\begin{itemize}
    \item \textit{Accretion collapse}: The perturbation is not massive enough and disperses post-collapse, but triggers a gravitational seed that accretes the (non-rotating) background matter, which then collapses into a black hole -- {PBH formation dominated by $\phi$}.
    \item \textit{Direct collapse}: The initial perturbation is massive enough to directly collapse into a black hole -- {PBH formation dominated by $\xi$}. 
\end{itemize}
We find that this threshold accurately predicts whether direct or accretion collapse will take place, even when angular momentum is added. Given that the initial angular momentum is contained in the collapsing massless field $\xi$, we focus on the latter mechanism -- direct collapse. This is the most optimistic scenario to form PBHs with spin, as in the accretion case we expect most of the angular momentum of $\xi$ to be radiated away when the initial perturbation disperses\footnote{The angular momentum transfer between the fields $\xi$ and $\phi$ is minimal if they only couple via gravity, as the transfer is suppressed by a factor of $G$ through the EFE. This dependence on $G$ is absent when the fields are coupled directly.}.
Consequently, for the perturbation's radial amplitude we use a range of $A \in [0.0825, 0.09]\mpl$. 
We vary $\omega \in [0, 12] H_0^{-1}$ to parameterize the amount of angular momentum in the system. 

\subsection{PBH mass} \label{Ssect::results_pbhmass}

We show PBH mass data for perturbations of three different initial sizes in the top row of Fig. \ref{Sfig:pbh_panel}. We conclude that the efficiency is $\sim 15\%$ and is only weakly sensitive to the initial angular momentum and size of the perturbation. Naively, one might expect that higher initial angular momentum will prevent matter from collapsing, as inward gravitational acceleration can to an extent be balanced by rotational motion, so an increased angular momentum could result in a decrease in the efficiency $(\mbh/\Mxi)$. Our results show that this effect is not dominant -- we suspect that this is related to the fact that much of the perturbation's angular momentum is concentrated away from the equatorial plane (unlike, say, that of an accretion disk), and this spinning matter does not orbit but instead spirals into the centre. We comment further on this effect in section \ref{Sapp::initial_data}.

Furthermore, we note that the mass accretion  \emph{rate} of the PBHs is equally insensitive to the angular momentum of the perturbation, at least for the spin values we probe. This means that the accreted mass can quickly surpass the initial seed mass, so that predictions for final PBH mass will depend heavily on assumptions made about the continued accretion rate. We note that the dependence of the accretion rate on the perturbation's angular momentum may well change if the initial angular momentum increases considerably, which is an interesting direction for future studies.

\subsection{PBH spin}\label{Ssect::results_pbhspin}

We investigate the evolution of the dimensionful PBH angular momentum around the $z$-axis $J_{\textrm{BH}}$. We plot the corresponding efficiency ($J_\mathrm{BH}/J^0_\xi$) in the middle row of Fig. \ref{Sfig:pbh_panel}, from which we see that $J_\mathrm{BH}$ is consistent with zero initially, meaning the non-spinning parts of the perturbation cause initial AH formation. The PBH then starts accreting matter with angular momentum until $J_\mathrm{BH}$ peaks at $2-4\%$ of the initial total angular momentum $J^0$ and finally asymptotes to a constant value.

There is a post-peak dip in angular momentum that appears consistently for all simulations, which we believe are the result of some components of the shell spinning up, meaning local angular momentum conservation requires other parts to spin in the opposite direction. This causes the formation of regions with net negative spin, whose accretion by the PBH causes the dip.

Finally, we show the efficiency of the dimensionless spin ($a_\mathrm{BH}/\axi$) in the bottom row of Fig. \ref{Sfig:pbh_panel}, which peaks at a maximum $\sim 25\%$. Overall, our results suggest that the peak efficiencies $(J_\mathrm{BH}/J^0_\xi)$ and $(a_\mathrm{BH} / \axi)$ increase with increasing perturbation radius, and whether or not this trend continues for even larger radii and values of $\axi$ needs to be investigated further. Lastly, $a_\mathrm{BH}$ decreases very quickly due to the accretion of the background matter, which is non-rotating. This demonstrates post-formation evolution and dynamics are important.
\section{Discussion} \label{Ssect:conclusions}

In this work, we use numerical relativity to show that sufficiently massive superhorizon perturbations with inherent angular momentum will generically collapse into a PBH with spin. This process is rather efficient: $\sim 10\%$ of the initial mass and $\sim 5\%$ of the initial angular momentum make up the PBH at AH formation. We show that the initial PBH mass and immediate post-formation accretion rate only depend weakly on the perturbation's initial size and angular momentum, for the parameters we explore. 
The PBH  spin $(a_\mathrm{BH}/\axi)$ efficiency peaks at $\sim 25\%$ but crucially decreases quickly during the subsequent evolution as during a matter-like era, black holes keep growing as they accrete non-rotating background matter. 

To illustrate this, we assume that the rapid accretion rate shown in Fig. \ref{Sfig:pbh_panel} levels off quickly and that the PBH continues to grow self-similarly\footnote{We emphasize that instead, the rapid accretion rate may be sustained until the PBH mass is proportional to the Hubble mass. Additionally, subsequent growth could be significantly slower than self-similar.}, i.e. $\mbh\propto 1/H$, so that the dimensionless spin evolves as $a_\mathrm{BH}\propto H^2$. Using the matter-dominated era scaling between the Hubble parameter and the temperature $H\propto T^{3/4}$, we obtain
\begin{equation}
    \frac{a_\mathrm{BH}(T)}{a_\mathrm{BH}(T_0)} \approx \left(\frac{T}{T_0}\right)^{3/2},
\end{equation}
where $a_\mathrm{BH}(T)$ is the dimensionless black hole spin at temperature $T$, and $T_0$ is the temperature at formation. In the most optimistic scenario where the PBH is the endpoint of a highly rotating initial seed and is near extremal $a_\mathrm{BH}^0\approx 1$, this implies that the typical leftover spin is $\leq \mathcal{O}(0.1)$ if the duration of the matter-dominated epoch is $\Delta T \gtrapprox 0.8T_0$. In this scenario, the final PBH mass depends on the temperature of the universe at the end of the matter-dominated epoch $T$ and can be approximated by 
\begin{equation}
    M_\textrm{BH} \approx 10^{36} \left(\frac{1~\mathrm{MeV}}{T}\right)^2 g \approx 10^3 \left(\frac{1~\mathrm{MeV}}{T}\right)^2 M_\odot~,
\end{equation}
where $T$ is taken relative to $T_\mathrm{BBN}=1~\mathrm{MeV}$ and $M_\odot \approx 10^{33}g$.

Computational cost limits us from tracking the PBH evolution beyond a few e-folds after formation, so the assumptions we make about the continued accretion rate are important. However, PBHs are expected to accrete significantly during a matter-dominated epoch, even if the exact rate at which they do so is unknown. Therefore, we argue that if PBHs with large spins are to form in a matter-dominated epoch, one needs two ingredients: firstly, the PBHs must be highly spinning at formation and secondly, the matter-dominated epoch's duration must be short.

%% file: Chapter5/chapter5.tex
\chapter{Dimensional reduction with matter fields}\label{Chapter5}

\graphicspath{{Chapter5/Figs/}}

In this chapter, we describe an extension of the modified cartoon method from \cite{Cook:2016soy}, which involves the addition of matter fields. We discuss the dimensional reduction of the initial condition solver used in chapter \ref{Chapter4} and the evolution equations used in chapters \ref{Chapter3} and \ref{Chapter4}. In this chapter we limit ourselves to the case of a real scalar field, but extending the framework to other matter fields, e.g. complex scalar fields or electromagnetism, is straightforward. Finally, we list an example of physical scenarios that could be efficiently modelled using these equations.

Since this section's focus is on the method's numerical performance and we use geometric units in the code, we will work in those units here, i.e. we set $G = c = 1$.

\section{Introduction}\label{sect::cartoon_intro}

NR simulations are notoriously computationally heavy and a main roadblock to densely covering a certain parameter space with NR, or even running few NR simulations at high resolution, is one of resources. To illustrate this, it helps to consider a $D$-dimensional uniform grid with $N$ grid points in every direction, in which case the computational cost of a simulation on this grid will roughly scale like $N^D$. This is the main reason 3+1D NR simulations are computationally so costly and this only becomes worse for higher-dimensional problems. This problem can be remedied partly by techniques such as adaptive meshes, but the computational cost of these simulations means that one is usually limited to numerically testing only a small part of parameter space, and this part is often only sparsely covered by simulations.
 
A well-known and currently very relevant example is the case of binary BH inspirals, for which the merger needs to be resolved by NR simulations. One is limited to inspirals with relatively small mass ratios (where we assume the mass ratio of any binary is taken as larger than 1), although head-on collisions with mass ratios up to 1000 have been performed \cite{Lousto:2022hoq}. High mass ratios require high resolutions and mesh refinement, which eventually effectively grinds simulations to a halt. Apart from being unable to probe this part of parameter space, it is another challenge to cover small mass ratio inspirals with misaligned BH spins. These issues are becoming more urgent with the prospect of next-generation GW detectors being built, such as the Laser Interferometer Space Antenna (LISA) \cite{Robson_2019}, the Einstein Telescope (ET) \cite{Punturo_2010} and Cosmic Explorer (CE) \cite{evans2021horizon}. which will detect many more GW inspirals than current detectors, meaning much more elaborate GW template banks are needed to interpret these GW signals for binary BH inspirals \cite{Ferguson_2021}. 

These issues are not unique to the area of GWs. In the specific case of PBHs, the issue of heavy simulations can sometimes be remedied by applying techniques of dimensional reduction, e.g. for the numerical determination of the collapse threshold of overdensities, already referenced in chapter \ref{Chapter3}, and the study of critical PBH collapse processes, see e.g. \cite{Choptuik:1993,Evans_1994,Koike_1995,Niemeyer_1998:nj,gundlach2000critical,Gundlach_2003,Musco_2009}.

\begin{figure}[t]
    \begin{subfigure}[t]{0.3\linewidth}
      \centering
      \includegraphics[width=1.\linewidth]{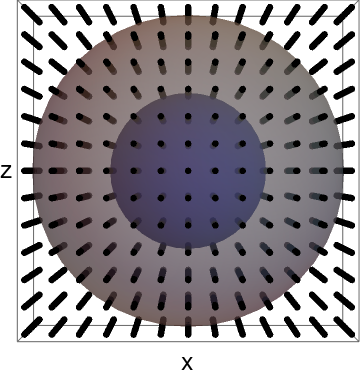}
      \caption{Coverage using a three-dimensional grid. The $y$ axis is not labeled but goes into the page. It is clear that there is much redundant information on the grid. }
      \label{fig::cartoon0}
    \end{subfigure}\hfill
    \begin{subfigure}[t]{0.3\linewidth}
      \centering
      \includegraphics[width=1.\linewidth]{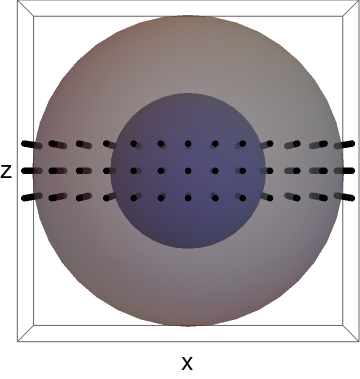}
      \caption{Here, the grid is reduced in the $z$-direction, as proposed in \cite{Alcubierre:1999ab}. Gradients in the $z=0$ plane can be computed using regular finite differencing, and grid points away from this plane are filled between timesteps using interpolation.}
      \label{fig::cartoon1}
    \end{subfigure}\hfill
    \begin{subfigure}[t]{0.3\linewidth}
      \centering
      \includegraphics[width=1.\linewidth]{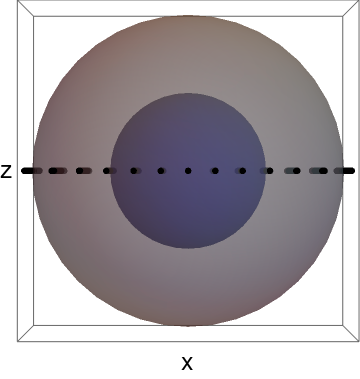}
      \caption{Here, the grid in the $z$-direction is reduced to a single layer, as proposed in \cite{Pretorius_2005}. Gradients in the $z=0$ plane are expressed in terms of gradients in the $y$-direction.}
      \label{fig::cartoon2}
    \end{subfigure}
    \caption{\textbf{Dimensional reduction with the cartoon method.} Schematic representation of a spherically symmetric physical setup covered by different types of grids.}
    \label{fig::cartoon}
\end{figure}

Within the framework of dimensional reduction, one assumes that the physical process one is simulating is symmetric to such an extent that it does not need to be simulated in its full dimensionality. For instance, one only needs to simulate the radial dimension of any spherically symmetric problem, making these investigations computationally comparatively light. The full solutions can then be recovered by rotating the obtained radial solution around the elevation and azimuthal angles. Dimensional reduction is not just useful to reduce the computational cost of 3+1D simulations, but can also be used to solve higher dimensional problems using existing two or three dimensional codes, e.g. higher dimensional black holes \cite{Emparan_2008}. One dimensional reduction paradigm is the so-called cartoon method \cite{Alcubierre:1999ab}, of which a modification specifically geared towards NR problems was published in \cite{Pretorius_2005}. A subsequent modification was published in \cite{Cook:2016soy}, in which the authers develop cartoon expressions for the BSSN formulation of GR. Fig. \ref{fig::cartoon} contains images of the different types of grids used in these publications. 

In this chapter, we will discuss how expressions and equations are adapted in the context of the cartoon method of dimensional reduction in section \ref{sect::cartoon_expressions}. We will discuss cartoon expressions for the CTTK method for solving the constraint equations in section \ref{sect::cartoon_init_conditions} and expressions for scalar field evolution equations in the context of NR in section \ref{sect::cartoon_evol_eqns}. Finally, we discuss vacuum bubble collision as an example of a physical process that can be investigated using these expressions in section \ref{sect::cartoon_applications}.

\section{Cartoon expressions}\label{sect::cartoon_expressions}
In the rest of this chapter, we will assume that we are dealing with 3+1D physical scenarios that are axisymmetric around the $x$-axis, i.e. there is an $SO(2)$-symmetry. This choice is inspired by the axisymmetry of the collision of two true vacuum bubbles in a first order phase transition, but the same symmetry is exhibited by any head-on collision of two spherically symmetric objects, e.g. black holes, or by any spherically symmetric object itself. Ideally, we have a formulation of the problem in which the only data we need is on the top half of the $x-y$ plane, i.e. in the region $x \in \mathbb{R}, y \in \mathbb{R}_{>0}, z = 0$. It is then possible to obtain the full three-dimensional solution at any point in time by simply rotating the solution in the $x-y$ plane around the $\phi$ direction. Therefore, one only needs to define and evolve initial conditions in this $x-y$ plane. The angle $\phi$ is defined by a new polar coordinate system in the $y-z$ plane, such that 
\begin{subequations}
\begin{align}
    \rho &= \sqrt{y^2 + z^2}, \quad \tan^{-1}(\frac{z}{y}),\\
    y &= \rho \cos{\phi}, \quad z = \rho \sin{\phi}.
\end{align}
\end{subequations}
We will focus on strong gravity problems that are being simulated using a BSSN formulation of the EFE. Additionally, we will assume that there is a reflective symmetry $\phi \to -\phi$, i.e. we consider only twist-free axisymmetric spacetimes, which makes dealing with the dimensional reduction more straightforward. These are all simplifications of the most general scenario of a $(D-1)+1$ dimensional spacetime with an $SO(d)$-symmetry, covered in full in \cite{Cook:2016soy} for the case of a vacuum spacetime. To be extra clear, we note that the processes studied in chapter \ref{Chapter3} are axisymmetric and twist-free and could therefore be studied using the methods described in this chapter. However, the processes in chapter \ref{Chapter4} have angular momentum and are therefore not twist-free, so that the methods described in this chapter do not apply and one has to use full 3+1D simulations instead, as we have done.

Our notation regarding indices will be as follows: early uppercase Latin indices' range is $A,B = 0,\ldots,3$, while middle uppercase Latin indices' range is $I,J = 1,\ldots,3$. Similarly, middle lowercase indices range is $i,j = 1,2$. Finally, early lowercase indices' range is $a,b = 3$, following the conventions in \cite{Cook:2016soy}, in which the indices $a,b$ range over the cartooned dimensions. 

When one assumes an $SO(d)$ symmetry, it can be shown that various components of several types of tensor vanish on the simulated plane. For an axisymmetry specifically, any $z$-component of any vector, such as $\beta^z$, vanishes on the $z=0$ plane, as does e.g. $\partial_z \phi$, where $\phi$ is some scalar quantity. Additionally, for any symmetric tensor $T_{AB}$, $T_{zi} = \partial_z T_{ij} = 0$. The price to pay for these simplified expressions is that one needs to keep track of a new cartoon index $w$, which in the context of numerical GR only appears in the additional tensor components $g_{ww}$ and $T_{ww}$ that one needs to evolve. The $ww$ components represent the diagonal components of the cartooned dimensions, e.g. in this case 
\begin{equation}
    \begin{pmatrix}
        g_{00} & g_{01} & g_{02} & g_{03} \\
        & g_{11} & g_{12} & g_{13} \\
        & & g_{22} & g_{23} \\
        & & & g_{33} \\ 
    \end{pmatrix} \quad\xrightarrow{SO(2)} \quad
    \begin{pmatrix}
        g_{00} & g_{01} & g_{02} & \\
        & g_{11} & g_{12} & \\
        & & g_{22} & \\
        & & & g_{ww} \\ 
    \end{pmatrix},
\end{equation}
where we have only included the unique non-vanishing components. When there is even more symmetry, more components vanish and the $g_{ww}$ component represents more than one original metric component, e.g. in the case of an $SO(3)$ symmetry 
\begin{equation}
    \begin{pmatrix}
        g_{00} & g_{01} & g_{02} & g_{03} \\
        & g_{11} & g_{12} & g_{13} \\
        & & g_{22} & g_{23} \\
        & & & g_{33} \\ 
    \end{pmatrix} \quad\xrightarrow{SO(3)} \quad
    \begin{pmatrix}
        g_{00} & g_{01} & & \\
        & g_{11} & & \\
        & & g_{ww} & \\
        & & & g_{ww} \\ 
    \end{pmatrix},
\end{equation}
It is worth mentioning that because in the cartoon formalism the metric becomes block diagonal, one finds that $g^{ww} = 1/g_{ww}$.

For instructive purposes, we will show why some specific tensor components vanish on the dimensionally reduced domain, following the discussion in \cite{Cook:2016soy}. Firstly, we consider a symmetric tensor $T$ that transforms as 
\begin{equation}
    \bar{T}_{AB} = \frac{\partial X^A}{\partial Y^C}\frac{\partial X^B}{\partial Y^D}T_{CD},
\end{equation}
where $X^A$ and $Y^C$ are different sets of coordinates. To show that $T_{iw} = 0$, we use that $\bar{T}_{i\varphi} = 0$ by symmetry and we transform to Cartesian coordinates as 
\begin{equation}
\begin{split}
    0\equiv \bar{T}_{i\varphi} &= \frac{\partial X^A}{\partial x}\frac{\partial X^B}{\partial \varphi}T_{AB},\\
    &= \frac{\partial x}{\partial x}\left(\frac{\partial y}{\partial \varphi}T_{xy} + \frac{\partial z}{\partial \varphi}T_{xz}\right),\\
    &= -zT_{xy}  + yT_{xz}\\
    T_{xz} &= \frac{z}{y}T_{xy}.
\end{split}
\end{equation}
Similarly, we may transform the component $\bar{T}_{\rho\varphi}$, which also vanishes by symmetry, as 
\begin{equation}
\begin{split}
    0\equiv \bar{T}_{\rho\varphi} &= \frac{\partial X^A}{\partial \rho}\frac{\partial X^B}{\partial \varphi}T_{AB},\\
    &=  \frac{\partial y}{\partial \rho}\frac{\partial y}{\partial \varphi}T_{yy} + 
        \frac{\partial y}{\partial \rho}\frac{\partial z}{\partial \varphi}T_{yz} + 
        \frac{\partial z}{\partial \rho}\frac{\partial y}{\partial \varphi}T_{zy} + 
        \frac{\partial z}{\partial \rho}\frac{\partial z}{\partial \varphi}T_{zz},\\
    &= -z\cos\varphi T_{yy} + y\cos\varphi T_{yz} - z\sin\varphi T_{zy} + y\sin\varphi T_{zz},\\
    \left(\frac{y^2}{\rho} - \frac{z^2}{\rho}\right)T_{yz} &= \frac{yz}{\rho}\left(T_{yy} - T_{zz}\right),\\
    T_{yz} &= \frac{yz}{y^2 + z^2}\left(T_{yy} - T_{zz}\right).
\end{split}
\end{equation}
Since we assume that we operate exclusively in the region where $z=0$, the two equations above imply that $T_{iz} = T_{iw} = 0$. We list a more complete list of modified cartoon expressions below, which can be derived in similar fashion. We stress that expressions hold in the $z=0$ plane for an axisymmetry around the $x$-axis, for any scalar field $\phi$, vector field $V$ and tensor field $T$. 
\begin{subequations}
\begin{align}
    \partial_z \phi &= \partial_i \partial_z \phi = 0,\\
    \partial_z^2 \phi &= \frac{\partial_y\phi}{y},\\
    V^z &= \partial_i V^z = \partial_z V^i = \partial_z^2 V^z = 0,\\
    \partial_z V^z &= \frac{V^y}{y},\\
    \partial_i \partial_z V^z &= \big(\frac{\partial_i V^y}{y} - \delta_{iy}\frac{V^y}{y^2}\big),\\
    \partial_z^2 V^i &= \big(\frac{\partial_y V^i}{y} - \delta^i_{y}\frac{V^y}{y^2}\big),\\
    T_{iz} &= \partial_z T_{zz} = \partial_i \partial_z T_{zz} = \partial_z^2 T_{iz} = \partial_z T_{ij} = \partial_i\partial_z T_{jk} = 0,\\
    T_{zz} &= T_{ww}.
\end{align}
\end{subequations}

To illustrate how these expressions are used in practice, let us consider a simple wave equation in 3+1D Minkowski space, i.e. 
\begin{equation}
    \Box\phi = g^{AB}\partial_A\partial_B\phi = -\partial_t^2\phi + 2g^{0I}\partial_I\partial_t\phi + g^{IJ}\partial_I\partial_J\phi = 0
\end{equation}
It is common to decompose this second order system into two first order equations, i.e. 
\begin{subequations}
\begin{align}
    \partial_t\phi &= \Pi\\
    \partial_t\Pi &= 2g^{0I}\partial_I\Pi + g^{IJ}\partial_I\partial_J\phi.
\end{align}
\end{subequations}
When one wants do dimensionally reduce the above set of equations, the second equation is the only non-trivial one. We can write the RHS as 
\begin{equation}
    \partial_t\Pi = 2g^{0i}\partial_i\Pi + 2g^{0a}\partial_a\Pi + g^{ij}\partial_i\partial_j\phi + 2g^{ia}\partial_i\partial_a\phi + g^{ab}\partial_a\partial_b\phi.
\end{equation}
The second and fourth term on the RHS must vanish, because they involve a $z$-derivative of $\Pi$ and a mixed $i,z$-derivative of $\phi$ respectively, and the first and third terms only involve derivatives with respect to dimensions that we have access to, so one only needs to rewrite the last term. Since we know that $\partial_a\partial_b\phi = \delta_{ab}\partial_y\phi/y$, this equation becomes 
\begin{equation}\label{eqn::pifinal}
    \partial_t\Pi = 2g^{0i}\partial_i\Pi + g^{ij}\partial_i\partial_j\phi + g^{ww}\frac{\partial_y\phi}{y},
\end{equation}
We stress that the computation of these terms only requires gradients in the simulated directions, and all derivatives in the $z$ direction are gone.

\section{Initial conditions}\label{sect::cartoon_init_conditions}

It is vital in numerical GR to be able to solve the constraint equations on the initial hyperslice, to ensure that the obtained solution obeys the full four-dimensional EFE. In this chapter, we will discuss dimensional reduction of the CTTK method \cite{Aurrekoetxea:2022mpw}, in which the Hamiltonian constraint is solved by solving an algebraic equation for the trace of the extrinsic curvature, \eqn{eqn:CTTKK} and an elliptic equation for its traceless part, \eqn{eqn::cttk5}, which we copy here for convenience, 
\begin{subequations}
\begin{align}
    K^2 &= 12\psi_0^{-5}\partial^j\partial_j\psi_0 + \frac{3}{2}\psi_0^{-12}\bar{A}_{ij}\bar{A}^{ij} + 24\pi\rho,\label{cttkcartK}\\
    \partial^j\partial_j V_i &= \frac{2}{3}\psi^6\partial_i K + 8\pi \psi^6S_i\label{cttkcartMom}.
\end{align}
\end{subequations}

When one is interested in solving these equations on a two-dimensional grid representing an axisymmetric three-dimensional spacetime, extra cartoon terms are introduced. Firstly, setting $\psi_0 = 1$ the Hamiltonian constraint \eqn{cttkcartK} becomes 
\begin{equation}\label{eqn::cttkhamconstraintcartoon}
    K^2 = \frac{24\pi\rho}{\mpl^2} + \frac{3}{2}(\bar{A}_{ij}\bar{A}^{ij} + \bar{A}_{ww}\bar{A}^{ww}).
\end{equation}
The $i=3$ equation of the momentum constraints in \eqn{cttkcartMom} vanishes on both sides trivially, while the $i=1,2$ equations are modified with cartoon terms as follows 
\begin{equation}
    \partial^j\partial_j V_i + \frac{\partial_yV_i}{y} - \frac{\delta^y_iV_y}{y^2} = \frac{2}{3}\psi^6\partial_iK + 8\pi \psi^6S_i.
\end{equation}
This equation can be solved numerically using standard techniques, e.g. by linearising and iterating to obtain a solution. 

Finally, when one reconstructs the extrinsic curvature tensor from the solutions $V_i$, more modified cartoon expressions come into play. Because $W_I = V_I + \partial_I U = 3V_I/4$, one reconstructs $\bar{A}_{IJ}$ as 
\begin{equation}
    \bar{A}^{IJ} = \frac{3}{4}\big(\partial^I W^J + \partial^J W^I\big) - \frac{1}{2}\gamma^{IJ}\partial_K W^K,
\end{equation}
where we assume explicitly that the metric is conformally flat and the conformal factor $\chi$ is normalized to one. This is cartoon modified as
\begin{subequations}
\begin{align}
    \bar{A}^{ij} &= \frac{3}{4}\big(\partial^i W^j + \partial^j W^i\big) - \frac{1}{2}\gamma^{ij}(\partial_k W^k + \partial_z W^z),\\
    &= \frac{3}{4}\big(\partial^i W^j + \partial^j W^i\big) - \frac{1}{2}\gamma^{ij}(\partial_k W^k + \frac{V^y}{y}) \nonumber \\
    \bar{A}^{ww} = \bar{A}^{zz} &= \frac{3}{2}\partial^z W^z - \frac{1}{2}\gamma^{zz}(\partial_k W^k + \partial_z W^z)\\
    &= \frac{3}{2}\frac{V^y}{y} - \frac{1}{2}\gamma^{ww}(\partial_k W^k + \frac{V^y}{y}). \nonumber
\end{align}
\end{subequations}

\section{Cartoon expressions for a scalar field in twist-free axisymmetry}\label{sect::cartoon_evol_eqns}

In this section, we list the cartoon expressions for the stress-energy tensor and evolution equations for a real scalar field minimally coupled to gravity in twist-free axisymmetry. The cartooned evolution equations for the metric variables can be found in \cite{Cook:2016soy}. To add and evolve non-trivial matter configurations, one needs to cartoon both the stress components and the matter evolution equations. Focusing on a real scalar field, the stress-energy components in the BSSN formalism are 
\begin{subequations}\label{eqn::scalarfieldemtensor}
\begin{align}
    \rho &= \frac{1}{2}\Big(\Pi^2 + \chi\gamma^{IJ}\partial_I\phi\partial_J\phi\Big) + V(\phi),\\
    S_I &= -\Pi \partial_I\phi,\\
    S_{IJ} &= \partial_I\phi\partial_J\phi - \frac{1}{2} \bar{\gamma}_{IJ}\bar{\gamma}^{KL}\partial_K\phi\partial_L\phi + \frac{1}{2}\gamma_{IJ}\Pi^2 - \gamma_{IJ}V(\phi)\\
    S &= \gamma^{IJ}S_{IJ},
\end{align}
\end{subequations}
while the evolution equations for the scalar field and its first time derivative are 
\begin{subequations}
\begin{align}
    \partial_t \phi &= \alpha \Pi + \beta^I\partial_I \phi, \\
    \partial_t \Pi &= \beta^I\partial_I \Pi + \alpha\partial^I\partial_I \phi + \partial^I \phi \partial_I \alpha\\ 
    &\quad+ \alpha\Big(K\Pi - \gamma^{IJ}\Gamma^K_{IJ}\partial_K\phi - \frac{dV}{d\phi}\Big)\nonumber.
\end{align}
\end{subequations}
In terms of the stress tensor, dimensional reduction introduces a new stress tensor component $S_{ww}$, which also changes the expression for the trace, so that the complete cartooned expressions are 
\begin{subequations}
\begin{align}
    \rho &= \frac{1}{2}\Big(\Pi^2 + \chi\gamma^{ij}\partial_i\phi\partial_j\phi\Big) + V(\phi),\\
    S_i &= -\Pi \partial_i\phi,\\
    S_{ij} &= \partial_i\phi\partial_j\phi - \frac{1}{2} \bar{\gamma}_{ij}\bar{\gamma}^{kl}\partial_k\phi\partial_l\phi + \frac{1}{2}\gamma_{ij}\Pi_\phi^2 - \gamma_{ij}V(\phi)\\
    S_{ww} &= \frac{1}{2}\gamma_{ww}\Pi^2\\
    S &= \gamma^{ij}S_{ij} + \gamma^{ww}S_{ww}.
\end{align}
\end{subequations}
In the evolution equations, the only expression with a double derivative is the second, so that the cartooned expressions become 
\begin{subequations}
\begin{align}
    \partial_t \phi &= \alpha \Pi + \beta^i\partial_i \phi, \\
    \partial_t \Pi_\phi &= \beta^i\partial_i \Pi_\phi + \alpha\partial^i\partial_i \phi + \partial^i \phi \partial_i \alpha\\ 
    &\quad+ \alpha\Big(K\Pi_\phi - \gamma^{ij}\Gamma^k_{ij}\partial_k\phi - \frac{dV}{d\phi}\Big)\nonumber \\
    &\quad+ \alpha g^{ww}\frac{\partial_y \phi}{y} - \alpha \gamma^{ww}\Gamma^k_{ww}\partial_k \phi,\nonumber
\end{align}
\end{subequations}
and for completeness we include the expression
\begin{equation}
    \Gamma^k_{ww} = \gamma^{kj}\frac{g_{jy} - \delta_{jy}g_{ww}}{y}.
\end{equation}

\section{Vacuum bubble collision with dimensional reduction}\label{sect::cartoon_applications}

\begin{figure}
    \centering
    \includegraphics[width=0.7\linewidth]{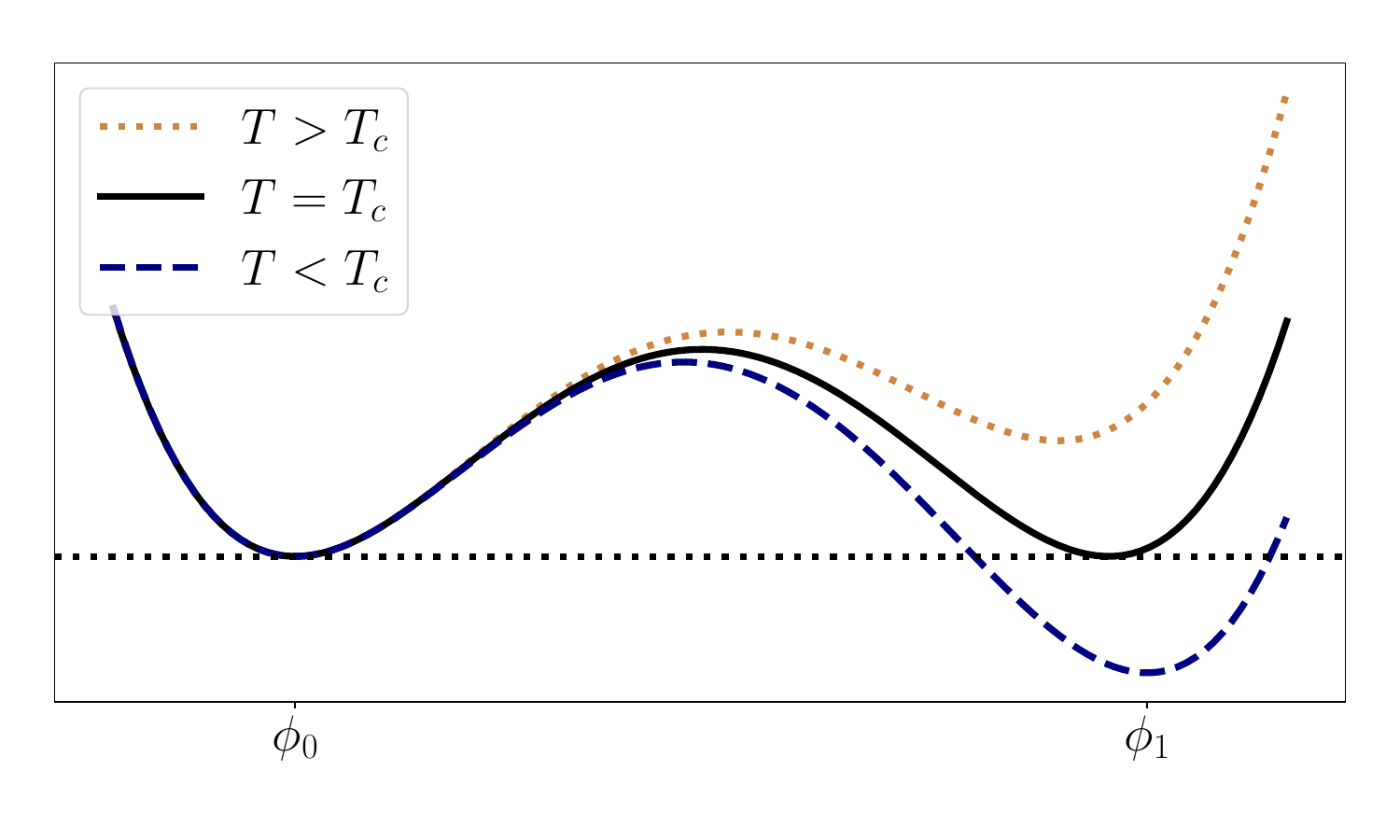}
    \caption{\textbf{Potential change} for a scalar field undergoing a first order phase transition. } 
    \label{fig::bubble_potential}
\end{figure}

We test the formalism described in the previous two sections by using it to simulate the collision of two true vacuum bubbles in a first-order phase transition. For completeness, we first discuss some of the relevant research context briefly. Readers that are only interested in the numerical results are encouraged to go straight to section \ref{sect:cartoon_nums}. Many extensions of the standard model introduce a first-order phase transition, in which as the temperature of the universe drops, a new energetically favoured state is formed at some critical temperature $T_c$. A simple example is that of a scalar field in a potential with two minima, as illustrated in Fig. \ref{fig::bubble_potential}. In such a scenario, while $T > T_c$, the scalar field $\phi$ has value $\phi_0$ throughout the universe, which is the potential's global minimum at that stage. The other minimum at $\phi_1$ is local, i.e. $V(\phi_0) < V(\phi_1)$. At $T=T_c$ however, we get $V(\phi_0) = V(\phi_1)$ and when the temperature drops further, a new global minimum is formed, i.e. 
\begin{equation}
\begin{split}
    V(\phi_0) < V(\phi_1),\quad &\textrm{for}\quad T > T_c\\
    V(\phi_0) = V(\phi_1),\quad &\textrm{for}\quad T = T_c\\
    V(\phi_0) > V(\phi_1),\quad &\textrm{for}\quad T < T_c.
\end{split}
\end{equation}
The field is now stuck in the false, local minimum at $\phi_0$ and even though there is no obvious classical mechanism for the field to reach the new global minimum, because the potential barrier is in the way, the field can quantum tunnel through the barrier to $\phi_1$ \cite{Coleman:1977,Callan:1977,Linde:1981zj}, which forms a true vacuum bubble. 

The energy of the inner region of this bubble is lower than the energy outside it, which causes a pressure difference that provides a force that drives expansion of the bubble. This outward force is counteracted by the surface tension of the bubble wall, which provides a force that drive bubble contraction. If the energy difference between true and false vacuum is $\Delta V$ and the surface tension is $\sigma$, one can express the total energy of a bubble with radius $R$ as  
\begin{equation}
    E_\textrm{bubble} = -\frac{4}{3}\pi R^3 \Delta V + 4\pi R^2 \sigma,
\end{equation}
and the bubble will expand if this energy is negative, i.e. for bubbles larger than the critical radius $R_c = 3\sigma/\Delta V$. The inner true vacuum region is surrounded by an energetic bubble wall, where the field values interpolate beween the true and false vacua. As the bubble grows, energy is released from regions that move from the false vacuum to the true vacuum. This energy accelerates the outward movement of the bubble walls, which can rapidly almost reach the speed of light. We note that this causes Lorentz contraction of the bubble walls, which is a challenge in simulations of the expansion of vacuum bubbles, since more resolution is needed to resolve the bubble wall accurately as the difference between the length scales of the simulation domain and the contracted bubble wall increases. Setups like the one described above can be found in many extensions of the standard model, e.g. ones related to dark/hidden sectors \cite{Espinosa_2008,Baldes_2019,Breitbach_2019,Fairbairn:2019xog} and electroweak baryogenesis \cite{No:2011fi,Vaskonen:2016yiu,Dorsch:2016nrg,Artymowski:2016tme,Beniwal:2017eik,Beniwal:2018hyi}.

If two vacuum bubbles form in distinct regions and grow, their walls may meet and collide. Given the high velocity of the bubble walls, these events can be violent and leave various imprints, e.g. a GW background \cite{Witten:1984} that, if strong enough, could be probed by future GW detectors such as LISA \cite{amaroseoane2017laser}. Because bubble collisions are such violent events, it is possible that matter at the collision point becomes dense enough to undergo gravitational collapse to a black hole. For instance, one can ask whether the immediate overdensities caused by the collision of one or multiple bubbles can lead to BH formation, as was first done by the authors of \cite{PhysRevD.26.2681}. In this situation however, it is unclear whether or not BH forms because the physics at the collision point are far from spherically symmetric, so the hoop conjecture \eqn{eqn::hoop_conjecture} cannot simply be applied, and the volumes in which enough energy for gravitational collapse is contained are much larger than the collision point. 

An alternative scenario is proposed by the authors of \cite{khlopov1998formation}, in which it is assumed that after the immediate collision of two bubble walls, the field can be reflected back to the false vacuum at the collision point, while most of the surroundings are already in the true vacuum and in fact, the field may oscillate between the two vacua several times at the collision point. The authors of \cite{Lewicki:2019gmv} show numerically that this process is realistic, and this would clearly create overdensities that could potentially be dense enough to collapse. The authors of \cite{Lewicki:2019gmv} argue that these regions are not spherically symmetric and that BH formation is therefore unlikely, but this has not been investigated with NR.

\subsection{Initial conditions}

In this section, we discuss in detail how we obtain initial conditions for dimensionally reduced simulations of vacuum bubble collisions. One must define initial conditions for two true vacuum bubbles seperated by a certain distance, which will subsequently expand and eventually collide. There are several ways to obtain initial conditions for the matter field and its first time derivative and we report on two specific examples here. Firstly, one may use the quartic potential from \cite{Dymnikova:2000dy}
\begin{equation}\label{eqn:potential_D}
    V_1(\phi) = \frac{1}{8}\lambda \big(\phi^2 - \phi_0^2\big)^2 + \epsilon\phi_0^3\big(\phi + \phi_0\big).
\end{equation}
In the thin-wall limit where $\frac{\epsilon}{\lambda}\ll 1$, the lowest-energy solution to the equation of motion that interpolates between the two vacua is given by 
\begin{equation}\label{eqn::bubblefield}
    \phi = \phi_0 \Big(\tanh{(\frac{\gamma m}{2}(r - R(t)))} - \frac{\epsilon}{\lambda}\Big),
\end{equation}
where $R(t) = vt + R_0$, $\gamma = \frac{1}{\sqrt{1-v^2}}$, $m = \sqrt{\lambda}\phi_0$ and $R_0 = \frac{2\lambda}{3\epsilon m}$ are the time-dependent radius, Lorentz factor, initial width and initial radius of the bubble respectively. The gamma factor is related to the radius via 
\begin{equation}\label{eqn::bubblegamma}
    \gamma = R\frac{\epsilon}{\lambda} + \frac{4}{27}\frac{\lambda^2}{\epsilon^2R^2}.
\end{equation}
One may take a first time derivative of \eqn{eqn::bubblefield} to obtain an expression for $\Pi = \partial_t\phi - \beta^i\partial_i\phi$. These expressions can thus be very simply obtained for any given radius $R$, from which one finds the corresponding value for $\gamma$ using \eqn{eqn::bubblegamma} and subsequently the profiles for $\phi$ and $\Pi$. Having found these configurations, constraint-solving configurations for the extrinsic curvature can be found using the method described in section \ref{sect::cartoon_init_conditions}. 

Secondly, we consider the potential given in \cite{Lewicki:2019gmv}
\begin{equation}
    V_2(\phi) = v^4 + av^2\phi^2 - (2a + 4)v\phi^3 + (a+3)\phi^4,
\end{equation}
which conveniently has a local minimum at $\phi = 0$ and a global minumum at $\phi = v$. We have no approximation of an analytic solution for this potential, but it is easy to employ a shooting method that numerically solves the equation of motion 
\begin{equation}\label{eqn::phi_shooting_eqn}
    \partial_r^2 + \frac{3}{r}\partial_r = V'(\phi),
\end{equation}
where $r^2 = t^2 + x^2 + y^2 + z^2$ and one imposes boundary conditions $\phi = v, \partial_r\phi = 0$ at $r = 0$ and $\phi \to 0$ as $r \to \infty$. This configuration minimizes the Euclidean action 
\begin{equation}
    S_\textrm{E} = \int ~d^4x \left[\frac{1}{2}\left(\partial_t\phi\right)^2 + \frac{1}{2}\left(\nabla\phi\right)^2 + V(\phi)\right]
\end{equation}
and will therefore dominate the tunneling probability. One may employ a shooting algorithm to solve \eqn{eqn::phi_shooting_eqn} and use this numerical solution directly on the initial hyperslice. Alternatively, one may fit a hyperbolic tangent function to the profile, i.e.
\begin{equation}
    \phi(r) = A \tanh\big(\frac{r - R_0}{\lambda}\big) + B,
\end{equation}
where $A, R_0, \lambda, B$ are the amplitude, initial radius, initial width and shift of the profile respectively. Obtaining such a fit can be useful for several reasons: firstly, it simplifies plugging the initial profile into the initial condition solver. Secondly, this initially stationary tangent hyperbolic profile may be extended to boosted solutions as in \cite{Dymnikova:2000dy}, although we have not yet attempted this and leave this to further studies.

\begin{figure}[t]
    \begin{subfigure}[t]{0.475\linewidth}
        \centering
        \includegraphics[width=1.\linewidth]{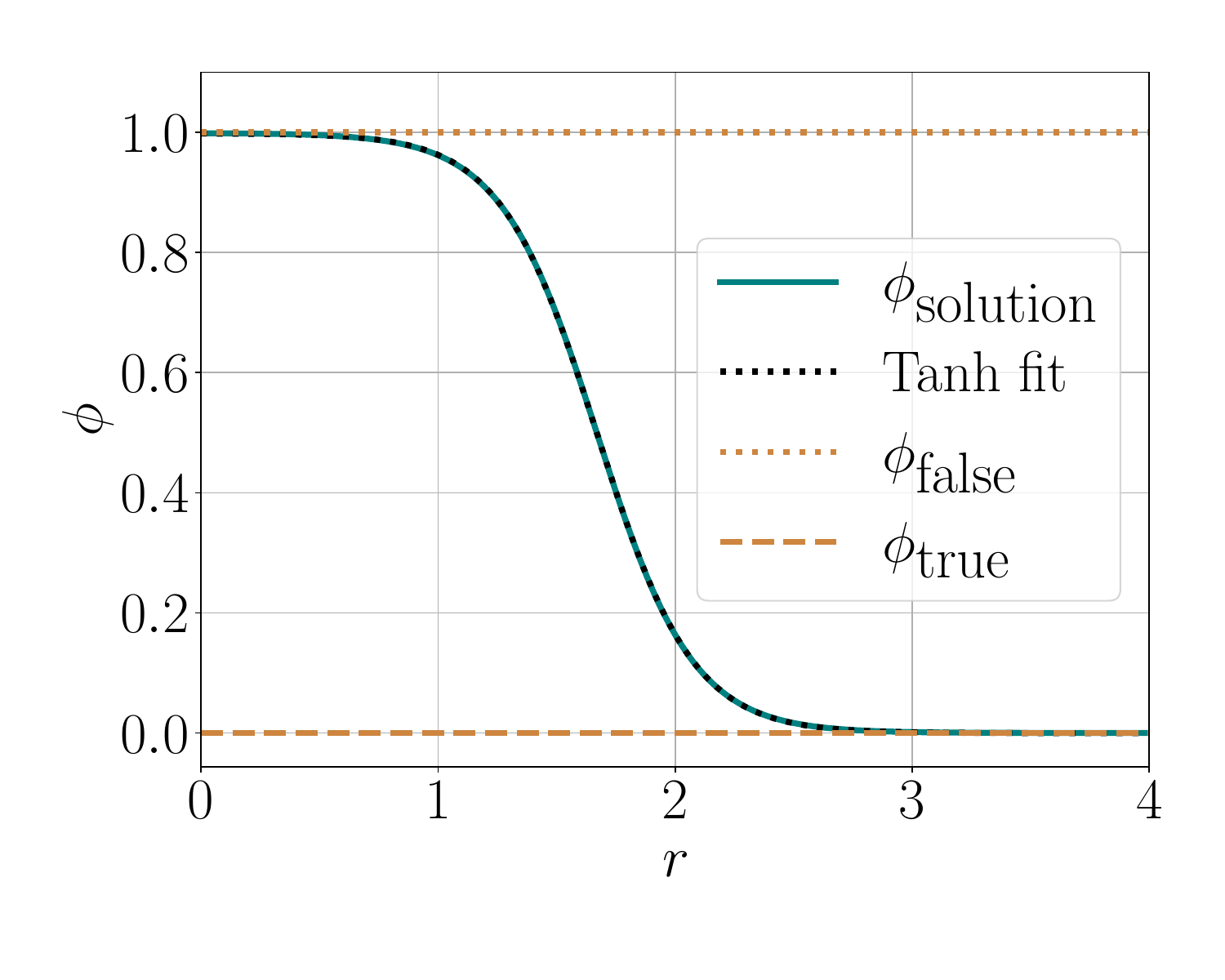}
        \caption{\textbf{Numerical solution for $\phi$} for equation \ref{eqn::phi_shooting_eqn}, for boundary conditions and parameters listed in the main text, obtained using a shooting algorithm. } 
        \label{fig::phi_shooting}
    \end{subfigure}\hfill
    \begin{subfigure}[t]{0.475\linewidth}
        \centering
        \includegraphics[width=\linewidth]{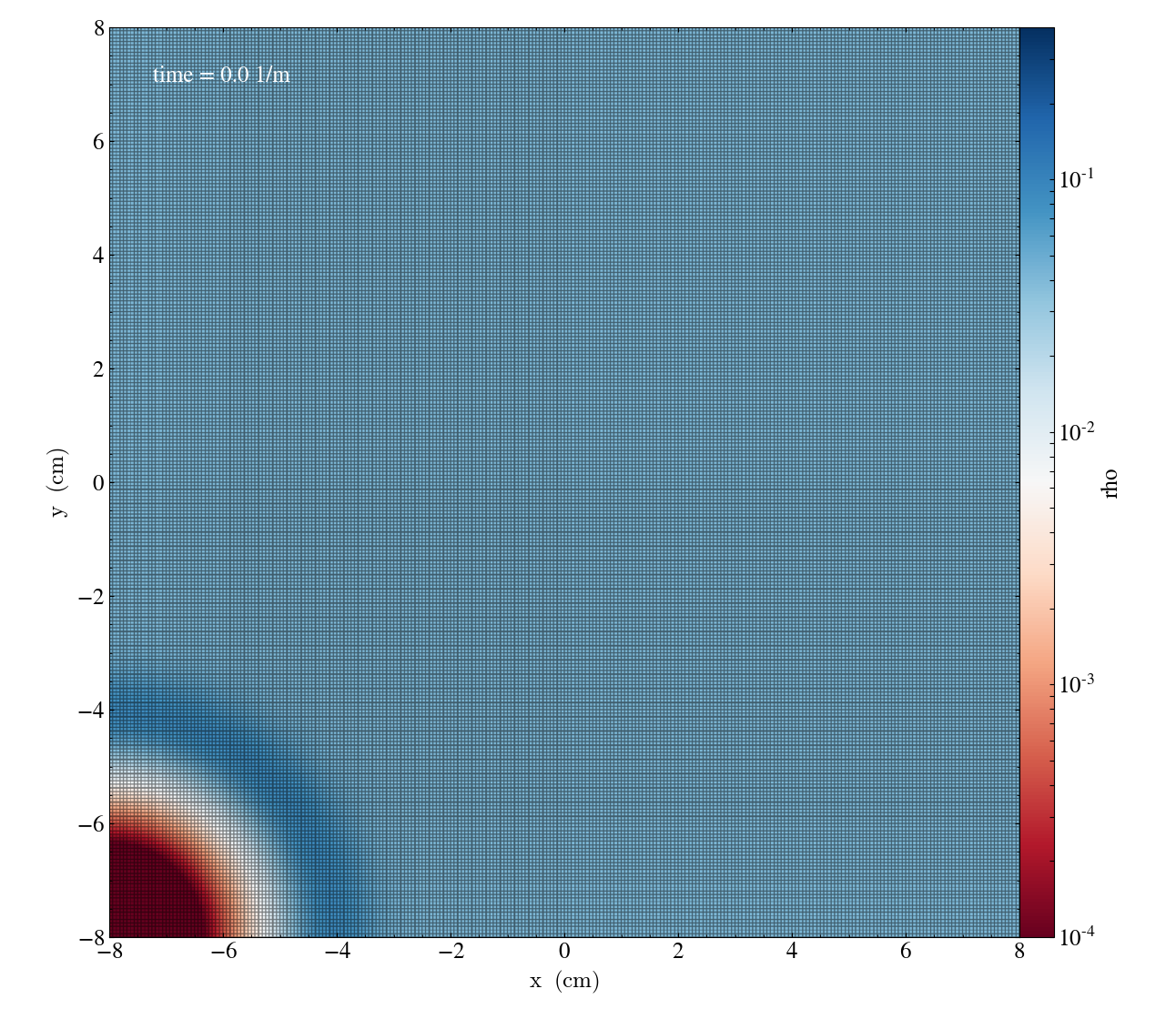}
        \caption{\textbf{Initial 2D spatial hyperslice} showing the energy density of the configuration shown in Fig. \ref{fig::phi_shooting}. } 
        \label{fig::bubble_slice_0}
    \end{subfigure}
    \caption{\textbf{Vacuum bubble initial field configuration} obtained using a shooting method.}
    \label{fig::bubble_init}
\end{figure}

The outcome of this procedure is illustrated in Fig. \ref{fig::phi_shooting}, for parameters $a=10$ and $v=1$. It is clear that the obtained numerical solution interpolates cleanly between the true vacuum at small radii and the false vacuum at larger distances. Furthermore, the hyperbolic tangent fit is accurate, with parameters $A = -0.499, R_0 = 1.67, \lambda=0.406$ and $B = 0.499$. A typical initial slice field configuration is shown in Fig. \ref{fig::bubble_slice_0}, where we have used the reflective $x\to -x$ symmetry to cut the size of the simulation box in half, and the axis of rotational symmetry is on the $x$-axis. We impose an additional reflective symmetry on the boundary on the right hand side, so that once the bubble wall reaches that boundary it will collide with its mirror image. 

To obtain valid initial conditions for NR, we use the CTTK method to find constraint-satisfying initial configurations for the extrinsic curvature, where we use the cartooned expressions from section \ref{sect::cartoon_init_conditions}. 

\subsection{Evolution}\label{sect:cartoon_nums}

Here, we present numerical results obtained dimensional reduction method described above. Our aim here is simply to show that the method is numerically robust and we will therefore in particular show good numerical convergence of our simulations.

\subsubsection{Numerical method}

For this test, we use the quartic potential from \eqn{eqn:potential_D}. In our simulations, we rescale all quantities so that $R_0 = 2\lambda/3\epsilon m = 1$ and we will report the values of other quantities with respect to $R_0$.

We choose $\lambda R_0^2 = 2.84\times 10^5, \epsilon R_0^2 = 1.42 \times 10^4, \phi_0 = 2.50 \times 10^{-2}$, after which the initial configuration for $\phi$ is given by \eqn{eqn::bubblefield}. We choose these parameters so that the ratio between the width of the bubble wall $\sigma$ and $R_0$ is $\sigma/R_0 = (2/m)/R_0 \approx 0.05$, so that these length scales are not too far apart, which makes resolving the entire system with a uniform grid more feasible. We let the bubble start from rest, so that $\Pi = \partial_t\phi = \beta^i\partial_i\phi$ vanishes on the initial slice, as well as the linear momentum densities, so that the total-gradient assumption mentioned in section \ref{sect:constraints} is valid.

The length of our simulation domain is $L/R_0 = 8$ ($L/R_0 = 4$) in the $x$($y$)-direction, resolved uniformly by 256 (128) grid points. To perform a convergence tests, we run two identical simulations with the number of grid points multiplied by 2 and 4. We do not use AMR and use a constant Courant factor of 0.125. We choose code units such that one code length unit $x_\textrm{code}$ equals $R_0$. 

We evolve the system using the cartooned evolution equations from section \ref{sect::cartoon_evol_eqns}. To quantify Hamiltonian and momentum constraint violation, we define variables $\mathcal{H} = \sqrt{H^2}$ and $\mathcal{M} = \sqrt{M_1^2 + M_2^2}$, where $H$ and $M$ are the Hamiltonian and momentum constraints respectively, and these values are integrated over the entire simulation domain. and tracked over the course of the simulation for the three resolutions $h_1, h_2$ and $h_3$ of $256, 512$ and $1024$ in the $x$-direction.

\subsubsection{Results}

\begin{figure}[t!]
    \centering
    \includegraphics[width=\linewidth]{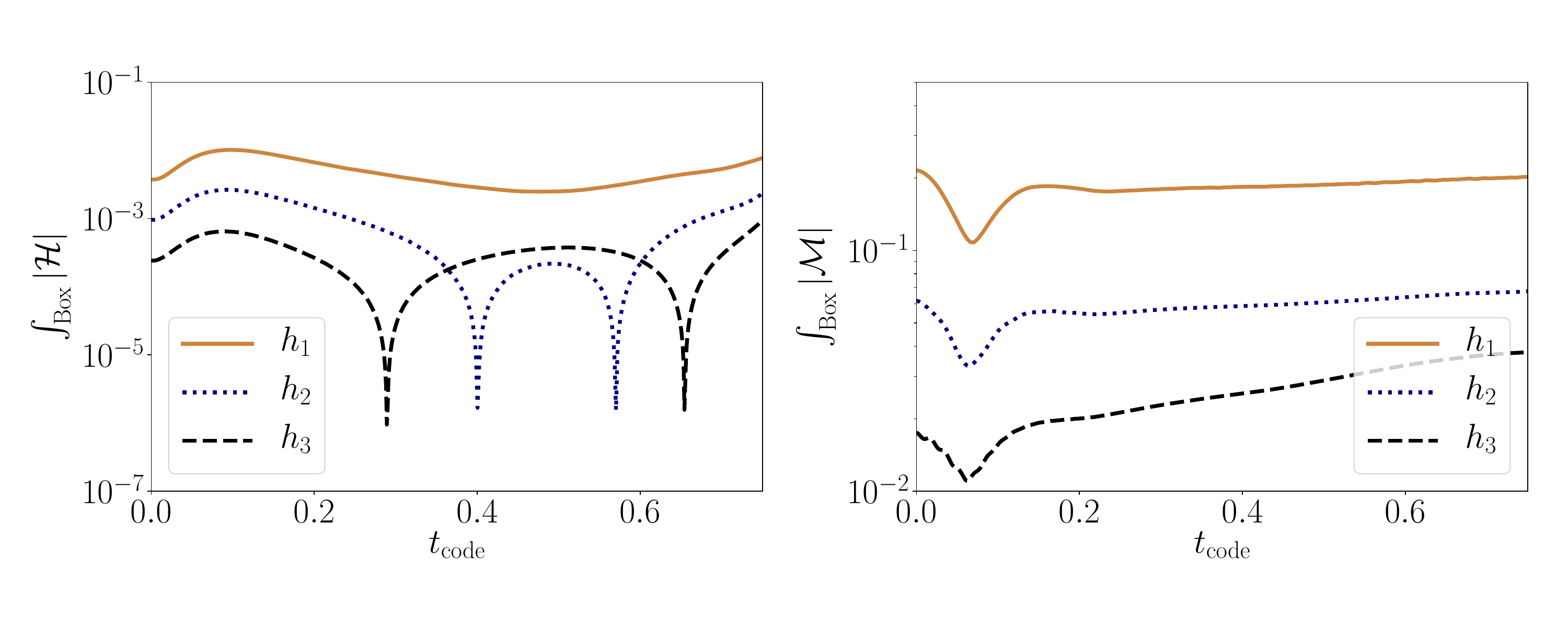}
    \caption{\textbf{Constraint violation for expanding vacuum bubbles} for three different resolutions. The panel on the left (right) hand side shows the absolute value of the Hamiltonian (momentum) constraint integrated over the entire simulation domain and it is clear that constraint violation decreases with increasing resolution over the course of the entire simulation, showing convergence. We checked that the ratio of absolute value of the constraints and the energy density around the bubble wall is $\sim 10^{-5}$ for this simulation, indicating that the constraints are indeed satisfied. }
    \label{fig::bubble_convergence}
\end{figure}

The results of these tests are shown in Fig. \ref{fig::bubble_convergence} and indicate good convergence of both the CTTK method and the evolution equations. We show the integrated moduli of the Hamiltonian and momentum constraints for identical simulations on uniform grids with grid spacings of $h_1 = 3.125 \times 10^{-2}/R_0, h_2 = h_1/2$ and $h_3 = h_1/4$. We expect that using the CCZ4 formalism in lieu of BSSN will further improve the accuracy of the simulations. 

With the grid being two-dimensional, it is feasible covering the simulation domain with a uniform resolution that is fine enough to resolve the accelerating bubble wall. Had the grid been three-dimensional, the computational cost would increase by roughly a factor 250 for the lightest simulation, dramatically increasing the computational resources and time required. This impact can be softened for a three-dimensional grid by leveraging $\grchombo$'s AMR capabilities and using fine regridding only around the bubble wall. However, we find that since field gradients around the bubble wall are large, one needs to cover a relatively wide region around the bubble wall with a fine resolution, meaning the simulations' computational cost is still significant. Furthermore, we find that interpolation errors associated to the boundaries of these. finer levels are significant and tend to cause serious constraint violation, which affects the convergence behaviour. Therefore, we stress that dimensional reduction is crucial in obtaining the performance shown in Fig. \ref{fig::bubble_convergence}. 

We expect that even in the dimensionally reduced setup, one will have to eventually use AMR to resolve the bubble wall, as it Lorentz contracts as it accelerates and its width will therefore become significantly smaller than the simulation domain length. Even though this will cause similar problems as described above, we expect that they can be dealt with more straightforwardly in the dimensionally reduced case. For instance, it is computationally cheaper to use a finer base resolution, so that one needs fewer additional AMR levels to resolve the bubble wall, introducing less interpolation and associated errors. Moreover, these errors are also decreased by increasing the distance between the boundaries of level $l$ and its subsequent level $l+1$, which comes at a computational cost that is much more manageable in the dimensionally reduced case. 

%% file: Chapter6/chapter6.tex
\chapter{Conclusions}\label{Chapter6}

\graphicspath{{Chapter6/Figs/}}

\section{Summary}

This thesis is an account of our research into PBH formation and dimensionally reduced simulations of matter fields in strong gravity using numerical relativity techniques. 

Chapters \ref{Chapter1} and \ref{Chapter2} make up the background material, i.e. in chapter \ref{Chapter1}, we review GR and its applications to BHs and cosmological spacetimes, after which we introduce PBHs. In chapter \ref{Chapter2}, we discuss NR, which is used for the research presented in this work. 

The research work in chapters \ref{Chapter3}, \ref{Chapter4} and \ref{Chapter5} discusses the study of formation processes of PBHs in a matter-dominated environment, as well as dimensionally reduced NR simulations with matter. In chapter \ref{Chapter3}, we study the collapse of spherically symmetric sub- and superhorizon overdensities in a matter-dominated early universe and the subsequent collapse into PBHs. In chapter \ref{Chapter4}, we study the collapse of overdensities with angular momentum in a matter-dominated early universe in order to deduce whether it is possible to form PBHs with spin in such a setting. In chapter \ref{Chapter5}, we discuss dimensionally reduced NR simulations with matter fields and its applications. 

Finally, the appendices contains supplementary material that has been omitted from the main text. 

\subsection{Primordial black holes in matter-domination}

In chapter \ref{Chapter3}, we develop a method to study PBH formation in a matter-dominated universe using real scalar fields and we apply this method to the collapse of spherically symmetric sub- and superhorizon non-linear overdensities. We find that in this setting, PBHs form via two distinctive mechanisms that we refer to as accretion and direct collapse, depending on the initial mass of the overdensity. We show that it is possible to predict via which mechanism a given overdensity collapses, with a criterion based on the hoop conjecture. This shows that in a matter-dominated universe, PBH formation is more general than in a radiation-dominated universe, in which an overdensity must collapse to a PBH directly because radiation pressure prevents efficient accretion onto the PBH. Our simulations confirm that the absence of pressure makes PBH formation more straightforward, as expected. Furthermore, we show that these PBHs have relatively small masses initially, i.e. $\mbh H_0\sim 10^{-2}\mpl^2$, as opposed to PBHs formed during a radiation-dominated period, whose mass is usually of the order of a Hubble mass. PBH formation is followed by rapid growth driven by accretion $M_\textrm{BH}\propto H^{-\beta}$ with $\beta\gg 1$, during which the PBH acquires most of its mass. This makes it harder to couple a PBH to a formation time based on its final mass and therefore potentially to a formation mechanism. On the other hand, it opens up the possibility that PBHs formed from overdensities with similar length scales end up having different masses, depending on how much each PBH accretes surrounding matter. Such an extended mass distribution can help to explain a larger part of DM today as PBHs. 

In chapter \ref{Chapter4}, we study the collapse of overdensities with angular momentum in a matter-dominated universe and we investigate how efficiently the overdensities' mass and angular momentum is transferred to the PBHs. We find mass and angular momentum transfer efficiencies of $\mathcal{O}(10\%)$ and $\mathcal{O}(5\%)$, respectively. The mass transfer efficiency is comparable to those found in chapter \ref{Chapter3}, which is to be expected since the overdensities' dimensionless spins are relatively small, peaking at $7.4 \times 10^{-2}$. The relatively low angular momentum transfer efficiency indicates that forming PBHs with large spins is non-trivial and involves the collapse of overdensities with large dimensionless spins, which would therefore deviate significantly from spherical symmetry. Such perturbations are more likely to be affected by other nonspherical effects that resist collapse, complicating the formation of high-spin PBHs further. Moreover, we show that subsequent evolution is important due to the seed PBH accreting non-rotating matter from the background, which decreases the dimensionless spin. This is expected, since it was clear from the spherically symmetric case from chapter \ref{Chapter3} that in the two-field setup it is the background field, which has no angular momentum by design, that is accreted post-formation. This picture may well be different in a changed setup, e.g. when the background field itself is perturbed, or when there are direct couplings between the two fields, in which case angular momentum could be transferred between the two fields more efficiently before PBH formation. In our case however, we argue that for PBHs with large spins to form during a matter-dominated epoch, the matter-dominated epoch cannot last too long to prevent the accretion of much non-spinning matter, and the PBHs must be highly spinning at formation, implying they are formed from highly spinning overdensities.

The fact that large overdensities are required to form PBHs with significant spins suggests that perturbative approaches are unable to tell the full story of the formation of spinning PBHs, as such large overdensities are in the nonlinear regime that perturbative calculations cannot accurately describe. Our work also shows that the evolution of the angular momentum of the collapsing region after it has decoupled from the Hubble flow cannot be ignored when determining the PBH spin at formation. Furthermore, post-formation accretion must be taken into account to determine the final PBH spin. These effects are not captured by perturbative work such as \cite{Harada:2017fjm}, but we show that NR takes them into account accurately and that NR can therefore be an important tool to uncover detailed physics in these settings. 

\subsection{Dimensional reduction with matter fields}

In chapter \ref{Chapter5}, we discuss the high computational cost of NR simulations and how these can be remedied using dimensional reduction. It should be clear that dimensional reduction comes at the cost of lessening the generality of the method, because it only allows for the simulation of processes that obey the assumes symmetries, axisymmetry around one of the Cartesian axes in our case. However, it is still useful especially for the study of processes that are conventionally studied with high degrees of symmetry, such as PBH collapse threshold or critical collapse research, as well as for simulations of head-on collisions of spherically symmetric objects, such as boson stars or vacuum bubbles. We extend the cartoon formalism presented in \cite{Cook:2016soy} by adding matter, specifically a real scalar field, and we give cartoon expressions for dimensional reduction of the CCTK method for solving the Einstein constraints. Additionally, we give explicit cartoon expressions for the evolution equations. It should be straightforward to import these expressions into working codes, so that other members of the community can immediately start building on the work presented here. Furthermore, our own cartoon codes for both the initial constraints and the evolution will become public in the near future, so that continuation of our work becomes even simpler. Finally, we discuss true vacuum bubble collision in the context of first-order phase transitions as a specific application of this work. We discuss two ways of obtaining initial field configurations for the field undergoing the phase transition and we present a convergence test showing continued stable numerical evolution of this system. We discuss why dimensional reduction is vital to obtain a code that is both numerically stable and computationally cheap. 

\section{Future work}

Regarding the work on PBH formation in a matter-dominated early universe in chapters \ref{Chapter3} and \ref{Chapter4}, it would be interesting to compare our results to those obtained from perfect pressureless fluid situations, to investigate whether the effective matter domination by oscillating scalar field method has an impact on the final results. This is of particular interest with respect to the role of angular momentum during the collapse, as it has been shown that it is important to know the equation of state precisely to predict the PBH spin \cite{saito2023spins}. Since the mass of the background scalar field introduces a length scale, it is reasonable to suspect that pressure support appears in the late stages of the collapse and the effects of such support should be investigated further. 

Moreover, the dynamics of the collapse may change if the only field present is the background field, which is perturbed itself to provide the overdensities collapsing to PBHs. We find that this can have large effects on the local expansion rate of the universe and how this influences the collapse and PBH formation requires further study. 

One could investigate whether a critical threshold $\delta_c$ for the transition between direct and accretion collapse can be defined and determined. The extent to which the presence of angular momentum affects the value of such a threshold is another open question, in matter-dominated universes as well as radiation-dominated ones. Our code is well-suited to investigate the questions in this paragraph and the previous, and would need major additions to simulate a perfect fluid coupled to GR. 

Finally, it is not clear from our simulations why there is a discrepancy between the high predicted PBH spins from perturbative calculations, see e.g. \cite{Harada:2017fjm}, and the low PBH spins that we find. In particular, further research is required to understand how an overdensity's angular momentum concentrates or disperses between the overdensity's decoupling from the Hubble flow and the formation of the PBH's horizon.

The work on dimensional reduction of NR simulations with matter fields in chapter \ref{Chapter5} is ready to be applied and we have already suggested one application, vacuum bubble collision studies in the context of first order phase transitions. The dimensional reduction framework can also be extended to other types of matter, e.g. complex scalar fields, in which case one opens up possibilities to efficiently study the behaviour of exotic compact objects, such as boson stars. For instance, boson star collisions obey the right symmetries to be studied in an axisymmetric setups, at least in the non-spinning case. In the spinning case, where both stars have identical spin, one loses the $\phi\to-\phi$ reflective symmetry and the method would need to be adapted accordingly, as discussed in \cite{Cook:2016soy}, but this can in principle still be done on a two-dimensional grid.

%% file: Appendix1/appendix1.tex

\chapter{Foliation of Einstein's equations}\label{Appendix1}

\section{Symmetries of the Riemann tensor}\label{app:riemann_symmetries}

It can be shown that the Riemann tensor $R^\mu_{\hspace*{2mm}\nu\rho\sigma}$ associated to the Levi-Civita connection $\nabla$ of metric $g_{\mu\nu}$ obeys the symmetries
\begin{subequations}
\begin{align}
    R^\mu_{\hspace*{2mm}\nu(\rho\sigma)} &= 0,\label{eqn::riem_symm_0}\\
    R^\mu_{\hspace*{2mm}[\nu\rho\sigma]} &= 0,\\
    R^\mu_{\hspace*{2mm}\nu[\rho\sigma;\lambda]} &= 0,\\
    R_{\mu\nu\rho\sigma} &= R_{\rho\sigma\mu\nu},\\
    R_{(\mu\nu)\rho\sigma} &= 0,
\end{align}
\end{subequations}
where round (square) brackets represent a symmetry (antisymmetry) operation, i.e. 
\begin{subequations}
\begin{align}
    T_{(\mu\nu)} &= \frac{1}{2}\left(T_{\mu\nu} + T_{\nu\mu}\right)\\
    T_{[\mu\nu]} &= \frac{1}{2}\left(T_{\mu\nu} - T_{\nu\mu}\right)
\end{align}
\end{subequations}
and a semicolon denotes a covariant derivative.

\section{The Lie derivative}\label{app::lie_derivative}

Taking derivatives of tensors must always be done with subtlety and care, as it necessarily involves comparing the components of tensors at two different points. The most straightforward thing that one can do is simply comparing the components of two tensors at different points in a given coordinate system, which leads to the definition of the partial derivative. However, we know that the partial derivative of a tensor does not itself possess tensor properties and is therefore not invariant under a coordinate transformation, so we need to do better. 

Let us say we are differentiating a tensor field $T$ and to this end, we are comparing its value $T_P$ at point $P$ and $T_Q$ at point $Q$. Somehow, we must find a way to map $T_P$ to a tensor $T'_Q$ at $Q$, so we can compare two tensors at $Q$ instead. One way to do so is by parallel transport, which leads to the definition of the covariant derivative, which is discussed in the main text of this thesis. 

However, there is another way to do so. Consider a nonzero vector field $X^\mu$ on manifold $\mathcal{M}$, whose integral curve can be found by integrating the ODE
\begin{equation}
    \frac{dx^\mu}{d\lambda} = X^\mu(\textbf{x}(\lambda)),
\end{equation}
where $\textbf{x}$ is used instead of $x^\mu$ for cleaner notation. These integral curves form a congruence of curves that fill the manifold, i.e. exactly one curve passes through each point on $\mathcal{M}$. 

One may also use the vector field $X^\mu$ to define the coordinate transformation 
\begin{equation}\label{eqn::lie_coord_transform}
    x'^\mu = x^\mu + \delta\lambda X^\mu(\textbf{x}),
\end{equation}
which maps $P$ to $Q$, assuming the integral curves of $X^\mu$ run from $P$ to $Q$ and the two points are infinitesimally close. This can be used to map the tensor $T_P$ to $T'_Q$, i.e. the tensor spaces at the two points are simply mapped through a coordinate transformation. We can also Taylor expand the tensor field $T$ around $P$ to find the value at $Q$. The difference between these two expressions will define the Lie derivative. 

Specifically, let us consider a rank two tensor field $T^\mu_{\hspace*{2mm}\nu}(\textbf{x})$, where $\textbf{x}$ are the coordinates of $P$, while $\textbf{x'}$ will be the coordinates of $Q$. We can coordinate transform to primed tensor $T^{\mu'}_{\hspace*{2mm}\nu'}(\textbf{x'})$ at $Q$ using the coordinate transformation generated by $X^\mu$, i.e. 
\begin{equation}
\begin{split}
    T^{\mu'}_{\hspace*{2mm}\nu'}(\textbf{x'}) &= \frac{\partial x^{\mu'}}{\partial x^\rho}\frac{\partial x^{\sigma}}{\partial x^{\nu'}}T^\rho_{\hspace*{2mm}\sigma}(\textbf{x})\\
    &= T^\mu_{\hspace*{2mm}\nu}(\textbf{x}) + \delta\lambda\left(\partial_\rho X^\mu T^\rho_{\hspace*{2mm}\nu} - \partial_\nu X^\rho T^\mu_{\hspace*{2mm}\rho} + \mathcal{O}\left(\delta\lambda^2\right)\right),
\end{split}
\end{equation}
where we have differentiated \eqn{eqn::lie_coord_transform} to obtain 
\begin{subequations}
\begin{align}
    \frac{\partial x^{\mu'}}{\partial x^\nu} &= \delta^\mu_{\hspace*{2mm}\nu} + \delta\lambda\partial_\nu X^\mu\\
    \frac{\partial x^{\mu}}{\partial x^{\nu'}} &= \delta^\mu_{\hspace*{2mm}\nu} - \delta\lambda\partial_\nu X^\mu.
\end{align}
\end{subequations}
We now apply a Taylor expansion to obtain 
\begin{equation}
    T^\mu_{\hspace*{2mm}\nu}(\textbf{x'}) = T^\mu_{\hspace*{2mm}\nu}(x^\rho + \delta\lambda X^\rho) = T^\mu_{\hspace*{2mm}\nu}(\textbf{x}) + \delta\lambda X^\sigma\partial_\sigma T^\mu_{\hspace*{2mm}\nu} + \mathcal{O}\left(\delta\lambda^2\right).
\end{equation}
These two expression together define the Lie derivative, i.e. 
\begin{equation}\label{eqn::Lie}
    \mathcal{L}_\textbf{X}T^\mu_{\hspace*{2mm}\nu} = \lim_{\delta\lambda\to 0}\frac{T^\mu_{\hspace*{2mm}\nu}(\textbf{x'}) - T^{\mu'}_{\hspace*{2mm}\nu'}(\textbf{x'})}{\delta\lambda}.
\end{equation}
We note that no connection is required for this mapping and therefore, this operation is in some sense more fundamental than the covariant derivative. For this rank two tensor, it is then easily shown that 
\begin{equation}\label{eqn::Lie_tworank}
    \mathcal{L}_\textbf{X}T^\mu_{\hspace*{2mm}\nu} = X^\rho\partial_\rho T^\mu_{\hspace*{2mm}\nu}- T^\rho_{\hspace*{2mm}\nu}\partial_\rho X^\mu + T^\mu_{\hspace*{2mm}\rho}\partial_\nu X^\rho.
\end{equation}
\eqn{eqn::Lie} is actually completely general and holds for any tensor of any rank. For a tensor of a specific rank, an expression analogous to \eqn{eqn::Lie_tworank} can be obtained through a procedure similar to the one described above. 

Finally, let's introduce a coordinate system in which the $0^\textrm{th}$ coordinate vector is aligned with $X^\mu$. In that case, the last two terms in \eqn{eqn::Lie_tworank} vanish and one obtains simply 
\begin{equation}
    \mathcal{L}_\textbf{X}T^\mu_{\hspace*{2mm}\nu} = \frac{\partial}{\partial x^0}\mathcal{L}_\textbf{X}T^\mu_{\hspace*{2mm}\nu}.
\end{equation}
This is explicitly used in section \ref{section::EFEprojection} of the main text to write the Lie derivatives of e.g. the metric and extrinsic curvature as partial time derivaties.

\section{BSSN formalism}\label{app::BSSN}

The ADM system can be adapted and reformulated to obtain a strongly hyperbolic system of equations. There are multiple ways that lead to Rome here but the formalism most relevant to the work described in this thesis is the BSSN formalism \cite{Nakamura:1987grc,Shibata:1995eot, Baumgarte:1998te}. We refer the reader to \cite{alcubierre,baumgarte_shapiro} for details on other formalisms.

The main aim of the reformulation is to create a system of PDEs that is in fact strongly hyperbolic and to this end, one must change the characteristic matrix. In the BSSN reformulation, this happens mainly in three ways. Firstly, one introduces the conformal connection coefficients $\bar{\Gamma}^i$, which will be defined below and promotes them to dynamical evolution variables. Secondly, one uses the constraint equations to replace certain terms in the evolution equations and lastly, one decomposes the ADM variables into conformal versions multiplied by a conformal factor, which will be detailed below. 

The conformal decomposition begins by writing the spatial metric as $\gamma_{ij} = \bar{\gamma}_{ij} / \chi$, i.e. as a conformal spatial metric multiplied by a conformal factor $1/\chi$. It is customary to choose the conformal factor such that the conformal metric has unit determinant, but the exact form of the conformal factor is not fixed and is sometimes taken to be $1/\chi^2$ or $\chi^4$. For $1/\chi$, it is useful to think about values of $\chi$ less than one as gravitational wells, as it can be related to the Newtonian gravitational potential for weak gravity cases by
\begin{equation}
    \chi = \big(\frac{1}{1-2\Psi}\big)^\frac{1}{4}.
\end{equation}
One separates the extrinsic curvature into its trace $K = \gamma^{ij}K_{ij}$ and traceless part $A_{ij} = K_{ij} - K\gamma_{ij}/3$ and then conformally decomposes the traceless part, $\bar{A}_{ij} = A_{ij}/\chi$. We also define the conformal connection functions $\bar{\Gamma}^i$ as contractions of the Christoffel symbols of the conformal spatial metric, i.e. 
\begin{equation}
    \bar{\Gamma}^{i}_{jk} = \frac{1}{2}\bar{\gamma}^{il}\big(\bar{\gamma}_{lj,k} + \bar{\gamma}_{lk,j} - \bar{\gamma}_{kj,l}\big),
\end{equation}
where indices after a comma indicate partial differentiation with respect to a coordinate, and 
\begin{equation}\label{eqn::conf_conn_functions}
    \bar{\Gamma}^i \equiv \bar{\gamma}^{jk}\bar{\Gamma}^i_{jk} = -\partial_j \bar{\gamma}^{ij}.
\end{equation}
The evolution equations for the conformal connections functions can be found by combining their definition with the evolution equation of the conformal spatial metric, after which the momentum constraints is used to eliminate a $\partial_j \bar{A}^{ij}$ term that appears. Finally, the $\bar{\Gamma}^i$ can be used to to calculate the three-dimensional Ricci tensor, which appears in the evolution equation of $\bar{A}_{ij}$ (it appears in the evolution equation for $K$ as well, but this is eliminated using the Hamiltonian constraint). One splits the Ricci tensor as 
\begin{equation}
    R_{ij} = R_{ij}^\chi + \bar{R}_{ij},
\end{equation}
where 
\begin{equation}
    R_{ij}^\chi = \frac{\bar{\gamma}_{ij}}{2\chi}(\bar{\gamma}^kl\bar{D}_k \bar{D}_l \chi - \frac{3}{2\chi}\bar{\gamma}^{kl}\partial_k\chi\partial_l\chi) + \frac{1}{2\chi}\big(\bar{D}_i\bar{D} - \frac{1}{2\chi}\partial_i\chi \partial_j\chi\big)
\end{equation}
and
\begin{equation}
    \bar{R}_{ij} = -\frac{1}{2}\bar{\gamma}^{kl}\partial_k\partial_l\bar{\gamma}_{ij} - \bar{\gamma}_{k(i}\partial_{j)}\bar{\Gamma}^k + \bar{\Gamma}^k\bar{\Gamma}_{(ij)k} + \bar{\gamma}^{mn}\big(2\bar{\Gamma}^k_{m(i}\bar{\Gamma}_{j)kn} + \bar{\Gamma}^k_{im}\bar{\Gamma}_{kjn}\big).
\end{equation}
In this form, it was shown that the BSSN formalism is strongly hyperbolic and therefore well-posed \cite{Sarbach:2002bt}.

It is sometimes useful to have an expression for $\Gamma^i_{jk}$ in terms of the conformal quantities, e.g. when one adds matter terms to a simulation with evolution equations include $\Gamma^i_{jk}$. For reference, if the conformal spatial metric is defined as $\gamma_{ij}\equiv \chi^{-n}\bar{\gamma}_{ij}$, then
\begin{equation}
    \Gamma^i_{jk} = \bar{\Gamma}^i_{jk} - \frac{n}{2\chi}\big(\bar{\gamma}^{il}\bar{\gamma}_{lj}\partial_k\chi + \bar{\gamma}^{il}\bar{\gamma}_{lk}\partial_j\chi - \chi\bar{\gamma}^{il}\bar{\gamma}_{kj}\partial_l\chi\big).
\end{equation}
The full set of BSSN evolution equations is then given by 
\begin{subequations}
\begin{align}
    \partial_t\chi=&\beta^k\partial_k\chi+\frac{2}{3}\chi (\alpha K-\partial_k\beta^k)
    \,, \label{eq:bssnchi}\\
    \partial_t\bar{\gamma}_{ij}
    =&\beta^k\partial_k\bar{\gamma}_{ij}-2\alpha\bar{A}_{ij}+\bar{\gamma}_{ik}
    \partial_j\beta^k+\bar{\gamma}_{jk}
    \partial_i\beta^k-\frac{2}{3}
    \bar{\gamma}_{ij}\partial_k\beta^k
    \,, \label{eq:bssngamma}\\
    \partial_t\bar{A}_{ij}=&\beta^k\partial_k\bar{A}_{ij}\chi\left\lbrace
    -D_iD_j\alpha
    +\alpha\left(\leftidx{^{(3)}}R_{ij}-
    8\pi\alpha S_{ij}\right)\right\rbrace^{TF}
    +\alpha\left(\bar{A}_{ij}K-2\bar{A}_{ik}
    \bar{A}^k_j\right) \nonumber\\
    &
    +\bar{A}_{ik}\partial_j\beta^k
    +\bar{A}_{jk}\partial_i\beta^k-\frac{2}{3}\bar{A}_{ij}\partial_k\beta^k
    \,, \label{eq:bssnA}\\
    \partial_t K=&\beta^k\partial_k K-D_i D^i\alpha+\alpha\left(\bar{A}_{ij}\bar{A}^{ij}+
    \frac{1}{3}K^2\right)+4\pi\alpha(\rho+S)
    \,, \label{eq:bssnK}\\
    \partial_t\bar{\Gamma}^i=&\beta^j\partial_j
    \bar{\Gamma}^i+\bar{\gamma}^{jk}+\bar{\gamma}^{jk}\partial_j\partial_k\beta^i+\frac{1}{3}\bar{\gamma}^{ij}
    \partial_j\partial_k\beta^k-\bar{\Gamma}^j\partial_j\beta^i+
    \frac{2}{3}\bar{\Gamma}^i\partial_j\beta^j\label{eq:bssnconf}\\
    &-2\bar{A}^{ij}\partial_j\alpha+2\alpha\left(
    \bar{\Gamma}^i_{jk}
    \bar{A}^{jk}-\frac{3}{2}\bar{A}^{ij}\frac{\partial_j\chi}{\chi}-\frac{2}{3}
    \bar{\gamma}^{ij}\partial_jK-8\pi\chi S^i
    \right). \nonumber
\end{align}
\end{subequations}

%% file: Appendix3/appendix3.tex

\chapter{Spinning primordial black holes in a matter-dominated universe}\label{Appendix3}

\graphicspath{{Appendix3/Figs/}}

\section{Apparent horizons in FLRW spacetimes}\label{Sapp::AH}

To find the AH of the formed PBH, we follow the procedure described in \cite{Thornburg:2003sf}. The AH is the outermost marginally outer-trapped surface on which the expansion of outgoing null geodesics $\Theta_+$ vanishes, i.e.
\begin{equation} \label{Spos_expansion_zero}
    \Theta_+\equiv D_i s^i + K_{ij} s^i s^j - K = 0~.
\end{equation}

In FLRW spacetimes one often encounters cosmological horizons, for which a local measure can be found in a similar manner using the expansion of ingoing null vectors $k_-^a = n^a - s^a$, equivalent to
\begin{equation} \label{Sneg_expansion_zero}
    \Theta_-\equiv -D_i s^i + K_{ij} s^i s^j - K = 0~.
\end{equation}
In the following, we will refer to $\Theta_+$ ($\Theta_{-}$) as outgoing (ingoing) expansion, and we note that it measures the fractional change in the area of an outward (inward) spherical flash of light \cite{baumgarte_shapiro}.

In general, these equations must be solved numerically, using a nonlinear root finder algorithm such as the Newton-Raphson method or quasi-Newton methods. However, to build some intuition for these quantities, it is useful to consider specific cases with a high degree of symmetry, which can be solved analytically. It can be shown \cite{baumgarte_shapiro} that for a spherically symmetric line element of the form
\begin{equation}
    ds^2 = -\alpha^2 dt^2 + \psi(t,r)^2 \left[dr^2 +r^2\left( d\theta^2+\sin^2\theta\, d\phi^2\right)\right]~,\nonumber
\end{equation}
and with $s^r=1/\psi(t,r)$, $s^\theta=s^\phi=0$, the outgoing/ingoing expansion simplifies to
\begin{equation}
    \Theta_\pm = \frac{2}{\psi}\left[\frac{\partial_t\psi}{\alpha} \pm \left(\frac{\partial_r\psi}{\psi} + \frac{1}{r}\right)\right].
\end{equation}
We find it instructive to treat some concrete examples here and we discuss the resulting expressions for a Schwarzschild, FLRW and McVittie spacetime below. We will only consider positive radius solutions to \eqn{Spos_expansion_zero} and \eqn{Sneg_expansion_zero}, regarding negative radius solutions as unphysical.

\begin{itemize}
    \item[(i)] For a Schwarzschild black hole, the conformal factor in isotropic coordinates is $\psi_\mathrm{BH}(r) = \left(1+GM/2r\right)^2$ and the lapse is $\alpha_\mathrm{BH}(r)=(1-GM/2r)/(1+GM/2r)$. The expansions are then given by
    \begin{equation}
        \Theta_\pm=\mp\frac{8\left(GM-2r\right)r}{(GM+2r)^3}~,
    \end{equation}
    which both vanish at $r=0$ and at the black hole horizon $r=GM/2$.
    
    In these coordinates, the areal radius for the Schwarzschild metric is given by $r \psi_\mathrm{BH}(t, r)$ and therefore, the area of a spherical surface is $A = 4\pi (r \psi_\mathrm{BH}(t, r))^2$ \footnote{We note that at $r=GM/2$ and with units restored, this area equals $16\pi G^2M^2$, as it should for this spacetime.}. The outgoing (ingoing) expansion can be thought of as measuring the fractional change in the area of an outward (inward) spherical flash of light. From the expression for $A$, this area decreases (increases) inside the apparent horizon and increases (decreases) outside the apparent horizon when the coordinate radius is increased (decreased). This justifies the fact that $r=GM/2$ solves both $\Theta_\pm = 0$.
    
    \item[(ii)] In an FLRW spacetime sliced by cosmic time $t$, $\psi_\mathrm{FLRW}(t) = a(t)$ and $\alpha_\mathrm{FLRW}=1$, so that 
    \begin{equation}
        \Theta_\pm=2\left(H \pm \frac{1}{ar}\right)~,
    \end{equation}
    and $\Theta_-$ vanishes at the comoving Hubble horizon $r=a^{-1}H^{-1}$. This is the point at which the universe begins to expand superluminally relative to the origin, and ingoing rays, that converge to the origin for smaller radii, end up receding from us.
    
    \item[(iii)] A black hole immersed in an FLRW background can be represented by the McVittie line element \cite{McVittie:1933zz}, for which $\alpha = \alpha_\mathrm{BH}\left(ar\right)$ and $\psi = \psi_\mathrm{FLRW}(t)\psi_\mathrm{BH}\left(ar\right)$, so that
    \begin{equation}
        \Theta_\pm=2\left(H\mp\frac{4ar\left(GM-2ar\right)}{\left(GM+2ar\right)^3}\right)~.
    \end{equation}
    When $GMH\ll 1$, as is the case in the situations we consider, $\Theta_-$ vanishes for $$r\approx\frac{1}{aH} - \frac{2GM}{a} \quad \text{and} \quad r\approx\frac{GM}{2a} + \frac{2H (GM)^2}{a}\,,$$ while $\Theta_+$ vanishes for $$r\approx\frac{GM}{2a} - \frac{2H (GM)^2}{a} \quad \text{and} \quad r\approx \frac{H(GM)^2}{4a}\,.$$ This means that an AH finder should in principle be able to find the PBH and cosmological horizon, as well as a perturbed $r=0$ solution. This is confirmed by our simulations. We are able to distinguish the two solutions for $\Theta_+ = 0$ because the physical area of the latter shrinks, rendering it unphysical.
\end{itemize}

\section{Initial data} \label{app:CTTK_check}

In section \ref{Ssect::init_data} of the main text, we allude to the fact that to solve the Einstein constraints on the initial hyperslice, we make an assumption through \eqn{eqn::cttk6} that the linear momentum density obeys $S_i = \partial_i F$ for some scalar function $F$ (the original method). Since the linear momentum densities of the initial data defined in section \ref{Ssect::init_data} do not obey this relation for any $F$, we must check that solving the constraints without this assumption, i.e. solving \eqn{eqn::cttk4} and \eqn{eqn::cttk5} as four coupled equations (the corrected method), results in the same initial configurations for the extrinsic curvature $K_{ij}$ and does not change our final results. 

\begin{figure}[t!]
    \centering
    \includegraphics[width=0.8\linewidth]{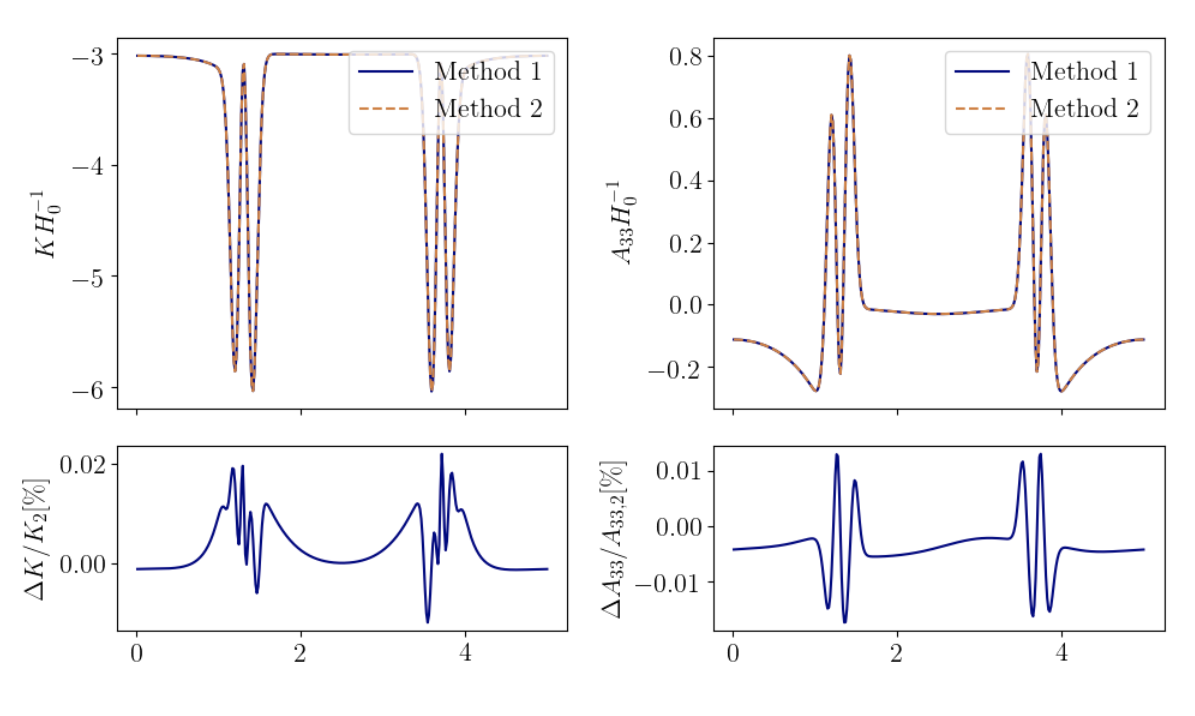}
    \vspace*{-5mm}
    \caption{\textbf{Comparison between initial data} obtained with method 1 (original) and method 2 (corrected) method. Shown in the top row are $K$ and one representative component of $A_{ij}$, along the $x$-direction through the entire simulation box and through the centre of the perturbation shell. The bottom row shows the normalised difference between the two configurations in the top row. }
    \label{fig:IC_comp}
\end{figure}

To show this, we firstly solve the Einstein constraints for a perturbation shell with $A = 0.09, R_0 = 1.2/H_0, \lambda = 0.15/H_0, B = 0.5, k = 2, \omega = 10H_0$, using both the original and the corrected method. The only initial configurations that then differ are the trace of the extrinsic curvature $K$ and the traceless part of the extrinsic curvature $A_{ij}$. We compare configurations obtained using both methods in Fig. \ref{fig:IC_comp}, which shows that the normalised difference between method 1 and 2 (i.e. the original and the corrected method respectively) is smaller than 0.02\%. Therefore, the initial conditions used in our production runs are virtually identical to the ones we would have obtained using the correct method.

\begin{figure}[t!]
    \centering
    \includegraphics[width=0.8\linewidth]{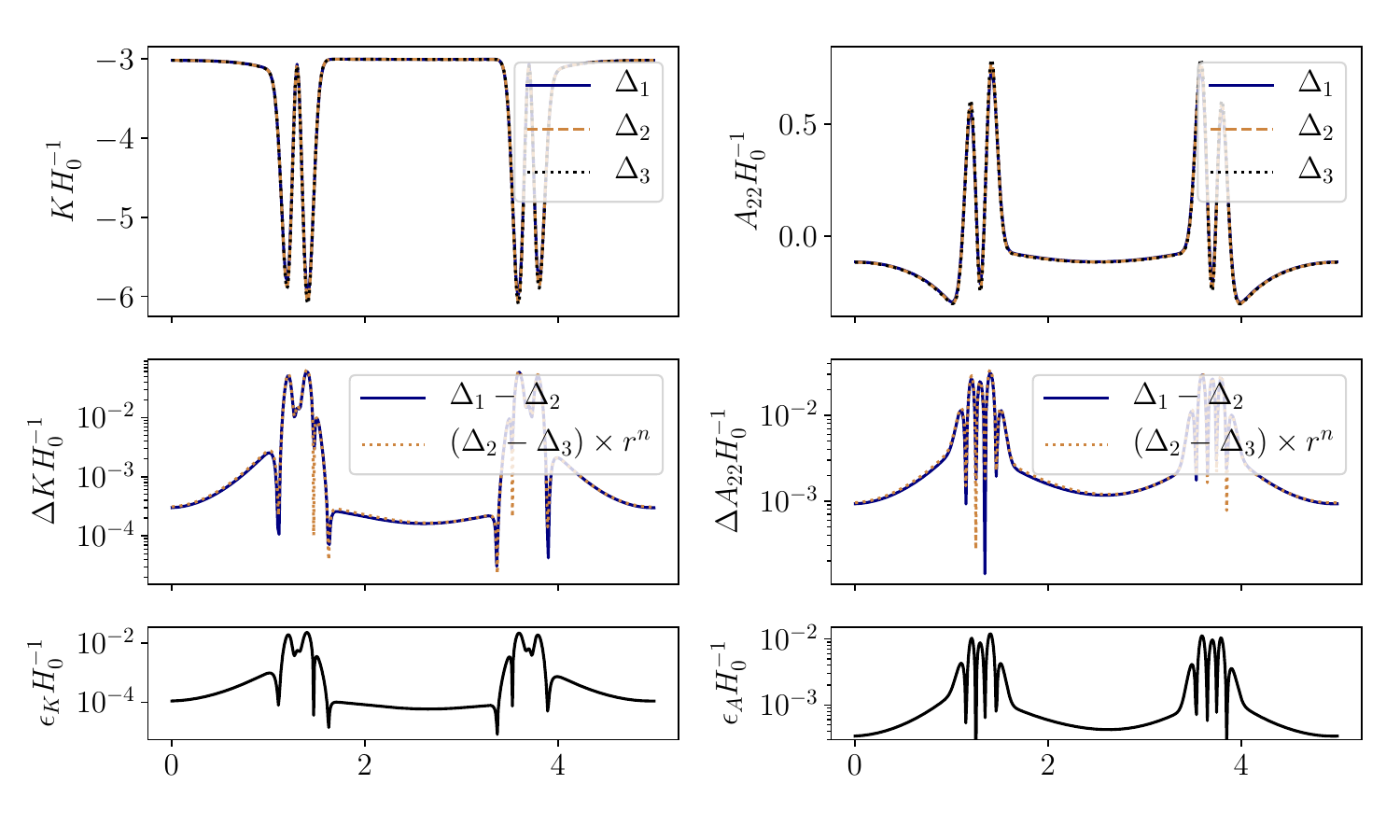}
    \vspace*{-5mm}
    \caption{\textbf{Convergence test for original method.} Shown in the top row are the configurations for $K$ and $A_{22}$ obtained at different resolutions. The middle row compares the difference between solutions at different resolutions, rescaled with respect to $r = 1.5$ and $n = 2$, while the bottom row shows an estimate of the remaining numerical solution error at the finest resolution.}
    \label{fig:IC_orig}
\end{figure}

\begin{figure}[t!]
    \centering
    \includegraphics[width=0.8\linewidth]{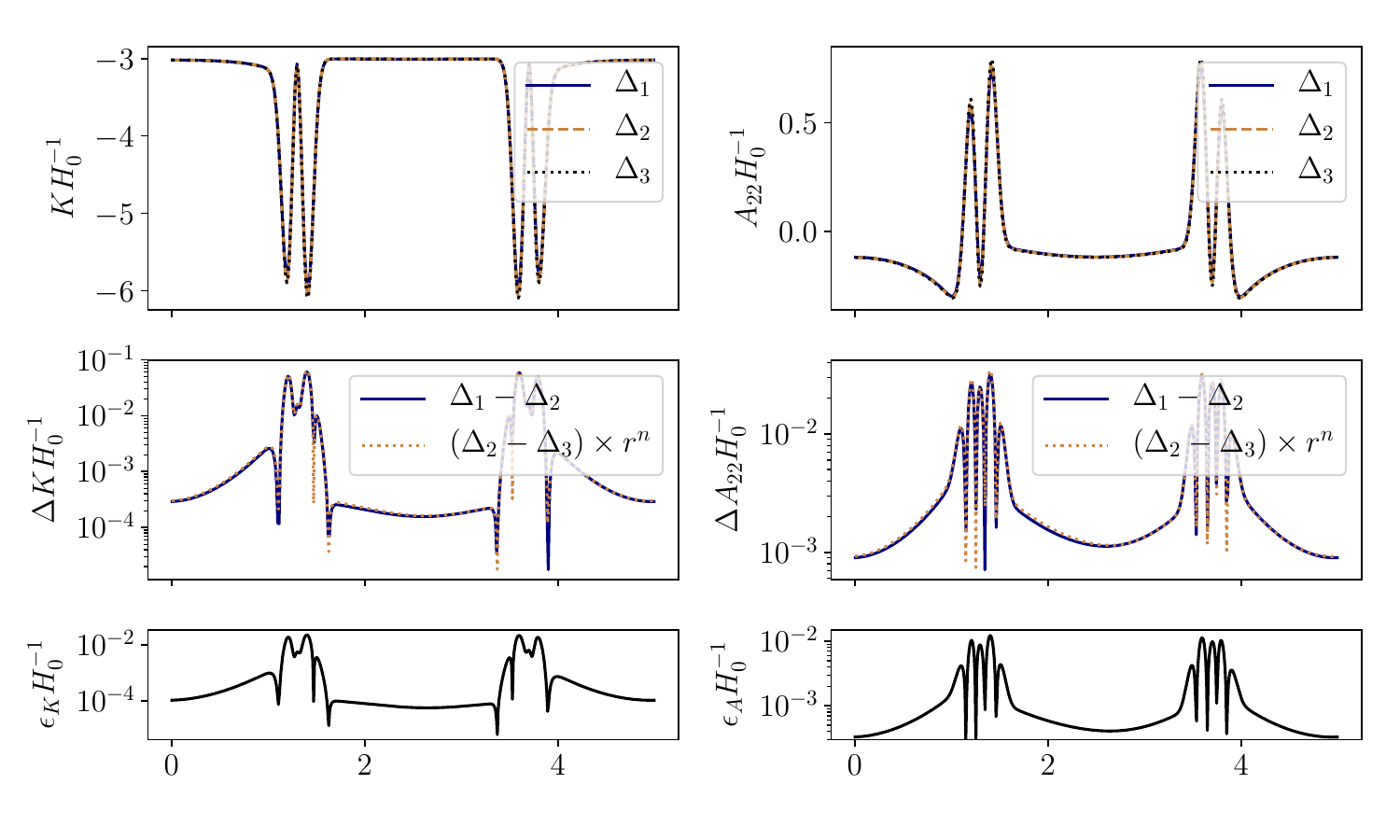}
    \vspace*{-5mm}
    \caption{\textbf{Convergence test for original method}, showing the same quantities as Fig. \ref{fig:IC_orig}.}
    \label{fig:IC_corr}
\end{figure}

Secondly, we have convergence tested both the original and corrected method. We use the same shell parameters as above, in a cubic box of length $L_\textrm{box} = 5/H_0$ and with base grid resolutions $N_1 = 256, N_2 = 384, N_3 = 576$, yielding resolutions $\Delta_1, \Delta_2$ and $\Delta_3$, so that the resolution ratio $r = \Delta_1/\Delta_2 = \Delta_2/\Delta_3 = 1.5$. Computational constraints keep us from going to higher resolutions. The results of these tests for the original (corrected) method are shown in Fig. \ref{fig:IC_orig} (\ref{fig:IC_corr}). 

It is clear from the top rows of both figures that the configurations obtained at different resolutions are very similar. In the middle rows of the two figures, we show the difference between solutions at $\Delta_1$ and $\Delta_2$ (solid blue line), as well as the difference between solutions at $\Delta_2$ and $\Delta_3$ rescaled by $r^n$ (dashed orange line), where $n$ is the desired order of convergence that we set to 2, given the stencils used in the solver. If the solver achieves second order convergence, the orange line stays on or below the blue line, which is the case for both methods. This shows that in this case, even though it is a mistake to use \eqn{eqn::cttk6}, it does not significantly affect the performance or rate of convergence achieved by the solver, suggesting that the total error is not dominated by the assumption of a curl-free $S_i$, at least at these resolutions. It would be interesting to see whether either method achieves this rate of convergence at arbitrarily high resolution, but we cannot draw conclusions concerning that question based on these results. 

The bottom panels of Fig \ref{fig:IC_orig} and Fig. \ref{fig:IC_corr} show the estimated numerical solution error at resolution $\Delta_3$, defined as $\epsilon = (u_{\Delta_2} - u_{\Delta_3})/(r^n - 1)$. These are identical between both methods and consistently small compared to the absolute values in the top rows. 

\begin{figure}[t!]
    \centering
    \includegraphics[width=0.7\linewidth]{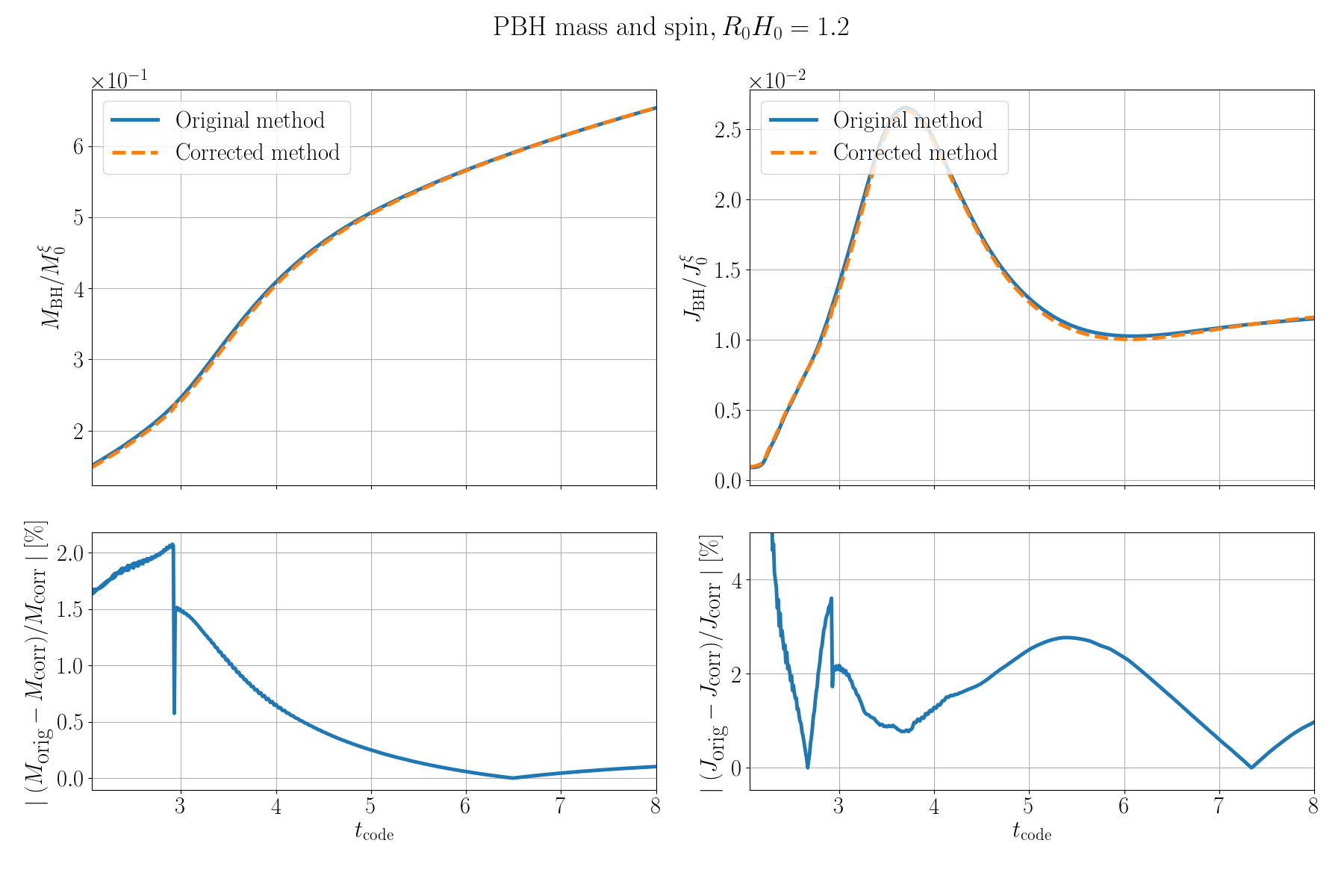}
    \vspace*{-5mm}
    \caption{Comparison between PBH mass and angular momentum, obtained with the original and corrected method. Top row show these two quantities, while the bottom row shows the normalised difference. } 
    \label{fig:evol_comp}
\end{figure}

Lastly, we have rerun a simulation using the corrected initial conditions, for the same initial conditions as above. In Fig. \ref{fig:evol_comp}, we compare the mass and angular momentum of the PBH formed from this perturbation with either method, and we show that the results are qualitatively identical and are quantitatively the same within 2\%. We note that this figure has code time on the $x$-axis. Because the $K$ configurations are initially slightly different and because our lapse depends dynamically on $K$, one cannot expect the same code time to correspond exactly to the same physical time with the original and corrected method. We suspect that this causes at least part of the differences shown in Fig. \ref{fig:evol_comp}.

We have not run new convergence tests with the corrected method, because these runs are computationally expensive. However, since the difference between the initial configurations and PBH mass and spin data is small, we are confident that new convergence tests would show the same level of convergence as the ones already present in chapter \ref{Chapter4}. These tests show convergence of PBH mass and spin, as well as convergence of the norms of the Hamiltonian and momentum constraint violation integrated over the simulation box. 

In summary, we show that in this setting, the original and corrected methods generate very similar initial configurations, that both methods show second order convergence and that using the corrected method does not significantly change the observed PBH mass and angular momentum. The conclusions presented in our paper do not change when using the corrected method for solving the Einstein constraints. 
